\documentclass[11pt,a4paper,dvipsnames]{article}

\usepackage{tikz}
\usetikzlibrary{calc,matrix,decorations.pathmorphing,decorations.markings,arrows,positioning,intersections,mindmap,backgrounds,patterns,arrows.meta,chains}

\usepackage{axodraw}

\usepackage{jstyle}
\usepackage{amsmath, amsfonts, amssymb}
\usepackage{mathtools}
\usepackage{graphicx, hyperref, verbatim}
\usepackage{xcolor}
\usepackage{float}
\usepackage{subfig}
\usepackage{caption}

\DeclareGraphicsExtensions{.pdf,.png,.jpg,.mps}
\usepackage{booktabs}
\usepackage{multirow}
\usepackage{cancel}
\usepackage{array}
\newcolumntype{P}[1]{>{\centering\arraybackslash}p{#1}}
\newcolumntype{M}[1]{>{\centering\arraybackslash}m{#1}}

\usepackage{physics}
\usepackage{diagbox}
\usepackage{pifont}
\usepackage{cleveref}
\crefname{equation}{}{}

\usetikzlibrary{calc}

\newcommand{\zb}{{\bar{z}}}

\newcommand{\bracketaa}[1]{\langle #1 \rangle}
\newcommand{\bracketas}[1]{\langle #1 ]}
\newcommand{\bracketsa}[1]{[ #1 \rangle}
\newcommand{\bracketss}[1]{[ #1 ]}

\renewcommand{\X}{\mathcal{X}}
\newcommand{\Y}{\mathcal{Y}}
\newcommand{\F}{\mathcal{F}}
\newcommand{\K}{\mathcal{K}}
\renewcommand{\H}{\mathcal{H}}

\newcommand{\Hb}{\bar{\mathcal{H}}}

\def\vb{{\bar{v}}} 

\def\a{{\alpha}} 
\def\b{{\beta}} 
\def\e{{\epsilon}} 

\def\t{\theta}

\def\da{{\dot{\alpha}}} 
\def\db{{\dot{\beta}}}

\def\be{\begin{equation}}
\def\ee{\end{equation}}
\def\ba{\begin{eqnarray}}
\def\ea{\end{eqnarray}}

\newcommand{\Ncal}{\mathcal{N}}

\newcommand{\SO}{\mathrm{SO}}
\newcommand{\SU}{\mathrm{SU}}
\newcommand{\U}{\mathrm{U}}
\newcommand{\USp}{\mathrm{USp}}

\newcommand{\midarrow}[3]{\draw [-Latex] (#1) -- ($(#1)!#2!(#3)$);}

\title{Meson correlators in 4d $\mathcal{N}=2$ SCFTs and hints for 8d structures at weak coupling}

\author[a,b]{Xi-Er Du,}
\author[c,d]{Zhongjie Huang,}
\author[c,d]{Bo Wang,}
\author[c,d]{Ellis Ye Yuan,}
\author[b]{Xinan Zhou.}

\affiliation[a]{School of Physical Sciences, University of Chinese Academy of Sciences,\\
Beijing 100049, China}
\affiliation[b]{Kavli Institute for Theoretical Sciences, University of Chinese Academy of Sciences, \\
Beijing 100190, China}
\affiliation[c]{Zhejiang Institute of Modern Physics, School of Physics, Zhejiang University, Hangzhou, \\
Zhejiang 310058, China }
\affiliation[d]{Joint Center for Quanta-to-Cosmos Physics, Zhejiang University,
Hangzhou, \\
Zhejiang 310058, China}

\emailAdd{duxier22@mails.ucas.ac.cn}
\emailAdd{zjhuang@zju.edu.cn}
\emailAdd{b\_w@zju.edu.cn}
\emailAdd{eyyuan@zju.edu.cn}
\emailAdd{xinan.zhou@ucas.ac.cn}

\abstract{We study correlators of $\frac{1}{2}$-BPS mesons in two examples of 4d SQCDs with $\mathcal{N}=2$ superconformal symmetry in the planar limit. We focus on the weakly coupled regime and obtain one-loop corrections to $n$-point meson correlators with arbitrary operator dimensions. We show that these corrections can be resumed into generating functions which exhibit emergent 8d structures similar to the ones previously observed at strong coupling via AdS/CFT.  These structures of the $\mathcal{N}=2$ theories also resemble the hidden 10d structures in 4d $\mathcal{N}=4$ SYM.}

\begin{document} 
\maketitle
\tableofcontents
	
\newpage

\section{Introduction}\label{sec:intro}
The bootstrap approach to holographic correlators, initiated in \cite{Rastelli:2016nze,Rastelli:2017udc}, has led to an impressive array of new results both at tree level \cite{Rastelli:2019gtj,Alday:2020lbp,Alday:2020dtb,Alday:2021odx} and loop levels \cite{Aprile:2017bgs,Aprile:2017xsp,Alday:2018kkw,Aprile:2019rep,Alday:2019nin,Alday:2021ajh,Huang:2021xws,Drummond:2022dxw}, for four points and higher points \cite{Goncalves:2019znr,Alday:2022lkk,Goncalves:2023oyx,Alday:2023kfm,Cao:2023cwa,Cao:2024bky,Huang:2024dxr} (see also \cite{Bissi:2022mrs} for a pedagogical review). From this rich set of examples, one observes many interesting emergent structures. These include hidden conformal symmetry \cite{Caron-Huot:2018kta}, AdS double copy \cite{Zhou:2021gnu} and Parisi-Sourlas dimensional reduction \cite{Zhou:2020ptb,Behan:2021pzk,Alday:2021odx}. However, the explorations of these structures are still very preliminary and a lot of work is needed to gain a better understanding. In this paper, we focus on the first emergent structure. 

This hidden conformal symmetry property was first discovered in four-point correlators of $\frac{1}{2}$-BPS operators in strongly coupled 4d $\mathcal{N}=4$ SYM where they are dual to tree-level scattering amplitudes of super gravitons of Type IIB supergravity in $AdS_5\times S^5$ \cite{Rastelli:2016nze,Rastelli:2017udc}. It was found in \cite{Caron-Huot:2018kta} that these correlators\footnote{More precisely, they are the ``reduced'' correlators where the consequence of superconformal symmetry has been automatically taken into account.}, quite surprisingly, can be resumed into a single generating function which is obtained by uplifting the simplest four-point correlator of operators with the lowest conformal dimension. This simplest four-point correlator depends only on $x_{ij}^2$ which are the $AdS_5$ boundary distances. The uplift is achieved by replacing $x_{ij}^2$ with $x_{ij}^2-t_{ij}$ where $t_{ij}$ are the distances on $S^5$. The combination $x_{ij}^2-t_{ij}$ can be interpreted as the conformally invariant distance in an emergent flat 10d spacetime, which gives the name to the structure. Soon after, a very similar hidden conformal symmetry was found in supergravity correlators in $AdS_3\times S^3\times K3$ \cite{Rastelli:2019gtj}, where the emergent higher dimensional spacetime is 6d. However, the origin of these hidden structures is unclear because the supergravity theories are not conformal. Adding to the mystery, it was found in \cite{Caron-Huot:2021usw} that such a 10d structure also exists in the opposite weakly coupled regime, albeit in a slightly different way. At weak coupling, loop corrections to four-point functions of $\frac{1}{2}$-BPS operators in $\mathcal{N}=4$ SYM can be computed by using the method of Lagrangian insertion \cite{Intriligator:1998ig,Eden:2011we}, which expresses these corrections as integrals. In contrast to the strong coupling case where $x_{ij}^2\to x_{ij}^2-t_{ij}$ is applied to the correlator, i.e., the integral, at weak coupling the same replacements act on the {\it integrand} given by the Lagrangian insertion method. 

It further appeared that such hidden conformal symmetry structures are not unique to theories which are gravitational in the dual bulk description. In \cite{Alday:2021odx}  hidden conformal symmetry was found in another class of strongly coupled theories which are physically very different. These are 4d $\mathcal{N}=2$ SCFTs which can be constructed either from D3 branes with probe D7 branes \cite{Karch:2002sh}, or D3 branes probing F-theory 7-brane singularities \cite{Fayyazuddin:1998fb,Aharony:1998xz}. At strong coupling and for large central charges, all these theories contain a decoupling sectors where correlators of $\frac{1}{2}$-BPS operators are dual to scattering amplitudes of super gluons in an $AdS_5\times S^3$ subspace. The important difference from the previous examples is that gravity decouples in this limit and the AdS theories are just SYM. Tree-level four-point functions in these theories were systematically computed in \cite{Alday:2021odx} and were found to be organized by a hidden 8d conformal symmetry.\footnote{Hints of 8d structures have recently also been found in tree-level five-point functions \cite{Huang:2024dxr}.} Note that some of these 4d $\mathcal{N}=2$ SCFTs admit Lagrangian descriptions as supersymmetric gauge theories just as $\mathcal{N}=4$ SYM. Therefore, it is natural to ask if similar 8d structures can be found at weak coupling for these theories as well. 

In this paper, we study correlation functions at weak coupling in two examples of such 4d $\mathcal{N}=2$ superconformal Lagrangian theories and find an affirmative answer at one loop. These two theories are 
\begin{itemize}
    \item $\mathcal{T}_1$: 4d $\mathcal{N}=4$ SYM with $\SU(N)$ gauge group coupled to $N_f$ $\mathcal{N}=2$ hypermultiplets in the fundamental representation. This theory is the low energy description of $N$ coincident D3 branes with $N_f$ D7 branes and has an $\SU(N_f)$ flavor symmetry. 
    \item $\mathcal{T}_2$: A 4d $\mathcal{N}=2$ $\USp(2N)$ gauge theory with four fundamental hypermultiplets and an antisymmetric hypermultiplet. This theory has an $\SO(8)$ flavor symmetry. It can be constructed with four D7 branes, an O7 plane and $N$ D3 branes on a $D_4$ singularity. We will often refer to this theory as the $D_4$ theory.  
\end{itemize}
A few comments are in order about these theories. The theory $\mathcal{T}_1$ clearly is the simplest but it is conformal only in the limit  $N_f/N\to 0$. This is because the beta function of the 't Hooft coupling is non-vanishing but is proportional to $N_f/N$. In the limit where $N\to \infty$ and $N_f$ is fixed (so that the D7 branes are probes), the beta function is zero and the theory is conformal. From the field theory perspective, the $\mathcal{N}=2$ matter are not allowed to run in Feynman diagram loops and are therefore ``quenched''. By contrast, the theory $\mathcal{T}_2$ is exactly marginal even when $N$ is finite. However, in this class of theories constructed from D3 branes on F-theory singularities, a marginal coupling is not always guaranteed and depends on the type of the singularity.\footnote{The other  singularities correspond to Argyres-Douglas type and Minahan-Nemeschansky type CFTs, and were recently discussed in \cite{Behan:2024vwg}.} For our question to make sense, we will focus on the case with a weak coupling limit which is the $D_4$ theory.\footnote{This theory has also been subject to recent studies in relation to stringy corrections at strong coupling using supersymmetric localization techniques \cite{Behan:2023fqq,Alday:2024yax}.} 

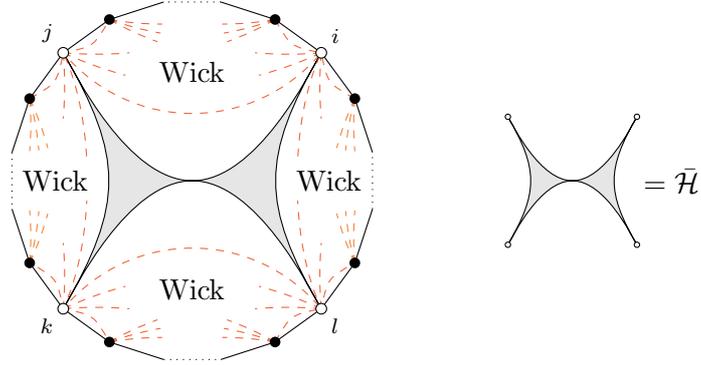
\begin{figure}[ht]
    \centering
    \begin{tikzpicture}
        \def\RR{2.4}
        \coordinate (m0) at (9:\RR);
        \coordinate (m1) at (27:\RR);
        \coordinate [label=45:{\scriptsize $i$}] (m2) at (45:\RR);
        \coordinate (m3) at (45+18:\RR);
        \coordinate (m4) at (45+2*18:\RR);
        \coordinate (m5) at (135-2*18:\RR);
        \coordinate (m6) at (135-18:\RR);
        \coordinate [label=135:{\scriptsize $j$}] (m7) at (135:\RR);
        \coordinate (m8) at (135+18:\RR);
        \coordinate (m9) at (135+2*18:\RR);
        \coordinate (m10) at (225-2*18:\RR);
        \coordinate (m11) at (225-18:\RR);
        \coordinate [label=225:{\scriptsize $k$}] (m12) at (225:\RR);
        \coordinate (m13) at (225+18:\RR);
        \coordinate (m14) at (225+2*18:\RR);
        \coordinate (m15) at (315-2*18:\RR);
        \coordinate (m16) at (315-18:\RR);
        \coordinate [label=315:{\scriptsize $l$}] (m17) at (315:\RR);
        \coordinate (m18) at (315+18:\RR);
        \coordinate (m19) at (315+2*18:\RR);
        \coordinate (l) at ($(m7)!.5!(m12)$);
        \coordinate (r) at ($(m2)!.5!(m17)$);
        \coordinate (o) at ($(l)!.5!(r)$);
        \draw (m0) -- (m1) -- (m2) -- (m3) -- (m4) (m5) -- (m6) -- (m7) -- (m8) -- (m9) (m10) -- (m11) -- (m12) -- (m13) -- (m14) (m15) -- (m16) -- (m17) -- (m18) -- (m19);
        \draw [dotted] (m4) -- (m5) (m9) -- (m10) (m14) -- (m15) (m19) -- (m0);
        \draw [thin,fill=gray!20] (m2) .. controls +(-120:.3) and ($(o)!.5!(r)$) .. (o) .. controls ($(o)!.5!(l)$) and +(-60:.3) .. (m7) .. controls +(-60:1.6) and +(60:1.6) .. (m12) .. controls +(60:.3) and ($(o)!.5!(l)$) .. (o) .. controls ($(o)!.5!(r)$) and +(120:.3) .. (m17) .. controls +(120:1.6) and +(-120:1.6) .. cycle;
        \draw [RedOrange,dashed] (m1) .. controls (45-9:\RR-.2) .. (m2) .. controls +(-115:1) and +(115:1) .. (m17) .. controls (-45+9:\RR-.2) .. (m18)  (m2) arc [start angle=-185,end angle=-175,radius=5]  (m17) arc [start angle=185,end angle=175,radius=5];
        \foreach \i in {155,165,175} \draw [Orange,dashed] (m1) arc [start angle=\i,end angle=\i+10,radius=4] (m18) arc [start angle=-\i,end angle=-\i-10,radius=4];
        \draw [RedOrange,dashed] (m8) .. controls (135+9:\RR-.2) .. (m7) .. controls +(-65:1) and +(65:1) .. (m12) .. controls (-135-9:\RR-.2) .. (m11)  (m7) arc [start angle=5,end angle=-5,radius=5]  (m12) arc [start angle=-5,end angle=5,radius=5];
        \foreach \i in {25,15,5} \draw [Orange,dashed] (m8) arc [start angle=\i,end angle=\i-10,radius=4] (m11) arc [start angle=-\i,end angle=-\i+10,radius=4];
        \node [fill=white] at (0.75*\RR,0) {Wick};
        \node [fill=white] at (-0.75*\RR,0) {Wick};
        \draw [RedOrange,dashed] (m3) .. controls (45+9:\RR-.2) .. (m2) .. controls +(-130:1.4) and +(-50:1.4) .. (m7) .. controls (135-9:\RR-.2) .. (m6);
        \foreach \i in {65,85} \draw [RedOrange,dashed] (m2) arc [start angle=-\i,end angle=-\i-10,radius=5] (m7) arc [start angle=-180+\i,end angle=-170+\i,radius=5];
        \foreach \i in {60,70,80} \draw [RedOrange,dashed] (m3) arc [start angle=-\i,end angle=-\i-10,radius=4] (m6) arc [start angle=-180+\i,end angle=-170+\i,radius=4];
        \draw [RedOrange,dashed] (m16) .. controls (-45-9:\RR-.2) .. (m17) .. controls +(130:1.4) and +(50:1.4) .. (m12) .. controls (-135+9:\RR-.2) .. (m13);
        \foreach \i in {65,85} \draw [RedOrange,dashed] (m17) arc [start angle=\i,end angle=\i+10,radius=5] (m12) arc [start angle=180-\i,end angle=170-\i,radius=5];
        \foreach \i in {60,70,80} \draw [RedOrange,dashed] (m16) arc [start angle=\i,end angle=\i+10,radius=4] (m13) arc [start angle=180-\i,end angle=170-\i,radius=4];
        \node [fill=white] at (0,0.6*\RR) {Wick};
        \node [fill=white] at (0,-0.6*\RR) {Wick};
        \foreach \i in {2,7,12,17} \draw [fill=white] (m\i) circle [radius=2pt];
        \foreach \i in {1,3,6,8,11,13,16,18} \fill (m\i) circle [radius=2pt];
        \begin{scope}[scale=.5,xshift=10cm]
            \coordinate (m2) at (45:\RR);
            \coordinate (m7) at (135:\RR);
            \coordinate (m12) at (225:\RR);
            \coordinate (m17) at (315:\RR);
            \coordinate (l) at ($(m7)!.5!(m12)$);
            \coordinate (r) at ($(m2)!.5!(m17)$);
            \coordinate (o) at ($(l)!.5!(r)$);
            \draw [thin,fill=gray!20] (m2) .. controls +(-120:.3) and ($(o)!.5!(r)$) .. (o) .. controls ($(o)!.5!(l)$) and +(-60:.3) .. (m7) .. controls +(-60:1.6) and +(60:1.6) .. (m12) .. controls +(60:.3) and ($(o)!.5!(l)$) .. (o) .. controls ($(o)!.5!(r)$) and +(120:.3) .. (m17) .. controls +(120:1.6) and +(-120:1.6) .. cycle;
            \foreach \i in {2,7,12,17} \draw [fill=white] (m\i) circle [radius=2pt];
        \end{scope}
        \node [anchor=west] at (5.8,0) {$=\Hb$};
    \end{tikzpicture}
    \caption{Structure of one-loop corrections to an $n$-point meson correlator.}
    \label{fig:intro}
\end{figure}

We will consider correlators of $\frac{1}{2}$-BPS ``mesons'' operators which are dual to the aforementioned AdS super gluons but now reinterpreted in the field theory parlance. They are of the form $q(\ldots)\bar{q}$, where $q$, $\bar{q}$ are ``scalar quarks'' coming from the matter hypermultiplets and in $\ldots$ we can insert arbitrarily many ``gluon fields'' $\phi$ from the vector multiplet. Therefore, there are infinitely many such operators with arbitrarily large conformal dimensions. These meson operators should be distinguished from the ``gluon states'' which are constructed by taking the product of gluon fields and their super partners and then taking the trace with respect to the color group. The latter are dual to super gravitons in AdS. However, the Feynman diagram computation of one-loop corrections to meson operators turns out to be quite similar to that of gluon state operators in the pure $\mathcal{N}=4$ SYM, as was done in \cite{Drukker:2008pi}. In fact, some of the techniques used there will also be useful in our problem. For the two theories $\mathcal{T}_1$ and $\mathcal{T}_2$, we find that one-loop corrections are essentially identical. The corrections can be written in a compact form diagrammatically illustrated in \Cref{fig:intro}.

Here the one-loop interaction, represented by the shaded region $\Hb$, involves four points at a time. The function $\Hb$ is proportional to the scalar one-loop box diagram in four dimensions which also appeared in the $\mathcal{N}=4$ SYM case. The insertions are within the polygon of which the vertices are the meson operators and divide the polygon into four white regions. The contributions from the other fields in the meson operators are effectively captured by the product of $q$, $\bar{q}$ contractions along the boundary and the planar Wick contractions of the $\phi$ field inside the white regions. It turns out that very remarkably the integrands of the one-loop corrections of correlators of different operators can be resumed into a generating function. In the generating function, the spacetime distances $x_{ij}^2$ and internal distances $t_{ij}$ again regroup into the combination $x_{ij}^2-t_{ij}$, thus manifesting the hidden 8d structure in the strong coupling limit. This gives rise to a parallel story to $\mathcal{N}=4$ SYM but now for $\mathcal{N}=2$ gauge theories. 

The rest of the paper is organized as follows. We start with the simpler theory of $\mathcal{N}=4$ SYM with $\mathcal{N}=2$ matter. In Section \ref{Sec:N=4withN=2general}, we review the basic ingredients and discuss the general properties of meson correlators. In Section \ref{Sec:4ptfun}, we consider four-point functions in $\mathcal{N}=4$ SYM with $\mathcal{N}=2$ matter and introduce the basic strategy for computing one-loop corrections. We find that these corrections can be resumed into a generating function manifesting 8d structures. We go to higher points in Section \ref{Sec:higherptfun} where we find similar higher dimensional structures persist. We also give the one-loop correction to general $n$-point correlators. In Section \ref{Sec:D4theory}, we study the $D_4$ theory and find the same result. We conclude in Section \ref{Sec:discussion} with a brief discussion of future directions.

\section{$\mathcal{N}=4$ SYM coupled to $\mathcal{N}=2$ matter and meson correlators}\label{Sec:N=4withN=2general}

\subsection{Lagrangian and Feynman rules}\label{subSec:N=4withN=2}

The theory we will first consider is $\mathcal{N}=4$ SYM coupled to $N_f$ copies of $\mathcal{N}=2$ hypermultiplets. These hypermultiplets are in the fundamental representation of the gauge group $\SU(N)$ and serve as matter fields coupled to SYM. They break the original R-symmetry group $\SU(4)_R$ into $\SU(2)_L \times \SU(2)_R \times \U(1)_r$, where the $\SU(2)_L$ part becomes a flavor group, and the remaining R-symmetry group is $\SU(2)_R \times \U(1)_r$. To be concrete, we list the component fields in the original $\mathcal{N}=4$ multiplet $V$ and the new $\mathcal{N}=2$ hypermultiplets $H$
\begin{align}  
    V =&\ (\phi^I, \lambda^i, \bar{\lambda}_i, A_\mu)\,,\qquad \qquad I=1,2,...,6,\qquad i=1,2,3,4, \\
    H =&\ (q^{n,a},\psi^n,\bar{\chi}^n)\,,\ \qquad\qquad n=1,2,...,N_f,\quad a=1,2,
\end{align}
where $q$ are complex scalars and $\psi$, $\bar{\chi}$ are Weyl fermions. We have also suppressed the gauge indices for notational simplicity.  
Since the R-symmetry is broken to $\SU(2)_L \times \SU(2)_R \times \U(1)_r$, the fields also split into smaller representations with respect to the subgroups
\begin{align}
    {\bf 4} \to {\bf 2}_R+ {\bf 2}_L,\qquad\qquad {\bf 6} \to {\bf 1}+{\bf 1}+{\bf 2}_R\times{\bf 2}_L.
\end{align}
Accordingly, the $\SU(4)_R$ and $\SO(6)_R$ indices of the original R-symmetry group split as   
\begin{align}\label{eq:labelsplit}
    i = {\overbracket[.5pt]{1,2}^a},{\overbracket[.5pt]{3,4}^{\bar{a}}}\;,\qquad\qquad I={\overbracket[.5pt]{1,2}^{\bar{A}}},{\overbracket[.5pt]{3,4,5,6}^A}\;.
\end{align}
Here $a$ and $\bar{a}$ are $\SU(2)_R$ and $\SU(2)_L$ indices respectively. The index $\bar{A}$ is not charged under the $\SU(2)_L\times \SU(2)_R$, while $A$ can be traded for the $a$ and $\bar{a}$ indices using Pauli matrices.

One can work out the explicit details of the Lagrangian of this theory using $\mathcal{N}=1$ superfields. As we will see later, only the bosonic part of the Lagrangian is relevant for the one-loop corrections to the conformal correlators. We record the result here (in Euclidean signature) and leave the details in Appendix \ref{derive N=2} 
\begin{align}\label{previousL}
    \mathcal{L} =&\ \frac{2N}{\lambda}\left\{ \tr\left[\frac{1}{4}F^{\mu\nu}F_{\mu\nu} + \frac{1}{2}D^\mu \phi^I D_\mu \phi^I - \frac{1}{4}{\tr}[\phi^I,\phi^J]^2 \right] + D^{\mu}\bar{q}_{n,a} D_\mu q^{n,a}\right. \nonumber\\[1mm]
    & +\bar{q}_{n,a} \{ \varphi,\varphi^\dagger \} q^{n,a} + \bar{q}_{n,a} [\phi^{a\bar{a}},\phi_{\bar{a}b}]q^{n,b} + (\bar{q}_{m,a} q^{n,a}) (\bar{q}_{n,b} q^{m,b})- \frac{1}{2}(\bar{q}_{m,a} q^{n,b})(\bar{q}_{n,b} q^{m,a})\nonumber\\
    & \left.  - \frac{1}{N}\left((\bar{q}_{m,a} q^{m,b})(\bar{q}_{n,b} q^{n,a}) - \frac{1}{2} (\bar{q}_{m,a} q^{m,a})(\bar{q}_{n,b} q^{n,b}) \right)+(\textrm{\small{fermionic part}})\right\},
\end{align}
where
$ \lambda \equiv g_{\rm YM}^2 N$ and we defined
\begin{align}
    \varphi = \frac{1}{\sqrt{2}}(\phi^1+i\phi^2)\, .
\end{align}
We also defined
\begin{equation}
    \phi^{a\bar{a}}=(\sigma^A)^{a\bar{a}}\phi^A\;,\quad A=1,2,3,4\;,
\end{equation}
where $\sigma^A$ are the (Euclidean) Pauli matrices and $a=1,2$, $\bar{a}=1,2$ are indices of $\SU(2)_R$ and $\SU(2)_L$ correspondingly.\footnote{Note we relabel the values for $\bar{a}$ and $A$ as compared to \eqref{eq:labelsplit}, for convenience.}  The transformation properties of the bosonic fields under various symmetry groups are summarized in Table  \ref{pfield}. We also list the Feynman rules in Table \ref{Feynmanp}.

\begin{table}[t]
\begin{center}
\begin{tabular}{|c|c|c|c|c|c|c|}
\hline
\textrm{Component} & Lorentz & $\SU(2)_R$ &$\SU(2)_L$ & $\U(1)_r$ & $\SU(N)$ & $\SU(N_f)$ \\ \hline
$A_\mu$ & vector & 1 & 1 & $0$ & $ N^2-1$ &1 \\
$\phi^A$& scalar & 2 & 2  & $0$ & $ N^2-1$ & 1\\
$\varphi$& scalar & 1 & 1  & $+2$ & $ N^2-1$ & 1\\ 
$q^{n,a}$& scalar & 2 & 1  & $0$ & $ N$ & $N_f$\\\hline
\end{tabular}
\caption{Bosonic component fields in the $\Ncal=2$ SCFT constructed from coupling $\mathcal{N}=4$ SYM to $\mathcal{N}=2$ matter. 
}
\label{pfield}
\end{center}
\end{table}

\begin{table}[t]\label{vertices}
\begin{center}
\begin{tabular}{|M{4cm}|M{8.5cm}|} \hline
    Schematic form & Rule \\ \hline
    \begin{tikzpicture}
        \path (-2,-0.6) rectangle +(4,1.2);
        \coordinate [label=left:{\small $\phi^{a\bar{a}}$}] (a) at (-1,0);
        \coordinate [label=right:{\small $\phi_{\bar{b}b}$}] (b) at (1,0);
        \draw [dashed,thick] (a) -- (b);
        \draw [thick,BrickRed,-Latex] ($(a)+(0.3,0.2)$) -- ($(b)+(-0.3,0.2)$);
        \draw [thick,BrickRed,-Latex] ($(b)+(-0.3,-0.2)$) -- ($(a)+(0.3,-0.2)$);
        \fill (a) circle [radius=2pt];
        \fill (b) circle [radius=2pt];
        \node [anchor=north] at (a) {\small $x_1$};
        \node [anchor=north] at (b) {\small $x_2$};
    \end{tikzpicture}
    &$\displaystyle {\lambda\over 2N} \frac{\delta_b^a \delta_{\bar{b}}^{\bar{a}}}{x_{12}^2}$\\
    \begin{tikzpicture}
        \path (-2,-0.6) rectangle +(4,1.2);
        \coordinate [label=left:{\small $q^{b}$}] (b) at (-1,0);
        \coordinate [label=right:{\small $\bar{q}_{a}$}] (a) at (1,0);
        \draw [thick] (b) -- (a);
        \draw [thick,-Latex] (b) -- ($(b)!0.6!(a)$);
        \draw [thick,RoyalBlue,-Latex] ($(b)+(0.3,0.2)$) -- ($(a)+(-0.3,0.2)$);
        \draw [thick,BrickRed,-Latex] ($(b)+(0.3,-0.2)$) -- ($(a)+(-0.3,-0.2)$);
        \fill (a) circle [radius=2pt];
        \fill (b) circle [radius=2pt];
        \node [anchor=north] at (a) {\small $x_1$};
        \node [anchor=north] at (b) {\small $x_2$};
    \end{tikzpicture}
    &$\displaystyle {\lambda\over 2 N} \frac{\delta_a^b}{x_{12}^2}$\\
    \begin{tikzpicture}
        \path (-2,-0.6) rectangle +(4,1.2);
        \coordinate [label=left:{\small $A_\mu$}] (a) at (-1,0);
        \coordinate [label=right:{\small $A_\nu$}] (b) at (1,0);
        \draw [thick,decorate,decoration={snake}] (a) -- (b);
        \draw [thick,BrickRed,-Latex] ($(a)+(0.3,0.2)$) -- ($(b)+(-0.3,0.2)$);
        \draw [thick,BrickRed,-Latex] ($(b)+(-0.3,-0.2)$) -- ($(a)+(0.3,-0.2)$);
        \fill (a) circle [radius=2pt];
        \fill (b) circle [radius=2pt];
        \node [anchor=north] at (a) {\small $x_1$};
        \node [anchor=north] at (b) {\small $x_2$};
    \end{tikzpicture}
    &$\displaystyle {\lambda\over 2 N} \frac{\eta_{\mu\nu}}{x_{12}^2}$\\
    \begin{tikzpicture}
        \path (-2,-0.6) rectangle +(4,1.2);
        \coordinate [label=left:{\small $\bar{q}_{a}$}] (a) at (120:1);
        \coordinate [label=left:{\small $q^{b}$}] (b) at (-120:1);
        \coordinate [label=right:{\small $A_\mu$}] (c) at (0:1);
        \coordinate (o) at (0,0);
        \draw [thick] (a) -- (o) -- (b);
        \draw [thick,-Latex] (b) -- ($(b)!0.7!(o)$);
        \draw [thick,-Latex] (o) -- ($(o)!0.7!(a)$);
        \draw [thick,decorate,decoration={snake}] (o) -- (c);
        \fill (o) circle [radius=2pt];
        \draw [thick,RoyalBlue,-Latex] (-130:0.9) .. controls +(60:0.75) and +(-60:0.75) .. (130:0.9);
        \draw [thick,BrickRed,-Latex] (10:0.9) .. controls +(180:0.75) and +(-60:0.75) .. (110:0.9);
        \draw [thick,BrickRed,-Latex] (-110:0.9) .. controls +(60:0.75) and +(180:0.75) .. (-10:0.9);
    \end{tikzpicture}
    &$\displaystyle i {2N\over\lambda} \int d^4 w \left(\partial^{\bar{q}}_\mu- \partial^q_\mu \right) \delta_a^b$\\ 
      \begin{tikzpicture}
        \path (-2,-0.6) rectangle +(4,1.2);
        \coordinate [label=left:{\small $\phi^{a\bar{a}}$}] (a) at (120:1);
        \coordinate [label=left:{\small $\phi_{\bar{b}b}$}] (b) at (-120:1);
        \coordinate [label=right:{\small $A_\mu$}] (c) at (0:1);
        \coordinate (o) at (0,0);
        \draw [thick,dashed] (a) -- (o) -- (b);
        \draw [thick,dashed] (b) -- ($(b)!0.7!(o)$);
        \draw [thick,dashed] (o) -- ($(o)!0.7!(a)$);
        \draw [thick,decorate,decoration={snake}] (o) -- (c);
        \fill (o) circle [radius=2pt];
        \draw [thick,BrickRed,-Latex] (130:0.9) .. controls +(-60:0.75) and +(60:0.75) .. (-130:0.9);
        \draw [thick,BrickRed,-Latex] (10:0.9) .. controls +(180:0.75) and +(-60:0.75) .. (110:0.9) node [anchor=south] {\scriptsize\color{black} $1$};
        \draw [thick,BrickRed,-Latex] (-110:0.9) node [anchor=north] {\scriptsize\color{black} $2$} .. controls +(60:0.75) and +(180:0.75) .. (-10:0.9);
    \end{tikzpicture}
    &$\displaystyle i {2N\over\lambda} \int d^4 w \left(\partial_{1,\mu}- \partial_{2,\mu} \right) \delta_a^b\delta^{\bar{a}}_{\bar{b}}$\\ 
    \begin{tikzpicture}
        \path (-2,-0.6) rectangle +(4,1.2);
        \coordinate [label=left:{\small $\bar{q}_{a}$}] (a) at (135:1);
        \coordinate [label=left:{\small $\phi^{b\bar{b}}$}] (b) at (-135:1);
        \coordinate [label=right:{\small $\phi_{\bar{c}c}$}] (c) at (-45:1);
        \coordinate [label=right:{\small $q^{d}$}] (d) at (45:1);
        \coordinate (o) at (0,0);
        \draw [thick,dashed] (b) -- (o) -- (c);
        \draw [thick] (a) -- (o) -- (d);
        \draw [thick,-Latex] (d) -- ($(d)!0.7!(o)$);
        \draw [thick,-Latex] (o) -- ($(o)!0.7!(a)$);
        \fill (o) circle [radius=2pt];
        \draw [thick,RoyalBlue,-Latex] (55:0.9) .. controls +(-135:0.7) and +(-45:0.7) .. (125:0.9);
        \draw [thick,BrickRed,-Latex] (-145:0.9) .. controls +(45:0.7) and +(-45:0.7) .. (145:0.9);
        \draw [thick,BrickRed,-Latex] (-55:0.9) .. controls +(135:0.7) and +(45:0.7) .. (-125:0.9);
        \draw [thick,BrickRed,-Latex] (35:0.9) .. controls +(-135:0.7) and +(135:0.7) .. (-35:0.9);
    \end{tikzpicture}
    & $\displaystyle {2N\over\lambda}  \int d^4 w \, \left( \delta_a^b\delta_c^d-\epsilon_{a c}\epsilon^{b d} \right) \delta_{\bar{c}}^{\bar{b}}$\\
    \begin{tikzpicture}
        \path (-2,-0.6) rectangle +(4,1.2);
        \coordinate [label=left:{\small $\bar{q}_{a}$}] (a) at (135:1);
        \coordinate [label=left:{\small $q^{b}$}] (b) at (-135:1);
        \coordinate [label=right:{\small $\bar{q}_{c}$}] (c) at (-45:1);
        \coordinate [label=right:{\small $q^{d}$}] (d) at (45:1);
        \coordinate (o) at (0,0);
        \draw [thick] (b) -- (o) -- (c);
        \draw [thick] (a) -- (o) -- (d);
        \draw [thick,-Latex] (d) -- ($(d)!0.7!(o)$);
        \draw [thick,-Latex] (o) -- ($(o)!0.7!(a)$);
        \draw [thick,-Latex] (b) -- ($(b)!0.7!(o)$);
        \draw [thick,-Latex] (o) -- ($(o)!0.7!(c)$);
        \fill (o) circle [radius=2pt];
        \draw [thick,RoyalBlue,-Latex] (55:0.9) .. controls +(-135:0.7) and +(-45:0.7) .. (125:0.9);
        \draw [thick,BrickRed,-Latex] (-145:0.9) .. controls +(45:0.7) and +(-45:0.7) .. (145:0.9);
        \draw [thick,RoyalBlue,-Latex] (-125:0.9) .. controls +(45:0.7) and +(135:0.7) .. (-55:0.9);
        \draw [thick,BrickRed,-Latex] (35:0.9) .. controls +(-135:0.7) and +(135:0.7) .. (-35:0.9);
    \end{tikzpicture}
    &\begin{tabular}{l}
        $\displaystyle {4N\over\lambda}  \int d^4 w \, \left(\delta_a^b\delta_c^d-\frac{1}{2}\delta_a^d \delta^b_c \right)$\\
    \end{tabular}\\
    \begin{tikzpicture}
        \path (-2,-0.6) rectangle +(4,1.2);
        \coordinate [label=left:{\small $\phi^{a\bar{a}}$}] (a) at (135:1);
        \coordinate [label=left:{\small $\phi^{b\bar{b}}$}] (b) at (-135:1);
        \coordinate [label=right:{\small $\phi_{\bar{c}c}$}] (c) at (-45:1);
        \coordinate [label=right:{\small $\phi_{\bar{d}d}$}] (d) at (45:1);
        \coordinate (o) at (0,0);
        \draw [thick,dashed] (b) -- (o) -- (c);
        \draw [thick,dashed] (a) -- (o) -- (d);
        \fill (o) circle [radius=2pt];
        \draw [thick,BrickRed,-Latex] (125:0.9) .. controls +(-45:0.7) and +(-135:0.7) .. (55:0.9);
        \draw [thick,BrickRed,-Latex] (-145:0.9) .. controls +(45:0.7) and +(-45:0.7) .. (145:0.9);
        \draw [thick,BrickRed,-Latex] (-55:0.9) .. controls +(135:0.7) and +(45:0.7) .. (-125:0.9);
        \draw [thick,BrickRed,-Latex] (35:0.9) .. controls +(-135:0.7) and +(135:0.7) .. (-35:0.9);
    \end{tikzpicture}
    &$\displaystyle -{2N\over\lambda}  \int d^4 w \, \left( 2\delta^a_c\delta^{\bar{a}}_{\bar{c}}\delta^b_d\delta^{\bar{b}}_{\bar{d}}-\delta^a_d\delta^{\bar{a}}_{\bar{d}}\delta^b_c\delta^{\bar{b}}_{\bar{c}}-\epsilon^{a b}\epsilon^{\bar{a}\bar{b}}\epsilon_{c d}\epsilon_{\bar{c}\bar{d}}\right)$\\
    \hline
\end{tabular}
\caption{A collection of position space Feynman rules for propagators and interaction vertices in $\mathcal{N}=4$ SYM coupled to $\mathcal{N}=2$ matter. Red lines represent color contractions and blue lines denote $\SU(N_f)$ flavor contractions (correspondingly we suppress all color and $\SU(N_f)$ flavor indices in the expressions).
} 
\label{Feynmanp}
\end{center}
\end{table}

\subsection{Meson correlators}
We focus on the following $\frac{1}{2}$-BPS operators which are ``mesons'' in the theory 
\begin{equation}\label{eq:SUNfmeson}
\mathcal{M}_p^{f;a_1\ldots a_p;\bar{a}_1\ldots \bar{a}_{p-2}}= \sqrt{\frac{2^pN}{\lambda^p}} (T^f)^m{}_n\, \bar{q}_m^{(a_1}\phi^{a_2(\bar{a}_1}\ldots \phi^{a_{p-1}\bar{a}_{p-2})}q^{a_p),n}\;.
\end{equation}
Here the color indices are implicitly contracted and the indices $a$ and $\bar{a}$ are symmetrized separately. The operators are so normalized such that the coefficients of their two-point functions are 1 at the leading order of $N$. The matrices $(T^f)^m{}_n$ with $f=1,\ldots, N_f^2-1$ are the generators of the flavor group $\SU(N_f)$. Therefore, the operators have protected conformal dimensions $\Delta=p$, and transform in the $(\frac{p}{2},\frac{p-2}{2})$ representation of $\SU(2)_R\times \SU(2)_L$ and the adjoint representation of $\SU(N_f)$. It is  convenient to work in an index-free fashion by introducing two-component polarization spinors $v_a$ and $\bar{v}_{\bar{a}}$ for $\SU(2)_R$ and $\SU(2)_L$ respectively
\begin{equation}
   \mathcal{M}_p^f(x;v,\bar{v})= \mathcal{M}_p^{f;a_1\ldots a_p;\bar{a}_1\ldots \bar{a}_{p-2}}(x)v_{a_1}\ldots v_{a_p}\bar{v}_{\bar{a}_1}\ldots \bar{v}_{\bar{a}_{p-2}}\;.
\end{equation}

\begin{figure}
    \centering
    \begin{tikzpicture}
    \subfloat{\label{fig:1a}
        \begin{scope}[thick]
            \coordinate [label=left:{\small $1$}] (p1) at (140:1.5);
            \coordinate [label=left:{\small $2$}] (p2) at (-140:1.5);
            \coordinate [label=right:{\small $3$}] (p3) at (-40:1.5);
            \coordinate [label=right:{\small $4$}] (p4) at (40:1.5);
            \coordinate (o) at (0,0);
            \draw (p1) -- (p3) -- (p2) -- (p4) -- cycle;
            \draw [-Latex] (o) -- ($(o)!.6!(p1)$);
            \draw [-Latex] (p1) -- ($(p1)!.6!(p4)$);
            \draw [-Latex] (p4) -- ($(p4)!.6!(o)$);
            \draw [-Latex] (o) -- ($(o)!.6!(p3)$);
            \draw [-Latex] (p3) -- ($(p3)!.6!(p2)$);
            \draw [-Latex] (p2) -- ($(p2)!.6!(o)$);
            \fill (o) circle [radius=2pt];
            \foreach \i in {p1,p2,p3,p4} \draw [fill=white] (\i) circle [radius=2pt];
            \node [anchor=center] at (0,-1.5) {\small (a)};
        \end{scope}
    }
    \subfloat{\label{fig:1b}
        \begin{scope}[thick,xshift=4cm]
            \coordinate [label=left:{\small $1$}] (p1) at (140:1.5);
            \coordinate [label=left:{\small $2$}] (p2) at (-140:1.5);
            \coordinate [label=right:{\small $3$}] (p3) at (-40:1.5);
            \coordinate [label=right:{\small $4$}] (p4) at (40:1.5);
            \coordinate (l) at (-0.5,0);
            \coordinate (r) at (0.5,0);
            \coordinate (t) at (0,0.75);
            \coordinate (tl) at (140:0.5);
            \coordinate (tr) at (40:0.5);
            \coordinate (b) at (0,-0.75);
            \coordinate (bl) at (-140:0.5);
            \coordinate (br) at (-40:0.5);
            \coordinate (o) at (0,0);
            \begin{scope}[RoyalBlue]
                \draw (p1) -- (p4) .. controls +(-140:0.8) and +(90:0.15) .. (r) .. controls +(-90:0.15) and +(140:0.8) .. (p3) -- (p2) .. controls +(40:0.8) and +(-90:0.15) .. (l) .. controls +(90:0.15) and +(-40:0.8) .. (p1);
                \draw [-Latex] (p1) -- ($(p1)!.6!(p4)$);
                \draw [-Latex] (p4) .. controls +(-140:0.8) and +(90:0.2) .. ($(r)+(0,-0.1)$);
                \draw [-Latex] (p3) -- ($(p3)!.6!(p2)$);
                \draw [-Latex] (p2) .. controls +(40:0.8) and +(-90:0.2) .. ($(l)+(0,0.1)$);
            \end{scope}
            \begin{scope}[BrickRed]
                \draw (t) .. controls +(180:0.7) and +(140:0.3) .. (tl) .. controls +(-40:0.3) and +(-140:0.3) .. (tr) .. controls +(40:0.3) and +(0:0.7) .. (t);
                \draw (b) .. controls +(180:0.7) and +(-140:0.3) .. (bl) .. controls +(40:0.3) and +(140:0.3) .. (br) .. controls +(-40:0.3) and +(0:0.7) .. (b);
                \draw [-Latex] (br) .. controls +(-40:0.3) and +(0:0.9) .. ($(b)+(-0.2,0)$);
                \draw [-Latex] (tl) .. controls +(140:0.3) and +(180:0.9) .. ($(t)+(0.2,0)$);
            \end{scope}
            \foreach \i in {p1,p2,p3,p4} \draw [fill=white] (\i) circle [radius=2pt];
            \node [anchor=center] at (0,-1.5) {\small (b)};
        \end{scope}
    }
    \subfloat{\label{fig:1c}
        \begin{scope}[thick,xshift=8cm]
            \coordinate [label=left:{\small $1$}] (p1) at (140:1.5);
            \coordinate [label=left:{\small $2$}] (p2) at (-140:1.5);
            \coordinate [label=right:{\small $3$}] (p3) at (-40:1.5);
            \coordinate [label=right:{\small $4$}] (p4) at (40:1.5);
            \coordinate (t) at (0,0.75);
            \coordinate (t2) at (0,0.4);
            \coordinate (tl) at (140:0.5);
            \coordinate (tr) at (40:0.5);
            \coordinate (b) at (0,-0.75);
            \coordinate (b2) at (0,-0.4);
            \coordinate (bl) at (-140:0.5);
            \coordinate (br) at (-40:0.5);
            \coordinate (o) at (0,0);
            \begin{scope}[RoyalBlue]
                \draw (p1) -- (p4) .. controls +(-140:0.8) and +(0:0.25) .. (t2) .. controls +(180:0.25) and +(-40:0.8) .. (p1);
                \draw [-Latex] (p1) -- ($(p1)!.6!(p4)$);
                \draw [-Latex] (p4) .. controls +(-140:0.8) and +(0:0.3) .. ($(t2)+(-0.1,0)$);
                \draw (p3) -- (p2) .. controls +(40:0.8) and +(180:0.25) .. (b2) .. controls +(0:0.25) and +(140:0.8) .. (p3);
                \draw [-Latex] (p3) -- ($(p3)!.6!(p2)$);
                \draw [-Latex] (p2) .. controls +(40:0.8) and +(180:0.3) .. ($(b2)+(0.1,0)$);
            \end{scope}
            \begin{scope}[BrickRed]
                \draw (t) .. controls +(180:1) and +(140:0.3) .. (tl) .. controls +(-40:0.3) and +(40:0.3) .. (bl) .. controls +(-140:0.3) and +(180:1) .. (b) .. controls +(0:1) and +(-40:0.3) .. (br) .. controls +(140:0.3) and +(-140:0.3) .. (tr) .. controls +(40:0.3) and +(0:1) .. (t);
                \draw [-Latex] (br) .. controls +(-40:0.3) and +(0:1.2) .. ($(b)+(-0.2,0)$);
                \draw [-Latex] (tl) .. controls +(140:0.3) and +(180:1.2) .. ($(t)+(0.2,0)$);
            \end{scope}
            \foreach \i in {p1,p2,p3,p4} \draw [fill=white] (\i) circle [radius=2pt];
            \node [anchor=center] at (0,-1.5) {\small (c)};
        \end{scope}
    }
    \end{tikzpicture}
    \caption{Color and flavor flows in the four-point function of $\mathcal{M}_2$, represented by the red and blue lines respectively. Each color loop gives a factor of $N$ and the correlator is dominated by the diagrams with the most color loops.}    \label{fig:qqqq order}
\end{figure}
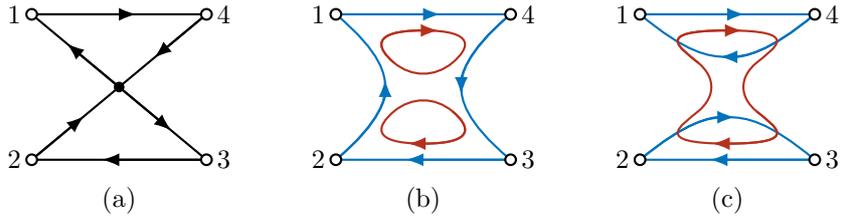

The target of this paper is the $n$-point correlation functions  of these meson operators
\begin{equation}
G_n(x_i;v_i,\bar{v}_i)=\langle \mathcal{M}_{p_1}^{f_1}(x_1;v_1,\bar{v}_1)\ldots \mathcal{M}_{p_n}^{f_n}(x_n;v_n,\bar{v}_n)\rangle\;.
\end{equation}
We will focus on the $N\to\infty$ limit with $\lambda$ fixed and small and compute the leading one-loop correction to the meson correlators. Since meson operators transform in the adjoint representation of the flavor group $\SU(N_F)$, the meson correlators have nontrivial flavor structures. It is natural to perform a decomposition of the correlator in terms of the traces of $\SU(N_f)$ generators
\begin{align}\label{eq:Nfdecompose}
    \langle \mathcal{M}_{p_1}^{f_1}\mathcal{M}_{p_2}^{f_2} \cdots \mathcal{M}_{p_n}^{f_n} \rangle = \sum_{\sigma \in S_n/Z_n}\tr(T^{f_{\sigma(1)}}\cdots T^{f_{\sigma(n)}}) G[\sigma] + \cdots.
\end{align}
We will refer to $G[\sigma]$ as the single-trace partial correlator and $\cdots$ stands for higher-trace contributions. A crucial point to note here is an interesting interplay between the color structures and the flavor structures: At the leading order in $1/N$, there are only single-trace flavor contributions in the meson correlators  while all higher-trace terms are suppressed. This statement can be illustrated by a simple example. Let us consider the contribution to the four-point correlator of $\mathcal{M}_2$ from inserting the vertex $\bar{q}q\bar{q}q$, where the contractions of $q$ and $\bar{q}$ are shown in \Cref{fig:1a}. There are two kinds of flavor flows and color flows as shown in \Cref{fig:1b} and \Cref{fig:1c}. In both cases, the flavor and color flows are orthogonal at the vertex, as dictated by the Feynman rules in \Cref{Feynmanp}. Since we consider the leading large $N$ contribution, we should only keep \Cref{fig:1b} which gives rise to the single-trace term in the flavor decomposition. 

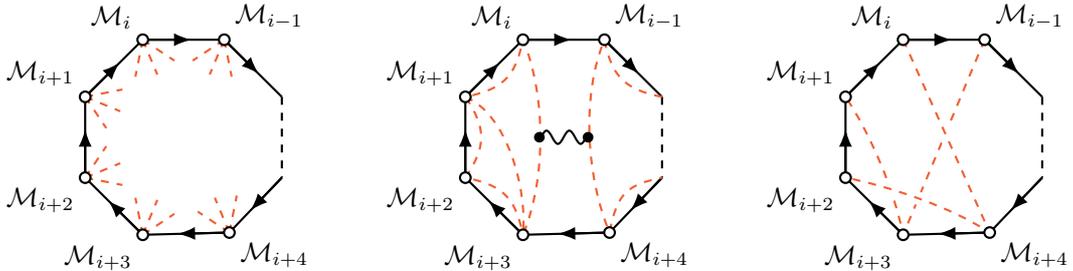
\begin{figure}[ht]
    \centering
    \begin{tikzpicture}[thick]
        \begin{scope}
            \coordinate (m0) at (22.5:1.4);
            \coordinate [label=67.5:{\small $\mathcal{M}_{i-1}$}] (m1) at (67.5:1.4);
            \coordinate [label=112.5:{\small $\mathcal{M}_{i}$}] (m2) at (112.5:1.4);
            \coordinate [label=157.5:{\small $\mathcal{M}_{i+1}$}] (m3) at (157.5:1.4);
            \coordinate [label=202.5:{\small $\mathcal{M}_{i+2}$}] (m4) at (202.5:1.4);
            \coordinate [label=247.5:{\small $\mathcal{M}_{i+3}$}] (m5) at (247.5:1.4);
            \coordinate [label=292.5:{\small $\mathcal{M}_{i+4}$}] (m6) at (295.5:1.4);
            \coordinate (m7) at (337.5:1.4);
            \draw (m0) -- (m1) -- (m2) -- (m3) -- (m4) -- (m5) -- (m6) -- (m7);
            \draw [-Latex] (m7) -- ($(m7)!.6!(m6)$);
            \draw [-Latex] (m6) -- ($(m6)!.6!(m5)$);
            \draw [-Latex] (m5) -- ($(m5)!.6!(m4)$);
            \draw [-Latex] (m4) -- ($(m4)!.6!(m3)$);
            \draw [-Latex] (m3) -- ($(m3)!.6!(m2)$);
            \draw [-Latex] (m2) -- ($(m2)!.6!(m1)$);
            \draw [-Latex] (m1) -- ($(m1)!.6!(m0)$);
            \draw [dashed] (m0) -- (m7);
            \foreach \i in {1,2,...,6} \foreach \j in {1,2,3} \draw [RedOrange,dashed] (m\i) -- +(135+45*\i+33.75*\j:0.5);
            \foreach \i in {1,2,...,6} \draw [fill=white] (m\i) circle [radius=2pt];
        \end{scope}
        \begin{scope}[xshift=5cm]
            \coordinate (m0) at (22.5:1.4);
            \coordinate [label=67.5:{\small $\mathcal{M}_{i-1}$}] (m1) at (67.5:1.4);
            \coordinate [label=112.5:{\small $\mathcal{M}_{i}$}] (m2) at (112.5:1.4);
            \coordinate [label=157.5:{\small $\mathcal{M}_{i+1}$}] (m3) at (157.5:1.4);
            \coordinate [label=202.5:{\small $\mathcal{M}_{i+2}$}] (m4) at (202.5:1.4);
            \coordinate [label=247.5:{\small $\mathcal{M}_{i+3}$}] (m5) at (247.5:1.4);
            \coordinate [label=292.5:{\small $\mathcal{M}_{i+4}$}] (m6) at (295.5:1.4);
            \coordinate (m7) at (337.5:1.4);
            \coordinate (l) at (-0.32,0);
            \coordinate (r) at (0.32,0);
            \draw (m0) -- (m1) -- (m2) -- (m3) -- (m4) -- (m5) -- (m6) -- (m7);
            \draw [-Latex] (m7) -- ($(m7)!.6!(m6)$);
            \draw [-Latex] (m6) -- ($(m6)!.6!(m5)$);
            \draw [-Latex] (m5) -- ($(m5)!.6!(m4)$);
            \draw [-Latex] (m4) -- ($(m4)!.6!(m3)$);
            \draw [-Latex] (m3) -- ($(m3)!.6!(m2)$);
            \draw [-Latex] (m2) -- ($(m2)!.6!(m1)$);
            \draw [-Latex] (m1) -- ($(m1)!.6!(m0)$);
            \draw [dashed] (m0) -- (m7);
            \draw [RedOrange,dashed] (m0) .. controls +(135+33.75:0.5) and +(-45-33.75:0.5) .. (m1) .. controls +(180+67.5:0.8) and +(45+67.5:0.8) .. (m6) .. controls +(45+33.75:0.5) and +(-135-33.75:0.5) .. (m7) (m3) .. controls +(45-67.5:0.6) and +(67.5+16.875:0.6) .. (m5) .. controls +(67.5:0.8) and +(-67.5:0.8) .. (m2) .. controls +(-67.5-33.75:0.5) and +(45-33.75:0.5) .. (m3) .. controls +(-90+33.75:0.5) and +(90-33.75:0.5) .. (m4) .. controls +(-45+33.75:0.5) and +(135-33.75:0.5) .. (m5);
            \draw [decorate,decoration={snake,segment length=3mm}] (l) -- (r);
            \foreach \i in {1,2,...,6} \draw [fill=white] (m\i) circle [radius=2pt];
            \foreach \i in {l,r} \fill (\i) circle [radius=2pt];
        \end{scope}
        \begin{scope}[xshift=10cm]
            \coordinate (m0) at (22.5:1.4);
            \coordinate [label=67.5:{\small $\mathcal{M}_{i-1}$}] (m1) at (67.5:1.4);
            \coordinate [label=112.5:{\small $\mathcal{M}_{i}$}] (m2) at (112.5:1.4);
            \coordinate [label=157.5:{\small $\mathcal{M}_{i+1}$}] (m3) at (157.5:1.4);
            \coordinate [label=202.5:{\small $\mathcal{M}_{i+2}$}] (m4) at (202.5:1.4);
            \coordinate [label=247.5:{\small $\mathcal{M}_{i+3}$}] (m5) at (247.5:1.4);
            \coordinate [label=292.5:{\small $\mathcal{M}_{i+4}$}] (m6) at (295.5:1.4);
            \coordinate (m7) at (337.5:1.4);
            \draw [RedOrange,dashed] (m1) -- (m5) .. controls +(67.5+33.75:0.6) and +(-90+33.75:0.6) .. (m3);
            \draw [white,line width=5pt] (m2) -- (m6) .. controls +(180-33.75:0.6) and +(-45+33.75:0.6) .. (m4);
            \draw [RedOrange,dashed] (m2) -- (m6) .. controls +(180-33.75:0.6) and +(-45+33.75:0.6) .. (m4);
            \draw (m0) -- (m1) -- (m2) -- (m3) -- (m4) -- (m5) -- (m6) -- (m7);
            \draw [-Latex] (m7) -- ($(m7)!.6!(m6)$);
            \draw [-Latex] (m6) -- ($(m6)!.6!(m5)$);
            \draw [-Latex] (m5) -- ($(m5)!.6!(m4)$);
            \draw [-Latex] (m4) -- ($(m4)!.6!(m3)$);
            \draw [-Latex] (m3) -- ($(m3)!.6!(m2)$);
            \draw [-Latex] (m2) -- ($(m2)!.6!(m1)$);
            \draw [-Latex] (m1) -- ($(m1)!.6!(m0)$);
            \draw [dashed] (m0) -- (m7);
            \foreach \i in {1,2,...,6} \draw [fill=white] (m\i) circle [radius=2pt];
        \end{scope}
    \end{tikzpicture}
    \caption{Examples of diagrams contributing to the partial correlator. The red dashed lines represent $\phi$ propagators. The second diagram is an example of planar diagram at leading large $N$, while the third diagram is not leading at large $N$.}
    \label{fig:externalq contract}
\end{figure}

In general, the orthogonality of color flows and flavor flows at vertices implies a competition between color and flavor loops, i.e., having more color loops means fewer flavor loops. Therefore, at the leading order in $1/N$ we only need to keep single-trace terms in the flavor decomposition. Without loss of generality, we focus on the partial correlator with canonical ordering 
\begin{align}
    G_n \equiv \langle p_1 p_2 \cdots p_n\rangle \equiv G[12\cdots n].
\end{align}
This gives rise to an $n$-gon with flavor flowing along the edges clockwisely, as \Cref{fig:externalq contract} shows. It is not hard to verify that the leading large $N$ contributions come from planar diagrams which can be drawn inside the polygon. In other words, they come from diagrams topologically equivalent to a disk (or its degenerate version, such as in \Cref{fig:1a}), whose boundary is given by the flavor flow polygon of the correlator. 

When we further consider superconformal symmetry, the chiral algebra construction in \cite{Beem:2013sza} puts additional constraint on the correlator. When all operators are inserted on a 2d plane with complex coordinates $(z_i,\bar{z}_i)$ and the R-polarization spinors $v_i$ are twisted as $v_i^\alpha=(1,\bar{z}_i)$, the correlator should become a meromorphic function
\begin{equation}\label{chiralalgebracond}
\partial_{\bar{z}_j}G_n\big(z_i,\bar{z}_i;v_i^\alpha=(1,\bar{z}_i)\big)=0\;.
\end{equation}
This will be a useful consistency check for the correlation functions computed by Feynman diagrams. In particular, for four-point functions, this condition can be solved and gives the following decomposition \cite{Nirschl:2004pa}
\begin{equation}\label{eq:G4decompose}
    G_4\left(x_i;v_{i},\bar{v}_i \right) = G_{\text{protected}}\left(x_i;v_{i},\bar{v}_i \right) +R \times H_4\left(x_i;v_{i},\bar{v}_i \right) \, ,
\end{equation}
where $G_{\text{protected}}$ is a rational function and is protected by superconformal symmetry. The factor $R$ is 
\begin{equation}\label{eq:Rdef}
    R=V_{1234} \, x^2_{13}x^2_{24} + V_{1342} \, x^2_{14}x^2_{23}+ V_{1423} \,  x^2_{12}x^2_{34}\;,\qquad  V_{ijk\ell} =v_{ij} v_{jk} v_{k\ell} v_{\ell i}\;,
\end{equation}
where
\begin{align}
    x^2_{ij} = (x_i-x_j)^2,\qquad   v_{ij}=\epsilon_{ab}v^a_i v^b_j \;.
\end{align}
For later convenience, we also define the cross-ratios in the usual way
\begin{equation}
    z \zb= \frac{x^2_{12}x^2_{34}}{x^2_{13}x^2_{24}}\, ,  \hspace{1em} (1-z)(1-\zb)= \frac{x^2_{23}x^2_{14}}{x^2_{13}x^2_{24}} \, .
\end{equation}

\section{Four-point functions}\label{Sec:4ptfun}

In this section, we compute the general four-point function for meson operators of arbitrary dimensions. We will follow a similar strategy as in $\mathcal{N}=4$ SYM  \cite{Drukker:2008pi}. We find that all four-point functions can be packaged into a generating function \cref{eq:4pointgenerating} which enjoys the hidden 8d symmetry at the integrand level.

\subsection{Building blocks at one-loop}

To compute one-loop corrections using Feynman diagrams, it is useful to introduce the following elementary Feynman integrals as the building blocks \cite{Beisert:2002bb,Drukker:2008pi}
\begin{align}\label{eq:XYdef}
    \X_{1234}&=\int \frac{d^4x_0}{x^2_{10}x^2_{20}x^2_{30}x^2_{40}}\equiv
    \parbox{2.2cm}{
    	\tikz{\begin{scope}[scale=0.6, >=Stealth]
    			\node [anchor=east] at (-1,1) {$1$};
    			\node [anchor=east] at (-1,-1.0) {$2$};
    			\node [anchor=west] at (1,-1.0) {$3$};
    			\node [anchor=west] at (1,1.0) {$4$};
    			\filldraw[black]  (0, 0) circle (2.5pt);
    			\draw[line width=0.75pt] {( 1,-1 ) -- ( -1, 1)};
    			\draw[line width=0.75pt] {( 1, 1 ) -- ( -1, -1)};
    			\draw[black, fill=white]  ( 1,-1 ) circle (2pt);
    			\draw[black, fill=white]  ( 1, 1 ) circle (2pt);
    			\draw[black, fill=white]  (-1,-1 ) circle (2pt);
    			\draw[black, fill=white]  (-1, 1 ) circle (2pt);
    \end{scope}}} \, ,  \qquad 
	\Y_{123}=\int  \frac{d^4x_0}{x^2_{10}x^2_{20}x^2_{30}} \, .
\end{align}
The integral $\mathcal{X}$ depends on four points while the other integral $\mathcal{Y}$ depends on only three points. Note that $\mathcal{X}_{1234}$ is a conformal integral coming from a  scalar contact vertex. 
These two integrals evaluate to
\begin{align}
    \X_{1234}&=\frac{\pi^2}{x^2_{13}x^2_{24}} \Phi(z,\zb) \, , \qquad   \Y_{123}=\lim_{x_4\to \infty} x^2_4 \, \X_{1234}\, ,
\end{align}
where 
\begin{equation}
    \Phi(z,\zb)=\frac{1}{z-\zb}\left( 2\text{Li}_{2}(z)-2\text{Li}_{2}(\zb)+\log(z\zb)\log\left(\frac{1-z}{1-\zb}\right)\right)\;,
\end{equation}
is the well-known scalar one-loop box function \cite{Usyukina:1992jd}. It turns out that any one-loop Feynman diagram that we encounter in the meson correlators can be expressed in terms of these building blocks. In particular, if we temporarily ignore any dependence on internal symmetries, then a typical s-channel vector exchange diagram
\begin{equation}\label{eq:exchange}
	\frac{\widetilde{\F}_{12,34}}{x^2_{12}x^2_{34}}= (\partial_1-\partial_2)(\partial_3-\partial_4)\int \frac{d^4x_i d^4x_j}{x^2_{1i}x^2_{2i}x^2_{ij}x^2_{3j}x^2_{4j}} \equiv  
	 \parbox{2.7cm}{\tikz{\begin{scope}[scale=0.6, >=Stealth]
			\draw [line width=0.8pt] (-1.5,  1) -- (-1.5, -1);
			\draw [line width=0.8pt] ( 1.5, -1) -- ( 1.5,  1);
			\draw [decorate,decoration={name=snake ,aspect=0, segment
					length=3.5mm,amplitude=2},line width=0.75pt]{( -1.5, 0) -- ( 1.5, 0)};
			\draw[] {( 1.5, 0 ) -- ( 1.5, 0)};
			\draw[] {( -1.5, 0.2 ) -- ( -1.5, 0.2)};
			\node [anchor=east] at (-1.5,1) {$1$};
			\node [anchor=east] at (-1.5,-1.0) {$2$};
			\node [anchor=west] at (1.5,-1.0) {$3$};
			\node [anchor=west] at (1.5,1.0) {$4$};
			\draw[black, fill=white]  ( 1.5,-1 ) circle (2pt);
			\draw[black, fill=white]  ( 1.5, 1 ) circle (2pt);
			\draw[black, fill=white]  (-1.5,-1 ) circle (2pt);
			\draw[black, fill=white]  (-1.5, 1 ) circle (2pt);
			\filldraw[black]  (-1.5, 0) circle (2.5pt);
			\filldraw[black]  ( 1.5, 0) circle (2.5pt);
			\end{scope}}} 
\end{equation}
evaluates to
\begin{align}\label{eq:tildeFexplicit}
    \widetilde{\F}_{12,34} & =x^2_{13}x^2_{24}\, \X_{1234}-x^2_{14}x^2_{23} \, \X_{1234}+\K_{1,34}-\K_{2,34}+\K_{3,12}-\K_{4,12} \, ,\\
    \K_{1,23}&=x^2_{13} \, \Y_{123}-x^2_{12} \, \Y_{123} \, .
\end{align}
Apart from the exchange and contact diagrams, there are also additional diagrams which contribute to the vertex corrections (corner corrections in \cite{Drukker:2008pi}) and the self-energy of propagators. One can easily observes that kinematically these diagrams only involve the integral $\mathcal{Y}$ and its pinching limit \cite{Drukker:2008pi}. By dressing these Feynman diagrams with proper R-structures and $\SU(2)_L$ structures we can work out the one-loop correlation function for any meson operators. The one-loop correction is therefore a linear combination of the two building blocks $\mathcal{X}$ and $\mathcal{Y}$ as defined in \cref{eq:XYdef}.

In the quenched limit $N_f/N\ll1$, the whole correlator is conformal and should be expressed purely in terms of conformal integrals. Because the building block $\mathcal{Y}$ explicitly breaks  conformal symmetry, it has to completely cancel out in the final result.\footnote{An explicit demonstration of the cancellation of $\mathcal{Y}$ can be found in \cite{Drukker:2008pi} for $\mathcal{N}=4$ SYM. We have also verified this cancellation in the two models that we study.} This means that in practice we are allowed to simply ignore all the diagrams containing corner corrections or self-energy corrections, and moreover, ignore any $\mathcal{Y}$ terms in the exchange diagrams \cref{eq:tildeFexplicit}. The latter leads us to define the effective vector exchange contribution
\begin{equation}\label{eq:exchange2}
    \F_{12,34} \equiv \widetilde{\F}_{12,34}\big|_{\Y=0}=\left(x^2_{13}x^2_{24}-x^2_{14}x^2_{23}\right) \X_{1234} \, ,
\end{equation}
which, at the risk of abusing the terminology,  we will refer to from now on as the exchange diagram. As a result, the one-loop meson correlators are linear combinations of $\mathcal{X}$. In particular, since  there is only a unique $\mathcal{X}_{1234}$ at four points,  four-point correlators with any meson operators always have to be proportional to this function.

\subsection{Low-lying examples}
We start by explicitly working out several examples with low-lying meson states, namely, $\langle 2222 \rangle $, $\langle 2233 \rangle $ and $\langle 3333 \rangle $. These examples cover all types of vertices that can possibly appear in the one-loop computation. From these examples we observe the emergence of patterns which can be directly generalized to four-point correlators of more general meson states later in this section, as well as to higher points in the next section. 

\begin{figure}[ht]
	\centering
	\begin{tikzpicture}
		\begin{scope}[thick]
			\coordinate [label=left:{\small $1$}] (p1) at (140:1.5);
			\coordinate [label=left:{\small $2$}] (p2) at (-140:1.5);
			\coordinate [label=right:{\small $3$}] (p3) at (-40:1.5);
			\coordinate [label=right:{\small $4$}] (p4) at (40:1.5);
			\coordinate (l) at ($(p1)!.5!(p2)$);
			\coordinate (r) at ($(p3)!.5!(p4)$);
			\draw (p1) -- (p2) -- (p3) -- (p4) -- cycle;
			\draw [-Latex] (p1) -- ($(p1)!.6!(p4)$);
			\draw [-Latex] (p4) -- ($(p4)!.6!(r)$);
			\draw [-Latex] (r) -- ($(r)!.6!(p3)$);
			\draw [-Latex] (p3) -- ($(p3)!.6!(p2)$);
			\draw [-Latex] (p2) -- ($(p2)!.6!(l)$);
			\draw [-Latex] (l) -- ($(l)!.6!(p1)$);
			\draw [decorate,decoration={snake}] (l) -- (r);
			\foreach \i in {p1,p2,p3,p4} \draw [fill=white] (\i) circle [radius=2pt];
			\foreach \i in {l,r} \fill (\i) circle [radius=2pt];
			\node [anchor=center] at (0,-1.5) {\small $E_{\langle2222\rangle}^1$};
		\end{scope}
		\begin{scope}[thick,xshift=3.6cm]
			\coordinate [label=left:{\small $1$}] (p1) at (140:1.5);
			\coordinate [label=left:{\small $2$}] (p2) at (-140:1.5);
			\coordinate [label=right:{\small $3$}] (p3) at (-40:1.5);
			\coordinate [label=right:{\small $4$}] (p4) at (40:1.5);
			\coordinate (t) at ($(p1)!.5!(p4)$);
			\coordinate (b) at ($(p2)!.5!(p3)$);
			\draw (p1) -- (p2) -- (p3) -- (p4) -- cycle;
			\draw [-Latex] (p1) -- ($(p1)!.6!(t)$);
			\draw [-Latex] (t) -- ($(t)!.6!(p4)$);
			\draw [-Latex] (p4) -- ($(p4)!.6!(p3)$);
			\draw [-Latex] (p3) -- ($(p3)!.6!(b)$);
			\draw [-Latex] (b) -- ($(b)!.6!(p2)$);
			\draw [-Latex] (p2) -- ($(p2)!.6!(p1)$);
			\draw [decorate,decoration={snake}] (t) -- (b);
			\foreach \i in {p1,p2,p3,p4} \draw [fill=white] (\i) circle [radius=2pt];
			\foreach \i in {t,b} \fill (\i) circle [radius=2pt];
			\node [anchor=center] at (0,-1.5) {\small $E_{\langle2222\rangle}^2$};
		\end{scope}
		\begin{scope}[thick,xshift=7.2cm]
			\coordinate [label=left:{\small $1$}] (p1) at (140:1.5);
			\coordinate [label=left:{\small $2$}] (p2) at (-140:1.5);
			\coordinate [label=right:{\small $3$}] (p3) at (-40:1.5);
			\coordinate [label=right:{\small $4$}] (p4) at (40:1.5);
			\coordinate (o) at (0,0);
			\draw (p1) -- (p4) -- (p2) -- (p3) -- cycle;
			\draw [-Latex] (p1) -- ($(p1)!.6!(p4)$);
			\draw [-Latex] (p4) -- ($(p4)!.6!(o)$);
			\draw [-Latex] (o) -- ($(o)!.6!(p3)$);
			\draw [-Latex] (p3) -- ($(p3)!.6!(p2)$);
			\draw [-Latex] (p2) -- ($(p2)!.6!(o)$);
			\draw [-Latex] (o) -- ($(o)!.6!(p1)$);
			\foreach \i in {p1,p2,p3,p4} \draw [fill=white] (\i) circle [radius=2pt];
			\foreach \i in {o} \fill (\i) circle [radius=2pt];
			\node [anchor=center] at (0,-1.5) {\small $C_{\langle2222\rangle}^1$};
		\end{scope}
		\begin{scope}[thick,xshift=10.8cm]
			\coordinate [label=left:{\small $1$}] (p1) at (140:1.5);
			\coordinate [label=left:{\small $2$}] (p2) at (-140:1.5);
			\coordinate [label=right:{\small $3$}] (p3) at (-40:1.5);
			\coordinate [label=right:{\small $4$}] (p4) at (40:1.5);
			\coordinate (o) at (0,0);
			\draw (p1) -- (p3) -- (p4) -- (p2) -- cycle;
			\draw [-Latex] (p1) -- ($(p1)!.6!(o)$);
			\draw [-Latex] (o) -- ($(o)!.6!(p4)$);
			\draw [-Latex] (p4) -- ($(p4)!.6!(p3)$);
			\draw [-Latex] (p3) -- ($(p3)!.6!(o)$);
			\draw [-Latex] (o) -- ($(o)!.6!(p2)$);
			\draw [-Latex] (p2) -- ($(p2)!.6!(p1)$);
			\foreach \i in {p1,p2,p3,p4} \draw [fill=white] (\i) circle [radius=2pt];
			\foreach \i in {o} \fill (\i) circle [radius=2pt];
			\node [anchor=center] at (0,-1.5) {\small $C_{\langle2222\rangle}^2$};
		\end{scope}
	\end{tikzpicture}
	\caption{Exchange diagrams and contact diagrams in the case of $\langle2222 \rangle$.}
	\label{fig:case2222}
\end{figure}

\paragraph{\underline{$\langle 2222 \rangle $}}~\\

\noindent The simplest example, $\langle2222\rangle$, contains four different diagrams, including two exchange diagrams and two contact diagrams, as listed in \Cref{fig:case2222}. The two exchange diagrams give\footnote{Here and after we will extract a factor of $N^{-\frac{n-2}{2}}$ in the $n$-point function.}
\begin{align}
    E^1_{\langle 2222 \rangle} &=\frac{\lambda}{2} \, d_{12} d_{23} d_{34} d_{41} \times \F_{12,34}  \, ,\\
    E^2_{\langle 2222 \rangle} &=\frac{\lambda}{2} \, d_{12} d_{23} d_{34} d_{41} \times \F_{41,23}  \, ,
\end{align}
and the two contact diagrams give
\begin{align}
    C^1_{\langle 2222 \rangle} &=\frac{\lambda}{2} \, d_{41} d_{23}(2v_{14}v_{32}-v_{12}v_{34}) \times \X_{1234} \, ,\\
    C^2_{\langle 2222 \rangle} &=\frac{\lambda}{2} \, d_{12} d_{34}(2v_{21}v_{43}-v_{41}v_{23}) \times \X_{1234} \, .
\end{align}
Here $d_{ij}$ are the scalar propagators between $q$ and $\bar{q}$ 
\begin{equation}
    d_{ij}=\frac{v_{ij}}{x^2_{ij}}\, .
\end{equation}
Adding up these diagrams, we obtain the final result
\begin{equation}\label{eq:res2222}
    \langle 2222 \rangle = E^1_{\langle 2222 \rangle}+E^2_{\langle 2222 \rangle}+C^1_{\langle 2222 \rangle}+C^2_{\langle 2222 \rangle} = \lambda \,  \frac{ R \, \X_{1234}}{x^2_{12}x^2_{23}x^2_{34}x^2_{41}} \, ,
\end{equation}
The factor $R$ is the same one in the solution to the superconformal Ward identity \cref{eq:G4decompose} and emerges only after summing up the four diagrams. The agreement with \cref{eq:G4decompose} provides a non-trivial consistency check of our computation.

\begin{figure}
	\centering
	\begin{tikzpicture}
		\begin{scope}[thick]
			\coordinate [label=left:{\small $1$}] (p1) at (140:1.5);
			\coordinate [label=left:{\small $2$}] (p2) at (-140:1.5);
			\coordinate [label=right:{\small $3$}] (p3) at (-40:1.5);
			\coordinate [label=right:{\small $4$}] (p4) at (40:1.5);
			\coordinate (l) at ($(p1)!.5!(p2)$);
			\coordinate (r) at (0.7,0);
			\draw (p1) -- (p2) -- (p3) -- (p4) -- cycle;
			\draw [-Latex] (p1) -- ($(p1)!.6!(p4)$);
			\draw [-Latex] (p4) -- ($(p4)!.6!(p3)$);
			\draw [-Latex] (p3) -- ($(p3)!.6!(p2)$);
			\draw [-Latex] (p2) -- ($(p2)!.6!(l)$);
			\draw [-Latex] (l) -- ($(l)!.6!(p1)$);
			\draw [dashed] (p4) .. controls +(-140:0.3) and +(90:0.4) .. (r) .. controls +(-90:0.4) and +(140:0.3) .. (p3);
			\draw [decorate,decoration={snake}] (l) -- (r);
			\foreach \i in {p1,p2,p3,p4} \draw [fill=white] (\i) circle [radius=2pt];
			\foreach \i in {l,r} \fill (\i) circle [radius=2pt];
			\node [anchor=center] at (0,-1.5) {\small $E_{\langle2233\rangle}^1$};
		\end{scope}
		\begin{scope}[thick,xshift=3.6cm]
			\coordinate [label=left:{\small $1$}] (p1) at (140:1.5);
			\coordinate [label=left:{\small $2$}] (p2) at (-140:1.5);
			\coordinate [label=right:{\small $3$}] (p3) at (-40:1.5);
			\coordinate [label=right:{\small $4$}] (p4) at (40:1.5);
			\coordinate (t) at ($(p1)!.5!(p4)$);
			\coordinate (b) at ($(p2)!.5!(p3)$);
			\coordinate (r) at (0.7,0);
			\draw (p1) -- (p2) -- (p3) -- (p4) -- cycle;
			\draw [-Latex] (p1) -- ($(p1)!.6!(t)$);
			\draw [-Latex] (t) -- ($(t)!.6!(p4)$);
			\draw [-Latex] (p4) -- ($(p4)!.6!(p3)$);
			\draw [-Latex] (p3) -- ($(p3)!.6!(b)$);
			\draw [-Latex] (b) -- ($(b)!.6!(p2)$);
			\draw [-Latex] (p2) -- ($(p2)!.6!(p1)$);
			\draw [RedOrange,dashed] (p4) .. controls +(-140:0.3) and +(90:0.4) .. (r) .. controls +(-90:0.4) and +(140:0.3) .. (p3);
			\draw [decorate,decoration={snake}] (t) -- (b);
			\foreach \i in {p1,p2,p3,p4} \draw [fill=white] (\i) circle [radius=2pt];
			\foreach \i in {t,b} \fill (\i) circle [radius=2pt];
			\node [anchor=center] at (0,-1.5) {\small $E_{\langle2233\rangle}^2$};
		\end{scope}
		\begin{scope}[thick,xshift=7.2cm]
			\coordinate [label=left:{\small $1$}] (p1) at (140:1.5);
			\coordinate [label=left:{\small $2$}] (p2) at (-140:1.5);
			\coordinate [label=right:{\small $3$}] (p3) at (-40:1.5);
			\coordinate [label=right:{\small $4$}] (p4) at (40:1.5);
			\coordinate (o) at (0,0);
			\draw (p1) -- (p4) -- (p3) -- (p2) -- (o) -- cycle;
			\draw [-Latex] (p1) -- ($(p1)!.6!(p4)$);
			\draw [-Latex] (p4) -- ($(p4)!.6!(p3)$);
			\draw [-Latex] (p3) -- ($(p3)!.6!(p2)$);
			\draw [-Latex] (p2) -- ($(p2)!.6!(o)$);
			\draw [-Latex] (o) -- ($(o)!.6!(p1)$);
			\draw [dashed] (p4) -- (o) -- (p3);
			\foreach \i in {p1,p2,p3,p4} \draw [fill=white] (\i) circle [radius=2pt];
			\foreach \i in {o} \fill (\i) circle [radius=2pt];
			\node [anchor=center] at (0,-1.5) {\small $C_{\langle2233\rangle}^1$};
		\end{scope}
		\begin{scope}[thick,xshift=10.8cm]
			\coordinate [label=left:{\small $1$}] (p1) at (140:1.5);
			\coordinate [label=left:{\small $2$}] (p2) at (-140:1.5);
			\coordinate [label=right:{\small $3$}] (p3) at (-40:1.5);
			\coordinate [label=right:{\small $4$}] (p4) at (40:1.5);
			\coordinate (o) at (0,0);
			\coordinate (r) at (0.8,0);
			\draw (p1) -- (p3) -- (p4) -- (p2) -- cycle;
			\draw [-Latex] (p1) -- ($(p1)!.6!(o)$);
			\draw [-Latex] (o) -- ($(o)!.6!(p4)$);
			\draw [-Latex] (p4) -- ($(p4)!.6!(p3)$);
			\draw [-Latex] (p3) -- ($(p3)!.6!(o)$);
			\draw [-Latex] (o) -- ($(o)!.6!(p2)$);
			\draw [-Latex] (p2) -- ($(p2)!.6!(p1)$);
			\draw [RedOrange,dashed] (p4) .. controls +(-120:0.3) and +(90:0.4) .. (r) .. controls +(-90:0.4) and +(120:0.3) .. (p3);
			\foreach \i in {p1,p2,p3,p4} \draw [fill=white] (\i) circle [radius=2pt];
			\foreach \i in {o} \fill (\i) circle [radius=2pt];
			\node [anchor=center] at (0,-1.5) {\small $C_{\langle2233\rangle}^2$};
		\end{scope}
	\end{tikzpicture}
	\caption{Exchange diagrams and contact diagrams in the case of $\langle2233 \rangle$.}
	\label{fig:case2233}
\end{figure}

\paragraph{\underline{$\langle 2233 \rangle $}}~\\

\noindent The case of $\langle 2233\rangle$ also involves four Feynman diagrams, as shown in \Cref{fig:case2233}. An explicit computation yields
\begin{subequations}\label{eq:diagrams2233}
    \begin{align}
        E^1_{\langle 2233 \rangle}&= \frac{\lambda}{2} \, d_{12} d_{23} d_{34} d_{41} \times \F_{12,34}  \times \left( \frac{t_{34}}{x^2_{34}}\right) \, ,\\
        E^2_{\langle 2233 \rangle}&= \frac{\lambda}{2} \, d_{12} d_{23} d_{34} d_{41} \times \F_{41,23}  \times \left( \frac{t_{34}}{x^2_{34}}\right) \, ,\\
        C^1_{\langle 2233 \rangle}
        &=\frac{\lambda}{2} \left(v_{14} v_{41} v_{23} v_{32} v_{34} \vb_{34} + v_{13} v_{41} v_{42} v_{23} v_{34} \vb_{43}\right)  \times  \frac{1}{x_{14}^2 x_{23}^2 x_{34}^2} \X_{1234} \nonumber \\
        &=\frac{\lambda}{2} \left(v_{14} v_{32}+ v_{13} v_{42} \right) d_{41} d_{23}  \times  \X_{1234}  \times \left( \frac{t_{34}}{x^2_{34}}\right) \, , \\
        C^2_{\langle 2233 \rangle}&=\frac{\lambda}{2} (2 v_{21}v_{43}-v_{41}v_{23})  d_{12} d_{34}  \times \X_{1234}  \times \left( \frac{t_{34}}{x^2_{34}}\right) \, ,
\end{align}
\end{subequations}
where $t_{ij}=v_{ij}\bar{v}_{ij}$ and $t_{ij}/x_{ij}^2$ is the propagator for the field $\phi$. Summing over all these diagrams, we obtain
\begin{equation}\label{res2233}
    \langle 2233 \rangle = E^1_{\langle 2233 \rangle}+E^2_{\langle 2233 \rangle}+C^1_{\langle 2233 \rangle}+C^2_{\langle 2233 \rangle} = \lambda \,  \frac{ R \X_{1234}}{x^2_{12}x^2_{23}x^2_{34}x^2_{41}} \left( \frac{t_{34}}{x^2_{34}}\right) \, .
\end{equation}
This result again includes $R$ as an overall factor, thus satisfying the superconformal Ward identity. It is also interesting that this result differs from $\langle 2222\rangle$ only by a simple factor $t_{34}/x_{34}^2$, which can be identified with the Wick contraction between the $\phi$ fields in the operators $\mathcal{M}_3$ inserted at $x_3$ and $x_4$. This appearance of the $\phi$ Wick contraction is manifested at the diagrammatic level in some of the contributions. We observe that two of the diagrams, $E_{\langle2233\rangle}^2$ and $C_{\langle2233\rangle}^2$, are explicitly related to the diagrams $E_{\langle2222\rangle}^2$ and $C_{\langle2222\rangle}^2$ by including an additional $\phi$ propagator (the orange dashed lines in \Cref{fig:case2233})
\begin{align}\label{eq:trivial2233}
    E_{\langle 2233\rangle}^2+C_{\langle2233\rangle}^2=\left(E_{\langle2222\rangle}^2+C_{\langle2222\rangle}^2\right)\,\frac{t_{34}}{x_{34}^2}\;.
\end{align}
We also note that in $\langle 2222\rangle$, the remaining diagrams $E_{\langle2222\rangle}^1$ and $C_{\langle2222\rangle}^1$ are related to $E_{\langle2222\rangle}^2$ and $C_{\langle2222\rangle}^2$ by relabeling the external operators. This motivates us to define the following combination
\begin{align}
    \H_{12,34} &= \frac{E^1_{\langle 2222 \rangle}+C^1_{\langle 2222 \rangle}}{\lambda}=\frac{d_{41}d_{23}}{2}\left(d_{12}d_{34}\mathcal{F}_{12,34}+(2v_{14}v_{32}-v_{12}v_{34})\mathcal{X}_{1234}\right)\nonumber\\
    &= \frac{d_{23}d_{41}\X_{1234}}{2} \,\left( v_{12}  v_{34}  \left(\frac{x^2_{13}x^2_{24}}{x^2_{12}x^2_{34}}-\frac{x^2_{14}x^2_{23}}{x^2_{12}x^2_{34}}-1\right) + 2v_{14}v_{32} \right) ,\label{eq:Feytop1}
\end{align}
such that
\begin{align}
    \langle2222\rangle=\lambda\left(\mathcal{H}_{12,34}+\mathcal{H}_{41,23}\right)\,.
\end{align}
The two diagrams $E_{\langle2233\rangle}^2$ and $C_{\langle2233\rangle}^2$ are trivially proportional to $\mathcal{H}_{41,23}$.

The appearance of the $\phi$ contraction from the remaining two diagrams $E_{\langle2233\rangle}^1$ and $C_{\langle2233\rangle}^1$, on the other hand, is not obvious, as the diagrams involve new interaction vertices $\phi A\phi$ and $\bar{q}\phi\phi q$ that are absent in $\langle2222\rangle$. Quite non-trivially, the explicit computation \cref{eq:diagrams2233} shows that the sum of these two diagrams turns out to be again proportional to the $\mathcal{H}$ defined in \cref{eq:Feytop1}
\begin{align}\label{eq:Feytop2}
    E_{\langle2233\rangle}^1+C_{\langle2233\rangle}^1=\lambda\,\mathcal{H}_{12,34}\,\frac{t_{34}}{x_{34}^2}\,.
\end{align}
This is achieved thanks to the delicate relations among different vertices which make the theory supersymmetric.

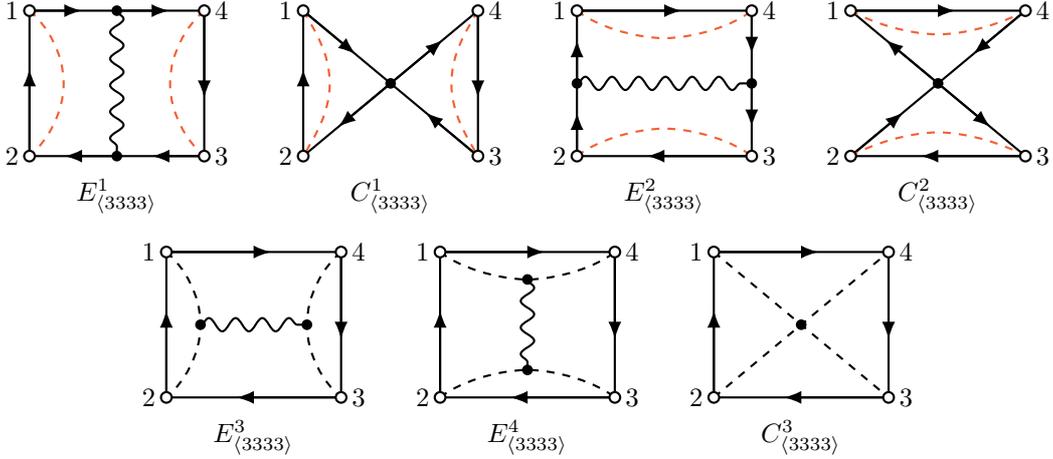
\begin{figure}[ht]
	\centering
	\begin{tikzpicture}
		\begin{scope}[thick,xshift=1.8cm,yshift=-3.2cm]
			\coordinate [label=left:{\small $1$}] (p1) at (140:1.5);
			\coordinate [label=left:{\small $2$}] (p2) at (-140:1.5);
			\coordinate [label=right:{\small $3$}] (p3) at (-40:1.5);
			\coordinate [label=right:{\small $4$}] (p4) at (40:1.5);
			\coordinate (l) at (-0.7,0);
			\coordinate (r) at (0.7,0);
			\draw (p1) -- (p2) -- (p3) -- (p4) -- cycle;
			\draw [-Latex] (p1) -- ($(p1)!.6!(p4)$);
			\draw [-Latex] (p4) -- ($(p4)!.6!(p3)$);
			\draw [-Latex] (p3) -- ($(p3)!.6!(p2)$);
			\draw [-Latex] (p2) -- ($(p2)!.6!(p1)$);
			\draw [dashed] (p4) .. controls +(-140:0.3) and +(90:0.4) .. (r) .. controls +(-90:0.4) and +(140:0.3) .. (p3) (p1) .. controls +(-40:0.3) and +(90:0.4) .. (l) .. controls +(-90:0.4) and +(40:0.3) .. (p2);
			\draw [decorate,decoration={snake}] (l) -- (r);
			\foreach \i in {p1,p2,p3,p4} \draw [fill=white] (\i) circle [radius=2pt];
			\foreach \i in {l,r} \fill (\i) circle [radius=2pt];
			\node [anchor=center] at (0,-1.5) {\small $E_{\langle3333\rangle}^3$};
		\end{scope}
		\begin{scope}[thick,xshift=5.4cm,yshift=-3.2cm]
			\coordinate [label=left:{\small $1$}] (p1) at (140:1.5);
			\coordinate [label=left:{\small $2$}] (p2) at (-140:1.5);
			\coordinate [label=right:{\small $3$}] (p3) at (-40:1.5);
			\coordinate [label=right:{\small $4$}] (p4) at (40:1.5);
			\coordinate (t) at (0,0.6);
			\coordinate (b) at (0,-0.6);
			\draw (p1) -- (p2) -- (p3) -- (p4) -- cycle;
			\draw [-Latex] (p1) -- ($(p1)!.6!(p4)$);
			\draw [-Latex] (p4) -- ($(p4)!.6!(p3)$);
			\draw [-Latex] (p3) -- ($(p3)!.6!(p2)$);
			\draw [-Latex] (p2) -- ($(p2)!.6!(p1)$);
			\draw [dashed] (p1) .. controls +(-40:0.3) and +(180:0.4) .. (t) .. controls +(0:0.4) and +(-140:0.3) .. (p4) (p2) .. controls +(40:0.3) and +(180:0.4) .. (b) .. controls +(0:0.4) and +(140:0.3) .. (p3);
			\draw [decorate,decoration={snake}] (t) -- (b);
			\foreach \i in {p1,p2,p3,p4} \draw [fill=white] (\i) circle [radius=2pt];
			\foreach \i in {t,b} \fill (\i) circle [radius=2pt];
			\node [anchor=center] at (0,-1.5) {\small $E_{\langle3333\rangle}^4$};
		\end{scope}
		\begin{scope}[thick,xshift=7.2cm]
    		\coordinate [label=left:{\small $1$}] (p1) at (140:1.5);
    		\coordinate [label=left:{\small $2$}] (p2) at (-140:1.5);
    		\coordinate [label=right:{\small $3$}] (p3) at (-40:1.5);
    		\coordinate [label=right:{\small $4$}] (p4) at (40:1.5);
    		\coordinate (l) at ($(p1)!.5!(p2)$);
    		\coordinate (r) at ($(p3)!.5!(p4)$);
    		\coordinate (t) at (0,0.6);
    		\coordinate (b) at (0,-0.6);
    		\draw (p1) -- (p2) -- (p3) -- (p4) -- cycle;
    		\draw [-Latex] (p1) -- ($(p1)!.6!(p4)$);
    		\draw [-Latex] (p4) -- ($(p4)!.6!(r)$);
    		\draw [-Latex] (r) -- ($(r)!.6!(p3)$);
    		\draw [-Latex] (p3) -- ($(p3)!.6!(p2)$);
    		\draw [-Latex] (p2) -- ($(p2)!.6!(l)$);
    		\draw [-Latex] (l) -- ($(l)!.6!(p1)$);
    		\draw [RedOrange,dashed] (p1) .. controls +(-40:0.3) and +(180:0.4) .. (t) .. controls +(0:0.4) and +(-140:0.3) .. (p4) (p2) .. controls +(40:0.3) and +(180:0.4) .. (b) .. controls +(0:0.4) and +(140:0.3) .. (p3);
    		\draw [decorate,decoration={snake}] (l) -- (r);
    		\foreach \i in {p1,p2,p3,p4} \draw [fill=white] (\i) circle [radius=2pt];
    		\foreach \i in {l,r} \fill (\i) circle [radius=2pt];
    		\node [anchor=center] at (0,-1.5) {\small $E_{\langle3333\rangle}^2$};
		\end{scope}
		\begin{scope}[thick]
			\coordinate [label=left:{\small $1$}] (p1) at (140:1.5);
			\coordinate [label=left:{\small $2$}] (p2) at (-140:1.5);
			\coordinate [label=right:{\small $3$}] (p3) at (-40:1.5);
			\coordinate [label=right:{\small $4$}] (p4) at (40:1.5);
			\coordinate (t) at ($(p1)!.5!(p4)$);
			\coordinate (b) at ($(p2)!.5!(p3)$);
			\coordinate (l) at (-0.7,0);
			\coordinate (r) at (0.7,0);
			\draw (p1) -- (p2) -- (p3) -- (p4) -- cycle;
			\draw [-Latex] (p1) -- ($(p1)!.6!(t)$);
			\draw [-Latex] (t) -- ($(t)!.6!(p4)$);
			\draw [-Latex] (p4) -- ($(p4)!.6!(p3)$);
			\draw [-Latex] (p3) -- ($(p3)!.6!(b)$);
			\draw [-Latex] (b) -- ($(b)!.6!(p2)$);
			\draw [-Latex] (p2) -- ($(p2)!.6!(p1)$);
			\draw [RedOrange,dashed] (p4) .. controls +(-140:0.3) and +(90:0.4) .. (r) .. controls +(-90:0.4) and +(140:0.3) .. (p3) (p1) .. controls +(-40:0.3) and +(90:0.4) .. (l) .. controls +(-90:0.4) and +(40:0.3) .. (p2);
			\draw [decorate,decoration={snake}] (t) -- (b);
			\foreach \i in {p1,p2,p3,p4} \draw [fill=white] (\i) circle [radius=2pt];
			\foreach \i in {t,b} \fill (\i) circle [radius=2pt];
			\node [anchor=center] at (0,-1.5) {\small $E_{\langle3333\rangle}^1$};
		\end{scope}
		\begin{scope}[thick,xshift=9cm,yshift=-3.2cm]
			\coordinate [label=left:{\small $1$}] (p1) at (140:1.5);
			\coordinate [label=left:{\small $2$}] (p2) at (-140:1.5);
			\coordinate [label=right:{\small $3$}] (p3) at (-40:1.5);
			\coordinate [label=right:{\small $4$}] (p4) at (40:1.5);
			\coordinate (o) at (0,0);
			\draw (p1) -- (p2) -- (p3) -- (p4) -- cycle;
			\draw [-Latex] (p1) -- ($(p1)!.6!(p4)$);
			\draw [-Latex] (p4) -- ($(p4)!.6!(p3)$);
			\draw [-Latex] (p3) -- ($(p3)!.6!(p2)$);
			\draw [-Latex] (p2) -- ($(p2)!.6!(p1)$);
			\draw [dashed] (p1) -- (p3) (p2) -- (p4);
			\foreach \i in {p1,p2,p3,p4} \draw [fill=white] (\i) circle [radius=2pt];
			\foreach \i in {o} \fill (\i) circle [radius=2pt];
			\node [anchor=center] at (0,-1.5) {\small $C_{\langle3333\rangle}^3$};
		\end{scope}
		\begin{scope}[thick,xshift=10.8cm]
    		\coordinate [label=left:{\small $1$}] (p1) at (140:1.5);
    		\coordinate [label=left:{\small $2$}] (p2) at (-140:1.5);
    		\coordinate [label=right:{\small $3$}] (p3) at (-40:1.5);
    		\coordinate [label=right:{\small $4$}] (p4) at (40:1.5);
    		\coordinate (t) at (0,0.65);
    		\coordinate (b) at (0,-0.65);
    		\coordinate (o) at (0,0);
    		\draw (p1) -- (p4) -- (p2) -- (p3) -- cycle;
    		\draw [-Latex] (p1) -- ($(p1)!.6!(p4)$);
    		\draw [-Latex] (p4) -- ($(p4)!.6!(o)$);
    		\draw [-Latex] (o) -- ($(o)!.6!(p3)$);
    		\draw [-Latex] (p3) -- ($(p3)!.6!(p2)$);
    		\draw [-Latex] (p2) -- ($(p2)!.6!(o)$);
    		\draw [-Latex] (o) -- ($(o)!.6!(p1)$);
    		\draw [RedOrange,dashed] (p1) .. controls +(-30:0.3) and +(180:0.4) .. (t) .. controls +(0:0.4) and +(-150:0.3) .. (p4) (p2) .. controls +(30:0.3) and +(180:0.4) .. (b) .. controls +(0:0.4) and +(150:0.3) .. (p3);
    		\foreach \i in {p1,p2,p3,p4} \draw [fill=white] (\i) circle [radius=2pt];
    		\foreach \i in {o} \fill (\i) circle [radius=2pt];
    		\node [anchor=center] at (0,-1.5) {\small $C_{\langle3333\rangle}^2$};
		\end{scope}
		\begin{scope}[thick,xshift=3.6cm]
			\coordinate [label=left:{\small $1$}] (p1) at (140:1.5);
			\coordinate [label=left:{\small $2$}] (p2) at (-140:1.5);
			\coordinate [label=right:{\small $3$}] (p3) at (-40:1.5);
			\coordinate [label=right:{\small $4$}] (p4) at (40:1.5);
			\coordinate (o) at (0,0);
			\coordinate (l) at (-0.8,0);
			\coordinate (r) at (0.8,0);
			\draw (p1) -- (p3) -- (p4) -- (p2) -- cycle;
			\draw [-Latex] (p1) -- ($(p1)!.6!(o)$);
			\draw [-Latex] (o) -- ($(o)!.6!(p4)$);
			\draw [-Latex] (p4) -- ($(p4)!.6!(p3)$);
			\draw [-Latex] (p3) -- ($(p3)!.6!(o)$);
			\draw [-Latex] (o) -- ($(o)!.6!(p2)$);
			\draw [-Latex] (p2) -- ($(p2)!.6!(p1)$);
			\draw [RedOrange,dashed] (p4) .. controls +(-120:0.3) and +(90:0.4) .. (r) .. controls +(-90:0.4) and +(120:0.3) .. (p3) (p1) .. controls +(-60:0.3) and +(90:0.4) .. (l) .. controls +(-90:0.4) and +(60:0.3) .. (p2);
			\foreach \i in {p1,p2,p3,p4} \draw [fill=white] (\i) circle [radius=2pt];
			\foreach \i in {o} \fill (\i) circle [radius=2pt];
			\node [anchor=center] at (0,-1.5) {\small $C_{\langle3333\rangle}^1$};
		\end{scope}
	\end{tikzpicture}
	\caption{Exchange diagrams and contact diagrams in the case of $\langle 3333 \rangle$.}
	\label{fig:case3333}
\end{figure}

\paragraph{\underline{$\langle 3333 \rangle $}}~\\

\noindent The correlator $\langle3333 \rangle$ consists of seven diagrams which are shown in \Cref{fig:case3333}. Following the same logic as in $\langle2233 \rangle$, we first isolate four of these diagrams which can be obtained from the diagrams in $\langle2222\rangle$  by adding two pairs of $\phi$ contractions. These diagrams naturally group into $\mathcal{H}$ functions
\begin{subequations}\label{eq:3333trivialdiagrams}
    \begin{align}
        E_{\langle3333\rangle}^1+C_{\langle3333\rangle}^1&=\lambda\,\mathcal{H}_{41,23}\times\left(\frac{t_{12}}{x_{12}^2}\frac{t_{34}}{x_{34}^2}\right),\\
        E_{\langle3333\rangle}^2+C_{\langle3333\rangle}^2&=\lambda\,\mathcal{H}_{12,34}\times\left(\frac{t_{14}}{x_{14}^2}\frac{t_{23}}{x_{23}^2}\right).
    \end{align}
\end{subequations}

The remaining three diagrams cannot be obtained in this way. In particular, the contact diagram $C_{\langle 3333\rangle}^3$ contains a new four-point vertex $\phi\phi\phi\phi$ that did not appear in $\langle2222\rangle$ and $\langle2233\rangle$. Explicitly evaluating these diagrams, we get
\begin{subequations}
    \begin{align}
        E^3_{\langle 3333 \rangle}&= \frac{\lambda}{2} \, d_{12} d_{23} d_{34} d_{41} \times \F_{12,34}  \times \left(\frac{t_{12}}{x^2_{12}} \frac{t_{34}}{x^2_{34}}\right) \, ,\\
        E^4_{\langle 3333 \rangle}&= \frac{\lambda}{2} \, d_{12} d_{23} d_{34} d_{41} \times \F_{41,23}  \times \left(\frac{t_{14}}{x^2_{14}} \frac{t_{23}}{x^2_{23}}\right) \, ,\\
        C^3_{\langle 3333 \rangle} &=\frac{\lambda}{2} \, d_{12} d_{23} d_{34} d_{41} \left(t_{14}t_{23}+t_{12}t_{34}-2t_{13}t_{24}\right)  \times \X_{1234} \,.
    \end{align}
\end{subequations}
Although not obvious, by carefully splitting the internal symmetry structure of the contact diagram $C_{\langle3333\rangle}^3$ into two parts, and combining them with the exchange diagrams $E_{\langle3333\rangle}^3$ and $E_{\langle3333\rangle}^4$ respectively, one can again make up two $\mathcal{H}$ functions. As a result, the sum of these three diagrams becomes
\begin{align}\label{eq:Feytop3}
    E^3_{\langle 3333 \rangle}+E^4_{\langle 3333 \rangle}+C^3_{\langle 3333 \rangle} &=\lambda\,\H_{12,34} \times \left(\frac{t_{12}}{x^2_{12}} \frac{t_{34}}{x^2_{34}}\right)+\lambda\, \H_{41,23} \times \left(\frac{t_{14}}{x^2_{14}} \frac{t_{23}}{x^2_{23}}\right) \, .
\end{align}
In a sense, we can interpret this as that the $\phi\phi\phi\phi$ contact interaction effectively splits into a sum of two possible Wick contractions between pairs of $\phi$ fields.

Adding up all the contributions \cref{eq:3333trivialdiagrams} and \cref{eq:Feytop3}, we find the full result factorizes 
\begin{equation}\label{res3333}
    \langle 3333 \rangle = \sum_{i}^{4} E^i_{\langle 3333 \rangle}+\sum_{j}^{3} C^j_{\langle 3333 \rangle} = \underbrace{\lambda \,  \left(\H_{12,34}+\H_{41,23}\right)}_{\langle2222\rangle} \left( \frac{t_{12}}{x^2_{12}} \frac{t_{34}}{x^2_{34}}+\frac{t_{14}}{x^2_{14}} \frac{t_{23}}{x^2_{23}}\right) \, ,
\end{equation}
which is again proportional to $\langle 2222 \rangle$. Note that the overall factor is given by all the possible Wick contractions among $\phi$ fields that do not violate planarity.

\subsection{Computational strategy and result for four arbitrary mesons}

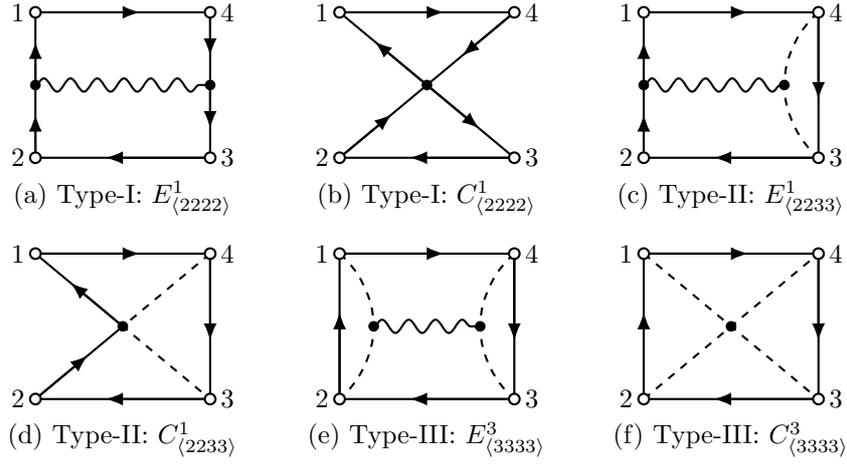
\begin{figure}[ht]
	\centering
	\begin{tikzpicture}
		\begin{scope}[thick]
			\coordinate [label=left:{\small $1$}] (p1) at (140:1.5);
			\coordinate [label=left:{\small $2$}] (p2) at (-140:1.5);
			\coordinate [label=right:{\small $3$}] (p3) at (-40:1.5);
			\coordinate [label=right:{\small $4$}] (p4) at (40:1.5);
			\coordinate (l) at ($(p1)!.5!(p2)$);
			\coordinate (r) at ($(p3)!.5!(p4)$);
			\draw (p1) -- (p2) -- (p3) -- (p4) -- cycle;
			\draw [-Latex] (p1) -- ($(p1)!.6!(p4)$);
			\draw [-Latex] (p4) -- ($(p4)!.6!(r)$);
			\draw [-Latex] (r) -- ($(r)!.6!(p3)$);
			\draw [-Latex] (p3) -- ($(p3)!.6!(p2)$);
			\draw [-Latex] (p2) -- ($(p2)!.6!(l)$);
			\draw [-Latex] (l) -- ($(l)!.6!(p1)$);
			\draw [decorate,decoration={snake}] (l) -- (r);
			\foreach \i in {p1,p2,p3,p4} \draw [fill=white] (\i) circle [radius=2pt];
			\foreach \i in {l,r} \fill (\i) circle [radius=2pt];
			\node [anchor=center] at (0,-1.5) {\small (a) Type-I: $E_{\langle2222\rangle}^1$};
		\end{scope}
		\begin{scope}[thick,xshift=4cm]
			\coordinate [label=left:{\small $1$}] (p1) at (140:1.5);
			\coordinate [label=left:{\small $2$}] (p2) at (-140:1.5);
			\coordinate [label=right:{\small $3$}] (p3) at (-40:1.5);
			\coordinate [label=right:{\small $4$}] (p4) at (40:1.5);
			\coordinate (o) at (0,0);
			\draw (p1) -- (p4) -- (p2) -- (p3) -- cycle;
			\draw [-Latex] (p1) -- ($(p1)!.6!(p4)$);
			\draw [-Latex] (p4) -- ($(p4)!.6!(o)$);
			\draw [-Latex] (o) -- ($(o)!.6!(p3)$);
			\draw [-Latex] (p3) -- ($(p3)!.6!(p2)$);
			\draw [-Latex] (p2) -- ($(p2)!.6!(o)$);
			\draw [-Latex] (o) -- ($(o)!.6!(p1)$);
			\foreach \i in {p1,p2,p3,p4} \draw [fill=white] (\i) circle [radius=2pt];
			\foreach \i in {o} \fill (\i) circle [radius=2pt];
			\node [anchor=center] at (0,-1.5) {\small (b) Type-I: $C_{\langle2222\rangle}^1$};
		\end{scope}
		\begin{scope}[thick,xshift=8cm]
			\coordinate [label=left:{\small $1$}] (p1) at (140:1.5);
			\coordinate [label=left:{\small $2$}] (p2) at (-140:1.5);
			\coordinate [label=right:{\small $3$}] (p3) at (-40:1.5);
			\coordinate [label=right:{\small $4$}] (p4) at (40:1.5);
			\coordinate (l) at ($(p1)!.5!(p2)$);
			\coordinate (r) at (0.7,0);
			\draw (p1) -- (p2) -- (p3) -- (p4) -- cycle;
			\draw [-Latex] (p1) -- ($(p1)!.6!(p4)$);
			\draw [-Latex] (p4) -- ($(p4)!.6!(p3)$);
			\draw [-Latex] (p3) -- ($(p3)!.6!(p2)$);
			\draw [-Latex] (p2) -- ($(p2)!.6!(l)$);
			\draw [-Latex] (l) -- ($(l)!.6!(p1)$);
			\draw [dashed] (p4) .. controls +(-140:0.3) and +(90:0.4) .. (r) .. controls +(-90:0.4) and +(140:0.3) .. (p3);
			\draw [decorate,decoration={snake}] (l) -- (r);
			\foreach \i in {p1,p2,p3,p4} \draw [fill=white] (\i) circle [radius=2pt];
			\foreach \i in {l,r} \fill (\i) circle [radius=2pt];
			\node [anchor=center] at (0,-1.5) {\small (c) Type-II: $E_{\langle2233\rangle}^1$};
		\end{scope}
		\begin{scope}[thick,yshift=-3.2cm]
		\coordinate [label=left:{\small $1$}] (p1) at (140:1.5);
		\coordinate [label=left:{\small $2$}] (p2) at (-140:1.5);
		\coordinate [label=right:{\small $3$}] (p3) at (-40:1.5);
		\coordinate [label=right:{\small $4$}] (p4) at (40:1.5);
		\coordinate (o) at (0,0);
		\draw (p1) -- (p4) -- (p3) -- (p2) -- (o) -- cycle;
		\draw [-Latex] (p1) -- ($(p1)!.6!(p4)$);
		\draw [-Latex] (p4) -- ($(p4)!.6!(p3)$);
		\draw [-Latex] (p3) -- ($(p3)!.6!(p2)$);
		\draw [-Latex] (p2) -- ($(p2)!.6!(o)$);
		\draw [-Latex] (o) -- ($(o)!.6!(p1)$);
		\draw [dashed] (p4) -- (o) -- (p3);
		\foreach \i in {p1,p2,p3,p4} \draw [fill=white] (\i) circle [radius=2pt];
		\foreach \i in {o} \fill (\i) circle [radius=2pt];
		\node [anchor=center] at (0,-1.5) {\small (d) Type-II: $C_{\langle2233\rangle}^1$};
		\end{scope}
		\begin{scope}[thick,xshift=4cm,yshift=-3.2cm]
		\coordinate [label=left:{\small $1$}] (p1) at (140:1.5);
		\coordinate [label=left:{\small $2$}] (p2) at (-140:1.5);
		\coordinate [label=right:{\small $3$}] (p3) at (-40:1.5);
		\coordinate [label=right:{\small $4$}] (p4) at (40:1.5);
		\coordinate (l) at (-0.7,0);
		\coordinate (r) at (0.7,0);
		\draw (p1) -- (p2) -- (p3) -- (p4) -- cycle;
		\draw [-Latex] (p1) -- ($(p1)!.6!(p4)$);
		\draw [-Latex] (p4) -- ($(p4)!.6!(p3)$);
		\draw [-Latex] (p3) -- ($(p3)!.6!(p2)$);
		\draw [-Latex] (p2) -- ($(p2)!.6!(p1)$);
		\draw [dashed] (p4) .. controls +(-140:0.3) and +(90:0.4) .. (r) .. controls +(-90:0.4) and +(140:0.3) .. (p3) (p1) .. controls +(-40:0.3) and +(90:0.4) .. (l) .. controls +(-90:0.4) and +(40:0.3) .. (p2);
		\draw [decorate,decoration={snake}] (l) -- (r);
		\foreach \i in {p1,p2,p3,p4} \draw [fill=white] (\i) circle [radius=2pt];
		\foreach \i in {l,r} \fill (\i) circle [radius=2pt];
		\node [anchor=center] at (0,-1.5) {\small (e) Type-III: $E_{\langle3333\rangle}^3$};
		\end{scope}
		\begin{scope}[thick,xshift=8cm,yshift=-3.2cm]
		\coordinate [label=left:{\small $1$}] (p1) at (140:1.5);
		\coordinate [label=left:{\small $2$}] (p2) at (-140:1.5);
		\coordinate [label=right:{\small $3$}] (p3) at (-40:1.5);
		\coordinate [label=right:{\small $4$}] (p4) at (40:1.5);
		\coordinate (o) at (0,0);
		\draw (p1) -- (p2) -- (p3) -- (p4) -- cycle;
		\draw [-Latex] (p1) -- ($(p1)!.6!(p4)$);
		\draw [-Latex] (p4) -- ($(p4)!.6!(p3)$);
		\draw [-Latex] (p3) -- ($(p3)!.6!(p2)$);
		\draw [-Latex] (p2) -- ($(p2)!.6!(p1)$);
		\draw [dashed] (p1) -- (p3) (p2) -- (p4);
		\foreach \i in {p1,p2,p3,p4} \draw [fill=white] (\i) circle [radius=2pt];
		\foreach \i in {o} \fill (\i) circle [radius=2pt];
		\node [anchor=center] at (0,-1.5) {\small (f) Type-III: $C_{\langle3333\rangle}^3$};
		\end{scope}
	\end{tikzpicture}
	\caption{Representatives of three types of elementary diagrams.}
	\label{fig:general4type}
\end{figure}

We now move onto correlators with arbitrary meson states $\langle p_1p_2p_3p_4\rangle$. It is not difficult to see that in the general case all diagrams can be generated from the ones already appearing in $\langle2222\rangle$, $\langle2233\rangle$ (plus its cyclic permutations) and $\langle3333\rangle$. Therefore, we distinguish those elementary diagrams where no further $\phi$ contractions can be removed. Such diagrams can be classified into three types, according to the number of regions cut out by the propagators
\begin{subequations}\label{eq:elementary4pt}
    \begin{align}
        \text{Type-I:}& \quad E^1_{\langle 2222 \rangle},\; C^1_{\langle 2222 \rangle},\; E^2_{\langle2222\rangle},\; C^2_{\langle2222\rangle}  \,.\label{elementary} \\
        \text{Type-II:} & \quad E^1_{\langle 2233 \rangle},\; {C^1_{\langle 2233 \rangle}} ,\; E^1_{\langle 2332 \rangle},\; {C^1_{\langle 2332 \rangle}} ,\; E^1_{\langle 3322 \rangle},\; {C^1_{\langle 3322 \rangle}} ,\; E^1_{\langle 3223 \rangle},\; {C^1_{\langle 3223 \rangle}} \,. \\
        \text{Type-III:}  & \quad E^3_{\langle 3333 \rangle},\; E^4_{\langle 3333 \rangle} ,\; C^3_{\langle 3333 \rangle} \,.
\end{align}
\end{subequations}
We list the representatives of these elementary diagrams in \Cref{fig:general4type}, up to cyclic permutation of the labels.

With this observation, we have the following strategy for computing general four-point functions. Going through the list of elementary diagrams, we enumerate all the possible Wick contractions for the remaining $\phi$ fields in the four operators. The Wick contractions are constrained by the requirement that the final diagram dressed by the $\phi$ contractions remains planar. The complete correlator is then obtained by summing over all the dressed diagrams. In dressing with $\phi$ contractions, the additional dashed lines must be drawn inside each individual regions of the elementary diagrams. Since diagrams with equivalent regions (like (a), (b) in \Cref{fig:general4type}) are always dressed in the same way, we can further regroup them together and work with the following combinations 
\begin{subequations}\label{eq:elementary4pteff}
    \begin{align}
        \text{Type-I:}& \quad \lambda\H_{12,34},\;\lambda\H_{41,23}  \,, \label{FeyntypeI}\\
        \text{Type-II:} & \quad \lambda\H_{12,34}\times\frac{t_{34}}{x_{34}^2},\;\lambda\H_{41,23}\times\frac{t_{14}}{x_{14}^2},\;\lambda\H_{34,12}\times\frac{t_{12}}{x_{12}^2},\;\lambda\H_{23,41}\times\frac{t_{23}}{x_{23}^2} \,, \label{FeyntypeII}\\
        \text{Type-III:}  & \quad \lambda \,  \left(\H_{12,34}+\H_{41,23}\right)\times \left( \frac{t_{12}}{x^2_{12}} \frac{t_{34}}{x^2_{34}}+\frac{t_{14}}{x^2_{14}} \frac{t_{23}}{x^2_{23}}\right) \,. \label{FeyntypeIII}
\end{align}
\end{subequations}
Here we note $\H_{34,12}=\H_{12,34}$ and $\H_{23,41}=\H_{41,23}$. Therefore, only two $\H$ functions appear in four-point functions. We further note that after regrouping the elementary diagrams, the building blocks (\ref{FeyntypeII}) for Type-II has a factorized propagator in addition to $\H$. This can be viewed as the effective reduction back to the case of Type-I together with the pair of $\phi$ being reinserted into the operators and then Wick contracted. The same also happens for Type-III where the reduction is accompanied by the Wick contraction of four $\phi$. All in all, we find that this procedure is equivalent to starting with only the two elementary combinations in Type-I and then sum over all possible dressings of $\phi$ contractions that are compatible with planarity. In particular, at four points, such $\phi$ contractions can only occur between operators that are adjacent in the planar ordering, hence there is no contraction between $13$ and $24$ in the canonically ordered partial correlators.

From this point of view of dressing, if we denote the number of $\phi$ contractions between point $i$ and $j$ as $b_{ij}$, the general result of the four-point function $\langle p_1p_2p_3p_4\rangle$ can be expressed as
\begin{align}\label{eq:anymeson4pt}
    \langle p_1 p_2 p_3 p_4 \rangle= \underbrace{\lambda \, (\mathcal{H}_{12,34}+\H_{41,23})}_{\langle2222\rangle} \times \sum_{b_{ij}} \prod_{ij\in\{12,23,34,41\}} \left(\frac{t_{ij}}{x_{ij}^2} \right)^{b_{ij}} \, ,
\end{align}
where the $b$ summation is performed over all solutions to
\begin{align}
    p_1=2+b_{41}+b_{12}, \qquad  &p_2=2+b_{12}+b_{23}, \nonumber \\
    p_3=2+b_{23}+b_{34}, \qquad  &p_4=2+b_{34}+b_{41} \, ,\label{eq:constrain}
\end{align}
with $b_{ij}\geq 0$. It is worth to note that, with the canonincal ordering in the flavor decomposition, partial correlators may vanish for some of the meson states configurations, e.g., $\langle2323\rangle$. They vanish because the corresponding values of $\{p_i\}$ do not allow any solutions to \cref{eq:constrain}. The non-vanishing correlators are always proportional to $\langle2222\rangle$, which contains the $R$ factor \cref{eq:Rdef}. Therefore the above results manifestly obey the superconformal Ward identities.

\subsection{Generating function and eight dimensional structures}

Given the structure of the general result \cref{eq:anymeson4pt} for arbitrary meson states, it is more convenient and intuitive to further perform a sum over all the correlators
\begin{align}\label{eq:Gbb4}
    \mathbb{G}_4=\sum_{p_1,p_2,p_3,p_4=2}^\infty\langle p_1p_2p_3p_4\rangle\,.
\end{align}
Substituting \cref{eq:anymeson4pt} into the RHS, we obtain a double summation first over $b_{ij}$ constrained by \cref{eq:constrain}, then over $p_i$. When we exchange the order of summation, the condition \cref{eq:constrain} becomes automatic and the sum over $b_{ij}$ becomes unconstrained. 
Since all these correlators are proportional to the one with the lowest weights, we can factor out $\langle 2222\rangle$ and rewrite \cref{eq:Gbb4} as
\begin{align}
    \mathbb{G}_4=\langle2222\rangle\times\sum_{b_{ij}=0}^\infty\prod_{ij\in\{12,23,34,41\}} \left(\frac{t_{ij}}{x_{ij}^2} \right)^{b_{ij}}\,,
\end{align}
where the sum over each $b_{ij}$ is just a geometric series. It is convenient to introduce the eight-dimensional distance $X_{ij}^2=x_{ij}^2-t_{ij}$, then the geometric series simply sums into
\begin{equation}
    \sum_{b_{ij}=0}^\infty\left(\frac{t_{ij}}{x_{ij}^2}\right)^{b_{ij}}=\frac{x_{ij}^2}{X_{ij}^2}\;.
\end{equation}
We can then rewrite (\ref{eq:Gbb4}) in a compact way as
\begin{equation}\label{eq:4pointgenerating}
    \mathbb{G}_4 = \lambda \frac{R \X_{1234}}{X_{12}^2 X_{23}^2 X_{34}^2 X_{41}^2 }  , \quad\quad  \X_{1234}= \int d^4 x_{0} \, \, \frac{1}{\prod_{i=1}^{4}{X_{i 0}^2}}\;, 
\end{equation}
where we have only used the eight-dimensional distances $X_{ij}^2$. Note here  $X_{i0}^2=x_{i0}^2$ and we should restrict the integrated point to the four-dimensional subspace. This rewriting makes manifest an eight-dimensional hidden symmetry at the level of integrands, which has been observed in the strongly coupled regime for the whole correlators  \cite{Alday:2021odx}. We also note that the structure in (\ref{eq:4pointgenerating}) parallels that of $\mathcal{N}=4$ SYM at weak coupling \cite{Caron-Huot:2021usw}, except that the distances $X_{ij}^2$ are ten-dimensional.

On the other hand, $\mathbb{G}_4$ can be viewed as a generating function of four-point correlators. Any four-point function $\langle p_1p_2p_3p_4\rangle$ can be extracted from this function by singling out the terms with the appropriate internal symmetry structures, which amounts to reinstating the constraints \cref{eq:constrain}. Explicitly, this is given by the operation
\begin{equation}\label{eq:4pointres}
    \langle p_1 p_2 p_3 p_4 \rangle=\left[   \mathbb{G}_4 \big|_{t_{ij} \to \beta_i \beta_j t_{ij}}\right]_{\prod_{i=1}^4 \beta_i^{p_i-2}} \,.
\end{equation}
It is also helpful to  view $\mathbb{G}_4$ as the four-point correlation of a master operator which is defined by summing up all the meson operators with different weights
\begin{equation}
    \mathbb{M}(x;v,\vb)=\sum_{p=2}^{\infty}\mathcal{M}_{p}(x;v,\vb) \;.
\end{equation}
We can then consider more generally correlation functions of these master operators for any number of points $n$
\begin{equation}
    \mathbb{G}_n = \langle \mathbb{M}(x_1;v_1,\vb_1) \mathbb{M}(x_2;v_2,\vb_1) \cdots \mathbb{M}(x_n;v_n,\vb_n)  \rangle \, ,
\end{equation}
which serves as generating function for $n$-point meson correlators in a similar spirit as \cref{eq:4pointres}.

\section{Higher-point functions}\label{Sec:higherptfun}

In this section, we compute the one-loop corrections to higher-point correlators and explore how the eight-dimensional structure is generalized. Our strategy is to extend the results from the four-point case where Feynman diagrams conspire to form the building block function $\mathcal{H}$. We expect that a similar phenomenon should also occur at higher points because the one-loop corrections essentially affects only four points at a time. To see explicitly how it arises, we analyze the five-point and six-point functions in detail which are also useful for illustrating a number of subtleties. From these examples, we extract the essential features of the one-loop corrections which hold also at higher points. This allow us to write down effective Feynman rules which are analogous to the ones for four-point functions. We then construct generating functions which package the $n$-point functions in a compact way.

\subsection{Five-point functions}
As in the four-point function case, we start with $\langle22222\rangle$ to see how the combination of diagrams works in the simplest example. We then look at the more complicated correlator $\langle43342\rangle$ which exhibits more features. These two examples turn out to be nontrivial enough for us to extract the effective Feynman rules which we will use to write down a generating function for five-point functions.

\begin{figure}[ht]
	\centering
	\begin{tikzpicture}
		\begin{scope}[thick]
			\coordinate [label=left:{\small $1$}] (p1) at (-140:1.5);
			\coordinate [label=right:{\small $2$}] (p2) at (-40:1.5);
			\coordinate [label=right:{\small $3$}] (p3) at (40:1.5);
			\coordinate [label=left:{\small $4$}] (p4) at (140:1.5);
			\coordinate [label=left:{\small $5$}] (p5) at (180:1.5);
			\coordinate (m12) at ($(p1)!.5!(p2)$);
			\coordinate (m34) at ($(p3)!.5!(p4)$);
			\draw (p1) -- (p2) -- (p3) -- (p4) -- (p5) -- cycle;
			\draw [-Latex] (p1) -- ($(p1)!.6!(p5)$);
			\draw [-Latex] (p5) -- ($(p5)!.6!(p4)$);
			\draw [-Latex] (p4) -- ($(p4)!.6!(m34)$);
			\draw [-Latex] (m34) -- ($(m34)!.6!(p3)$);
			\draw [-Latex] (p3) -- ($(p3)!.6!(p2)$);
			\draw [-Latex] (p2) -- ($(p2)!.6!(m12)$);
			\draw [-Latex] (m12) -- ($(m12)!.6!(p1)$);
			\draw [decorate,decoration={snake}] (m12) -- (m34);
			\foreach \i in {p1,p2,p3,p4,p5} \draw [fill=white] (\i) circle [radius=2pt];
			\foreach \i in {m12,m34} \fill (\i) circle [radius=2pt];
			\node [anchor=center] at (0,-1.5) {\small $E_{\langle22222\rangle}$};
		\end{scope}
		\begin{scope}[thick,xshift=4cm]
    		\coordinate [label=left:{\small $1$}] (p1) at (-140:1.5);
    		\coordinate [label=right:{\small $2$}] (p2) at (-40:1.5);
    		\coordinate [label=right:{\small $3$}] (p3) at (40:1.5);
    		\coordinate [label=left:{\small $4$}] (p4) at (140:1.5);
    		\coordinate [label=left:{\small $5$}] (p5) at (180:1.5);
    		\coordinate (o) at (0,0);
    		\draw (p1) -- (p5) -- (p4) -- (p2) -- (p3) -- cycle;
    		\draw [-Latex] (p1) -- ($(p1)!.6!(p5)$);
    		\draw [-Latex] (p5) -- ($(p5)!.6!(p4)$);
    		\draw [-Latex] (p4) -- ($(p4)!.6!(o)$);
    		\draw [-Latex] (o) -- ($(o)!.6!(p3)$);
    		\draw [-Latex] (p3) -- ($(p3)!.6!(p2)$);
    		\draw [-Latex] (p2) -- ($(p2)!.6!(o)$);
    		\draw [-Latex] (o) -- ($(o)!.6!(p1)$);
    		\foreach \i in {p1,p2,p3,p4,p5} \draw [fill=white] (\i) circle [radius=2pt];
    		\foreach \i in {o} \fill (\i) circle [radius=2pt];
    		\node [anchor=center] at (0,-1.5) {\small $C_{\langle22222\rangle}$};
		\end{scope}
	\end{tikzpicture}
	\caption{Exchange diagrams and contact diagrams in the case of $\langle22222 \rangle$.}
	\label{fig:case22222}
\end{figure}

\paragraph{\underline{$\langle 22222 \rangle $}}~\\

\noindent The one-loop correction to $\langle22222\rangle$ consists of the two diagrams shown in \Cref{fig:case22222} and their cyclic permutations. By comparing them with the diagrams for $\langle2222\rangle$ in \Cref{fig:case2222}, one immediately notices the relations
\begin{align}
    E_{\langle 22222 \rangle} &=\frac{d_{45}d_{51}}{d_{41}}\times E_{\langle2222\rangle}^1=\frac{\lambda}{2} \, d_{12} d_{23} d_{34} d_{45}d_{51} \times \F_{12,34}  \, \\
    C_{\langle 22222 \rangle} &=\frac{d_{45}d_{51}}{d_{41}}\times C_{\langle2222\rangle}^1=\frac{\lambda}{2} \, d_{23} d_{45} d_{51} (2v_{14}v_{32}-v_{12}v_{34}) \times \X_{1234} \, .
\end{align}
In other words, the exchange and  contact diagrams in the five-point function are both proportional to their counterparts in $\langle2222\rangle$ with the same  factor. It is then straightforward to combine these two five-point diagrams and we get
\begin{align}
    E_{\langle22222\rangle}+C_{\langle22222\rangle}=\lambda\,\H_{12,34}\times\frac{d_{45}d_{51}}{d_{41}}\,.
\end{align}
The full result of $\langle22222\rangle$ is then obtained by summing over its cyclic permutations.

Note the original definition of $\mathcal{H}$ includes a product of $q\bar{q}$ contraction from the boundaries of the square, which makes it difficult to be generalized to higher points. We improve it by defining a new function $\bar{\mathcal{H}}$ where this factor is extracted
\begin{align}
    \Hb_{ij,kl}=\frac{\H_{ij,kl}}{d_{ij}d_{jk}d_{kl}d_{li}}\;.
\end{align}
In terms of $\bar{\mathcal{H}}$,  the combinations at four and five points read
\begin{subequations}\label{eq:EC1toHb}
    \begin{align}
        E_{\langle2222\rangle}^1+C_{\langle2222\rangle}^1&=\lambda\,d_4\,\Hb_{12,34}\,,\\
        E_{\langle 22222\rangle}+C_{\langle22222\rangle}&=\lambda\,d_5\,\Hb_{12,34}\,,
    \end{align}
\end{subequations}
where we have introduced the shorthand notation $d_n$ to denote the product of $\bar{q}q$ contractions along the $n$-gon
\begin{align}
    d_n=d_{12}d_{23}d_{34}\cdots d_{(n-1)n}d_{n1}\;.
\end{align}
The relations \cref{eq:EC1toHb} can in fact be understood from a more general perspective: Whenever a gluon exchange is inserted between two $\bar{q}q$ lines, it combines with the contact diagram into an insertion of the $\Hb$ function
\begin{align}\label{eq:Hbrule1}
    \parbox{2.3cm}{\tikz[thick]{
        \def\RR{1}
        \coordinate (m0) at (22.5:\RR);
        \coordinate [label=67.5:{\scriptsize $i$}] (m1) at (67.5:\RR);
        \coordinate [label=112.5:{\scriptsize $i\!+\!1$}] (m2) at (112.5:\RR);
        \coordinate (m3) at (157.5:\RR);
        \coordinate (m4) at (202.5:\RR);
        \coordinate [label=247.5:{\scriptsize $\!j\!$}] (m5) at (247.5:\RR);
        \coordinate [label=292.5:{\scriptsize $\!j+1\!$}] (m6) at (295.5:\RR);
        \coordinate (m7) at (337.5:\RR);
        \coordinate (t) at ($(m1)!.5!(m2)$);
        \coordinate (b) at ($(m5)!.5!(m6)$);
        \draw (m0) -- (m1) -- (m2) -- (m3) (m4) -- (m5) -- (m6) -- (m7);
        \draw [dotted] (m3) -- (m4) (m7) -- (m0);
        \midarrow{m7}{0.75}{m6}\midarrow{m5}{0.75}{m4}\midarrow{m3}{.75}{m2}\midarrow{m1}{.75}{m0}
        \draw [decorate,decoration={snake}] (t) -- (b);
        \foreach \i in {1,2,5,6} \draw [fill=white] (m\i) circle [radius=2pt];
        \foreach \i in {t,b} \fill (\i) circle [radius=2pt];
    }}\;+\;
    \parbox{2.3cm}{\tikz[thick]{
        \def\RR{1}
        \coordinate (m0) at (22.5:\RR);
        \coordinate [label=67.5:{\scriptsize $i$}] (m1) at (67.5:\RR);
        \coordinate [label=112.5:{\scriptsize $i\!+\!1$}] (m2) at (112.5:\RR);
        \coordinate (m3) at (157.5:\RR);
        \coordinate (m4) at (202.5:\RR);
        \coordinate [label=247.5:{\scriptsize $\!j\!$}] (m5) at (247.5:\RR);
        \coordinate [label=292.5:{\scriptsize $\!j+1\!$}] (m6) at (295.5:\RR);
        \coordinate (m7) at (337.5:\RR);
        \coordinate (o) at (0,0);
        \draw (m0) -- (m1) -- (o) -- (m2) -- (m3) (m4) -- (m5) -- (o) -- (m6) -- (m7);
        \draw [dotted] (m3) -- (m4) (m7) -- (m0);
        \midarrow{m7}{.7}{m6}\midarrow{m6}{.7}{o}\midarrow{o}{.7}{m5}\midarrow{m5}{.7}{m4}\midarrow{m3}{.7}{m2}\midarrow{m2}{.7}{o}\midarrow{o}{.7}{m1}\midarrow{m1}{.7}{m0}
        \foreach \i in {1,2,5,6} \draw [fill=white] (m\i) circle [radius=2pt];
        \foreach \i in {o} \fill (\i) circle [radius=2pt];
    }}\;=\;
    \parbox{2.3cm}{\tikz[thick]{
        \def\RR{1}
        \coordinate (m0) at (22.5:\RR);
        \coordinate [label=67.5:{\scriptsize $i$}] (m1) at (67.5:\RR);
        \coordinate [label=112.5:{\scriptsize $i\!+\!1$}] (m2) at (112.5:\RR);
        \coordinate (m3) at (157.5:\RR);
        \coordinate (m4) at (202.5:\RR);
        \coordinate [label=247.5:{\scriptsize $\!j\!$}] (m5) at (247.5:\RR);
        \coordinate [label=292.5:{\scriptsize $\!j+1\!$}] (m6) at (295.5:\RR);
        \coordinate (m7) at (337.5:\RR);
        \coordinate (t) at ($(m1)!.5!(m2)$);
        \coordinate (b) at ($(m5)!.5!(m6)$);
        \coordinate (o) at (0,0);
        \draw (m0) -- (m1) -- (m2) -- (m3) (m4) -- (m5) -- (m6) -- (m7);
        \draw [dotted] (m3) -- (m4) (m7) -- (m0);
        \midarrow{m7}{.7}{m6}\midarrow{m6}{.7}{m5}\midarrow{m5}{.7}{m4}\midarrow{m3}{.7}{m2}\midarrow{m2}{.7}{m1}\midarrow{m1}{.7}{m0}
        \draw [thin,fill=gray!20] (m1) .. controls +(-120:.2) and ($(o)!.5!(t)$) .. (o) .. controls ($(o)!.5!(b)$) and +(120:.2) .. (m6) .. controls +(120:.5) and +(60:.5) .. (m5) .. controls +(60:.2) and ($(o)!.5!(b)$) .. (o) .. controls ($(o)!.5!(t)$) and +(-60:.2) .. (m2) .. controls +(-60:.5) and +(-120:.5) .. cycle;
        \foreach \i in {1,2,5,6} \draw [fill=white] (m\i) circle [radius=2pt];
    }}\;,\quad
    \parbox{1.5cm}{\tikz[thick]{
        \def\RR{1}
        \coordinate [label=67.5:{\scriptsize $i$}] (m1) at (67.5:\RR);
        \coordinate [label=112.5:{\scriptsize $j$}] (m2) at (112.5:\RR);
        \coordinate [label=247.5:{\scriptsize $k$}] (m5) at (247.5:\RR);
        \coordinate [label=292.5:{\scriptsize $l$}] (m6) at (295.5:\RR);
        \coordinate (t) at ($(m1)!.5!(m2)$);
        \coordinate (b) at ($(m5)!.5!(m6)$);
        \coordinate (o) at (0,0);
        \draw [thin,fill=gray!20] (m1) .. controls +(-120:.2) and ($(o)!.5!(t)$) .. (o) .. controls ($(o)!.5!(b)$) and +(120:.2) .. (m6) .. controls +(120:.5) and +(60:.5) .. (m5) .. controls +(60:.2) and ($(o)!.5!(b)$) .. (o) .. controls ($(o)!.5!(t)$) and +(-60:.2) .. (m2) .. controls +(-60:.5) and +(-120:.5) .. cycle;
        \foreach \i in {1,2,5,6} \draw [fill=white] (m\i) circle [radius=2pt];
    }}\;=\;\lambda\,\Hb_{ij,kl}\,.
\end{align}
Here we use a shaded region with smooth boundaries to diagrammatically represent $\Hb_{ij,kl}$, and the region is pinched in the middle separating the external points into $i$, $j$ and $k$, $l$, to indicate the channel of the exchange.

\begin{figure}[t]
    \centering
    \begin{tikzpicture}[thick]
    	\begin{scope}[xshift=2cm,yshift=-3.8cm]
			\coordinate [label=left:{\small $1$}] (p1) at (-140:1.5);
			\coordinate [label=right:{\small $2$}] (p2) at (-40:1.5);
			\coordinate [label=right:{\small $3$}] (p3) at (40:1.5);
			\coordinate [label=left:{\small $4$}] (p4) at (140:1.5);
			\coordinate [label=left:{\small $5$}] (p5) at (180:1.5);
			\coordinate (l) at (-0.8,0);
			\coordinate (r) at (0.8,0);
			\draw (p1) -- (p2) -- (p3) -- (p4) -- (p5) -- cycle;
			\draw [-Latex] (p1) -- ($(p1)!.6!(p5)$);
			\draw [-Latex] (p5) -- ($(p5)!.6!(p4)$);
			\draw [-Latex] (p4) -- ($(p4)!.6!(p3)$);
			\draw [-Latex] (p3) -- ($(p3)!.6!(p2)$);
			\draw [-Latex] (p2) -- ($(p2)!.6!(p1)$);
			\draw [dashed] (p4) .. controls +(-60:0.3) and +(90:0.4) .. (l) .. controls +(-90:0.4) and +(60:0.3) .. (p1) (p3) .. controls +(-120:0.3) and +(90:0.4) .. (r) .. controls +(-90:0.4) and +(120:0.3) .. (p2);
			\draw [RedOrange,dashed] (p1) -- (p4);
			\draw [decorate,decoration={snake}] (l) -- (r);
			\foreach \i in {p1,p2,p3,p4,p5} \draw [fill=white] (\i) circle [radius=2pt];
			\foreach \i in {l,r} \fill (\i) circle [radius=2pt];
			\node [anchor=center] at (0,-1.5) {\small $E_{\langle43342\rangle}^3$};
		\end{scope}
		\begin{scope}[xshift=6cm,yshift=-3.8cm]
			\coordinate [label=left:{\small $1$}] (p1) at (-140:1.5);
			\coordinate [label=right:{\small $2$}] (p2) at (-40:1.5);
			\coordinate [label=right:{\small $3$}] (p3) at (40:1.5);
			\coordinate [label=left:{\small $4$}] (p4) at (140:1.5);
			\coordinate [label=left:{\small $5$}] (p5) at (180:1.5);
			\coordinate (t) at (0,0.6);
			\coordinate (b) at (0,-0.6);
			\draw (p1) -- (p2) -- (p3) -- (p4) -- (p5) -- cycle;
			\draw [-Latex] (p1) -- ($(p1)!.6!(p5)$);
			\draw [-Latex] (p5) -- ($(p5)!.6!(p4)$);
			\draw [-Latex] (p4) -- ($(p4)!.6!(p3)$);
			\draw [-Latex] (p3) -- ($(p3)!.6!(p2)$);
			\draw [-Latex] (p2) -- ($(p2)!.6!(p1)$);
			\draw [dashed] (p3) .. controls +(-140:0.3) and +(0:0.4) .. (t) .. controls +(180:0.4) and +(-40:0.3) .. (p4) (p1) .. controls +(40:0.3) and +(180:0.4) .. (b) .. controls +(0:0.4) and +(140:0.3) .. (p2);
			\draw [RedOrange,dashed] (p1) -- (p4);
			\draw [decorate,decoration={snake}] (t) -- (b);
			\foreach \i in {p1,p2,p3,p4,p5} \draw [fill=white] (\i) circle [radius=2pt];
			\foreach \i in {t,b} \fill (\i) circle [radius=2pt];
			\node [anchor=center] at (0,-1.5) {\small $E_{\langle43342\rangle}^4$};
		\end{scope}
		\begin{scope}[xshift=8cm]
			\coordinate [label=left:{\small $1$}] (p1) at (-140:1.5);
			\coordinate [label=right:{\small $2$}] (p2) at (-40:1.5);
			\coordinate [label=right:{\small $3$}] (p3) at (40:1.5);
			\coordinate [label=left:{\small $4$}] (p4) at (140:1.5);
			\coordinate [label=left:{\small $5$}] (p5) at (180:1.5);
			\coordinate (t) at (0,0.6);
			\coordinate (b) at (0,-0.6);
			\coordinate (m14) at ($(p1)!.5!(p4)$);
			\coordinate (m23) at ($(p2)!.5!(p3)$);
			\draw (p1) -- (p2) -- (p3) -- (p4) -- (p5) -- cycle;
			\draw [-Latex] (p1) -- ($(p1)!.6!(p5)$);
			\draw [-Latex] (p5) -- ($(p5)!.6!(p4)$);
			\draw [-Latex] (p4) -- ($(p4)!.6!(p3)$);
			\draw [-Latex] (p3) -- ($(p3)!.6!(m23)$);
			\draw [-Latex] (m23) -- ($(m23)!.6!(p2)$);
			\draw [-Latex] (p2) -- ($(p2)!.6!(p1)$);
			\draw [dashed] (p1) -- (p4);
			\draw [RedOrange,dashed] (p3) .. controls +(-140:0.3) and +(0:0.4) .. (t) .. controls +(180:0.4) and +(-40:0.3) .. (p4) (p1) .. controls +(40:0.3) and +(180:0.4) .. (b) .. controls +(0:0.4) and +(140:0.3) .. (p2);
			\draw [decorate,decoration={snake}] (m14) -- (m23);
			\foreach \i in {p1,p2,p3,p4,p5} \draw [fill=white] (\i) circle [radius=2pt];
			\foreach \i in {m14,m23} \fill (\i) circle [radius=2pt];
			\node [anchor=center] at (0,-1.5) {\small $E_{\langle43342\rangle}^2$};
		\end{scope}
		\begin{scope}
			\coordinate [label=left:{\small $1$}] (p1) at (-140:1.5);
			\coordinate [label=right:{\small $2$}] (p2) at (-40:1.5);
			\coordinate [label=right:{\small $3$}] (p3) at (40:1.5);
			\coordinate [label=left:{\small $4$}] (p4) at (140:1.5);
			\coordinate [label=left:{\small $5$}] (p5) at (180:1.5);
			\coordinate (l) at (-0.8,0);
			\coordinate (r) at (0.8,0);
			\coordinate (m12) at ($(p1)!.5!(p2)$);
			\coordinate (m34) at ($(p3)!.5!(p4)$);
			\draw (p1) -- (p2) -- (p3) -- (p4) -- (p5) -- cycle;
			\draw [-Latex] (p1) -- ($(p1)!.6!(p5)$);
			\draw [-Latex] (p5) -- ($(p5)!.6!(p4)$);
			\draw [-Latex] (p4) -- ($(p4)!.6!(m34)$);
			\draw [-Latex] (m34) -- ($(m34)!.6!(p3)$);
			\draw [-Latex] (p3) -- ($(p3)!.6!(p2)$);
			\draw [-Latex] (p2) -- ($(p2)!.6!(m12)$);
			\draw [-Latex] (m12) -- ($(m12)!.6!(p1)$);
			\draw [RedOrange,dashed] (p1) -- (p4) .. controls +(-60:0.3) and +(90:0.4) .. (l) .. controls +(-90:0.4) and +(60:0.3) .. (p1) (p3) .. controls +(-120:0.3) and +(90:0.4) .. (r) .. controls +(-90:0.4) and +(120:0.3) .. (p2);
			\draw [decorate,decoration={snake}] (m12) -- (m34);
			\foreach \i in {p1,p2,p3,p4,p5} \draw [fill=white] (\i) circle [radius=2pt];
			\foreach \i in {m12,m34} \fill (\i) circle [radius=2pt];
			\node [anchor=center] at (0,-1.5) {\small $E_{\langle43342\rangle}^1$};
		\end{scope}
		\begin{scope}[xshift=10cm,yshift=-3.8cm]
			\coordinate [label=left:{\small $1$}] (p1) at (-140:1.5);
			\coordinate [label=right:{\small $2$}] (p2) at (-40:1.5);
			\coordinate [label=right:{\small $3$}] (p3) at (40:1.5);
			\coordinate [label=left:{\small $4$}] (p4) at (140:1.5);
			\coordinate [label=left:{\small $5$}] (p5) at (180:1.5);
			\coordinate (o) at (0,0);
			\draw (p1) -- (p2) -- (p3) -- (p4) -- (p5) -- cycle;
			\draw [-Latex] (p1) -- ($(p1)!.6!(p5)$);
			\draw [-Latex] (p5) -- ($(p5)!.6!(p4)$);
			\draw [-Latex] (p4) -- ($(p4)!.6!(p3)$);
			\draw [-Latex] (p3) -- ($(p3)!.6!(p2)$);
			\draw [-Latex] (p2) -- ($(p2)!.6!(p1)$);
			\draw [dashed] (p3) -- (p1) (p4) -- (p2);
			\draw [RedOrange,dashed] (p1) -- (p4);
			\foreach \i in {p1,p2,p3,p4,p5} \draw [fill=white] (\i) circle [radius=2pt];
			\foreach \i in {o} \fill (\i) circle [radius=2pt];
			\node [anchor=center] at (0,-1.5) {\small $C_{\langle43342\rangle}^3$};
		\end{scope}
		\begin{scope}[xshift=12cm]
			\coordinate [label=left:{\small $1$}] (p1) at (-140:1.5);
			\coordinate [label=right:{\small $2$}] (p2) at (-40:1.5);
			\coordinate [label=right:{\small $3$}] (p3) at (40:1.5);
			\coordinate [label=left:{\small $4$}] (p4) at (140:1.5);
			\coordinate [label=left:{\small $5$}] (p5) at (180:1.5);
			\coordinate (t) at (0,0.6);
			\coordinate (b) at (0,-0.6);
			\coordinate (o) at (0,0);
			\draw (p1) -- (p2) -- (o) -- (p3) -- (p4) -- (p5) -- cycle;
			\draw [-Latex] (p1) -- ($(p1)!.6!(p5)$);
			\draw [-Latex] (p5) -- ($(p5)!.6!(p4)$);
			\draw [-Latex] (p4) -- ($(p4)!.6!(p3)$);
			\draw [-Latex] (p3) -- ($(p3)!.6!(o)$);
			\draw [-Latex] (o) -- ($(o)!.6!(p2)$);
			\draw [-Latex] (p2) -- ($(p2)!.6!(p1)$);
			\draw [dashed] (p1) -- (o) -- (p4);
			\draw [RedOrange,dashed] (p3) .. controls +(-140:0.3) and +(0:0.4) .. (t) .. controls +(180:0.4) and +(-40:0.3) .. (p4) (p1) .. controls +(40:0.3) and +(180:0.4) .. (b) .. controls +(0:0.4) and +(140:0.3) .. (p2);
			\foreach \i in {p1,p2,p3,p4,p5} \draw [fill=white] (\i) circle [radius=2pt];
			\foreach \i in {o} \fill (\i) circle [radius=2pt];
			\node [anchor=center] at (0,-1.5) {\small $C_{\langle43342\rangle}^2$};
		\end{scope}
		\begin{scope}[xshift=4cm]
			\coordinate [label=left:{\small $1$}] (p1) at (-140:1.5);
			\coordinate [label=right:{\small $2$}] (p2) at (-40:1.5);
			\coordinate [label=right:{\small $3$}] (p3) at (40:1.5);
			\coordinate [label=left:{\small $4$}] (p4) at (140:1.5);
			\coordinate [label=left:{\small $5$}] (p5) at (180:1.5);
			\coordinate (l) at (-0.8,0);
			\coordinate (r) at (0.8,0);
			\coordinate (o) at (0,0);
			\draw (p1) -- (o) -- (p2) -- (p3) -- (o) -- (p4) -- (p5) -- cycle;
			\draw [-Latex] (p1) -- ($(p1)!.6!(p5)$);
			\draw [-Latex] (p5) -- ($(p5)!.6!(p4)$);
			\draw [-Latex] (p4) -- ($(p4)!.6!(o)$);
			\draw [-Latex] (o) -- ($(o)!.6!(p3)$);
			\draw [-Latex] (p3) -- ($(p3)!.6!(p2)$);
			\draw [-Latex] (p2) -- ($(p2)!.6!(o)$);
			\draw [-Latex] (o) -- ($(o)!.6!(p1)$);
			\draw [RedOrange,dashed] (p1) -- (p4) .. controls +(-60:0.3) and +(90:0.4) .. (l) .. controls +(-90:0.4) and +(60:0.3) .. (p1) (p3) .. controls +(-120:0.3) and +(90:0.4) .. (r) .. controls +(-90:0.4) and +(120:0.3) .. (p2);
			\foreach \i in {p1,p2,p3,p4,p5} \draw [fill=white] (\i) circle [radius=2pt];
			\foreach \i in {o} \fill (\i) circle [radius=2pt];
			\node [anchor=center] at (0,-1.5) {\small $C_{\langle43342\rangle}^1$};
		\end{scope}
	
		\draw [dotted] (-2,-1.8) rectangle (5.6,1.3);
		\node [anchor=north] at (1.8,-1.8) {\small (a)};
		\draw [dotted] (6,-1.8) rectangle (13.6,1.3);
		\node [anchor=north] at (9.8,-1.8) {\small (b)};
		\draw [dotted] (0,-2.5) rectangle (12,-5.6);
		\node [anchor=north] at (6,-5.6) {\small (c)};
    \end{tikzpicture}
    \caption{Exchange diagrams and contact diagrams in the case of $\langle43342 \rangle$. }
    \label{fig:case43342}
\end{figure}

\paragraph{\underline{$\langle 43342 \rangle $}}~\\

There are seven different diagrams in the one-loop correction to $\langle43342 \rangle$, which we list in \Cref{fig:case43342}. As in the case of $\langle3333\rangle$, it is useful to divide these diagrams into several groups. 

Let us first look at group (a), which is the situation already analyzed in \cref{eq:Hbrule1}. The additional dashed lines are factorized Wick contractions and we get 
\begin{align}
    E_{\langle43342\rangle}^1+C_{\langle43342\rangle}^1=\lambda\,\Hb_{12,34}\,d_5\times\left(\frac{t_{14}}{x_{14}^2}\right)^2\frac{t_{23}}{x_{23}^2}\,.
\end{align}

Next we examine the diagrams in group (b). Note here one end of the gluon propagator terminates on a $\bar{q}q$ line and the other end is on a $\phi\phi$ line. On the other hand, the contact interaction also involves two $\phi$ fields. This combination is similar to its counterpart in $\langle2233\rangle$ and we have two additional pairs of $\phi$ contraction. The sum of these two diagrams gives
\begin{align}
    E_{\langle43342\rangle}^2=E_{\langle3223\rangle}^1\times\frac{d_{45}d_{51}}{d_{41}}\times\frac{t_{12}}{x_{12}^2}\frac{t_{34}}{x_{34}^2}\,,\quad
    C_{\langle43342\rangle}^2=C_{\langle3223\rangle}^1\times\frac{d_{45}d_{51}}{d_{41}}\times\frac{t_{12}}{x_{12}^2}\frac{t_{34}}{x_{34}^2}\,.
\end{align}
Written in terms of $\Hb$, we have
\begin{subequations}
    \begin{align}
        E_{\langle3223\rangle}^1+C_{\langle3223\rangle}^1&=\lambda\,\Hb_{41,23}\,d_4\times\frac{t_{14}}{x_{14}^2}\,,\\
        E_{\langle43342\rangle}^2+C_{\langle43342\rangle}^2&=\lambda\,\Hb_{41,23}\,d_5\times\frac{t_{12}}{x_{12}^2}\frac{t_{14}}{x_{14}^2}\frac{t_{34}}{x_{34}^2}\,.
    \end{align}
\end{subequations}
As in \cref{eq:EC1toHb}, we can extract from these relations another rule which we diagrammatically represent as
\begin{align}\label{eq:Hbrule2}
    \parbox{2.3cm}{\tikz[thick]{
        \def\RR{1}
        \coordinate (m0) at (0:\RR);
        \coordinate [label=36:{\scriptsize $i$}] (m1) at (36:\RR);
        \coordinate [label=72:{\color{white}\scriptsize $i\!+\!1$}] (m2) at (72:\RR);
        \coordinate (m3) at (108:\RR);
        \coordinate [label=144:{\scriptsize $j$}] (m4) at (144:\RR);
        \coordinate (m5) at (180:\RR);
        \coordinate (m6) at (216:\RR);
        \coordinate [label=252:{\scriptsize $k$}] (m7) at (252:\RR);
        \coordinate [label=288:{\scriptsize $k\!+\!1$}] (m8) at (288:\RR);
        \coordinate (m9) at (324:\RR);
        \coordinate (t) at ($(m1)!.5!(m4)+(0,-0.2)$);
        \coordinate (b) at ($(m7)!.5!(m8)$);
        \draw (m0) -- (m1) -- (m2) (m3) -- (m4) -- (m5) (m6) -- (m7) -- (m8) -- (m9);
        \draw [dotted] (m2) -- (m3) (m5) -- (m6) (m9) -- (m0);
        \midarrow{m9}{.75}{m8}\midarrow{m7}{.75}{m6}\midarrow{m5}{.75}{m4}\midarrow{m4}{.75}{m3}\midarrow{m2}{.75}{m1}\midarrow{m1}{.75}{m0}
        \draw [dashed] (m1) .. controls +(-150:.5) and +(-30:.5) .. (m4);
        \draw [decorate,decoration={snake}] (t) -- (b);
        \foreach \i in {1,4,7,8} \draw [fill=white] (m\i) circle [radius=2pt];
        \foreach \i in {t,b} \fill (\i) circle [radius=2pt];
    }}\;+\;
    \parbox{2.3cm}{\tikz[thick]{
        \def\RR{1}
        \coordinate (m0) at (0:\RR);
        \coordinate [label=36:{\scriptsize $i$}] (m1) at (36:\RR);
        \coordinate [label=72:{\color{white}\scriptsize $i\!+\!1$}] (m2) at (72:\RR);
        \coordinate (m3) at (108:\RR);
        \coordinate [label=144:{\scriptsize $j$}] (m4) at (144:\RR);
        \coordinate (m5) at (180:\RR);
        \coordinate (m6) at (216:\RR);
        \coordinate [label=252:{\scriptsize $k$}] (m7) at (252:\RR);
        \coordinate [label=288:{\scriptsize $k\!+\!1$}] (m8) at (288:\RR);
        \coordinate (m9) at (324:\RR);
        \coordinate (t) at ($(m1)!.5!(m4)$);
        \coordinate (b) at ($(m7)!.5!(m8)$);
        \coordinate (o) at (0,0);
        \draw (m0) -- (m1) -- (m2) (m3) -- (m4) -- (m5) (m6) -- (m7) -- (o) -- (m8) -- (m9);
        \draw [dotted] (m2) -- (m3) (m5) -- (m6) (m9) -- (m0);
        \midarrow{m9}{.75}{m8}\midarrow{m8}{.75}{o}\midarrow{o}{.75}{m7}\midarrow{m7}{.75}{m6}\midarrow{m5}{.75}{m4}\midarrow{m4}{.75}{m3}\midarrow{m2}{.75}{m1}\midarrow{m1}{.75}{m0}
        \draw [dashed] (m1) -- (o) -- (m4);
        \foreach \i in {1,4,7,8} \draw [fill=white] (m\i) circle [radius=2pt];
        \foreach \i in {o} \fill (\i) circle [radius=2pt];
    }}\;=\;
    \parbox{2.3cm}{\tikz[thick]{
        \def\RR{1}
        \coordinate (m0) at (0:\RR);
        \coordinate [label=36:{\scriptsize $i$}] (m1) at (36:\RR);
        \coordinate [label=72:{\color{white}\scriptsize $i\!+\!1$}] (m2) at (72:\RR);
        \coordinate (m3) at (108:\RR);
        \coordinate [label=144:{\scriptsize $j$}] (m4) at (144:\RR);
        \coordinate (m5) at (180:\RR);
        \coordinate (m6) at (216:\RR);
        \coordinate [label=252:{\scriptsize $k$}] (m7) at (252:\RR);
        \coordinate [label=288:{\scriptsize $k\!+\!1$}] (m8) at (288:\RR);
        \coordinate (m9) at (324:\RR);
        \coordinate (t) at ($(m1)!.5!(m4)$);
        \coordinate (b) at ($(m7)!.5!(m8)$);
        \coordinate (o) at ($(t)!.5!(b)$);
        \draw (m0) -- (m1) -- (m2) (m3) -- (m4) -- (m5) (m6) -- (m7) -- (m8) -- (m9);
        \draw [dotted] (m2) -- (m3) (m5) -- (m6) (m9) -- (m0);
        \midarrow{m9}{.75}{m8}\midarrow{m7}{.75}{m6}\midarrow{m5}{.75}{m4}\midarrow{m4}{.75}{m3}\midarrow{m2}{.75}{m1}\midarrow{m1}{.75}{m0}
        \draw [thin,fill=gray!20] (m1) .. controls +(-155:.4) and ($(o)!.5!(t)$) .. (o) .. controls ($(o)!.5!(b)$) and +(120:.2) .. (m8) .. controls +(120:.4) and +(60:.4) .. (m7) .. controls +(60:.2) and ($(o)!.5!(b)$) .. (o) .. controls ($(o)!.5!(t)$) and +(-25:.4) .. (m4) .. controls +(-25:.6) and +(-155:.6) .. cycle;
        \draw [RedOrange,dashed] (m1) .. controls +(170:.5) and +(10:.5) .. (m4);
        \foreach \i in {1,4,7,8} \draw [fill=white] (m\i) circle [radius=2pt];
    }}\;.
\end{align}
The important difference from the previous case \cref{eq:Hbrule1} is that the point $i$ and $j$ are not necessarily adjacent in the planar ordering.

Finally, we focus on the diagrams in group (c). Apart from the three contractions among points $\{1,4,5\}$, these diagrams are structurally the same as the diagrams $E_{\langle3333\rangle}^3$, $E_{\langle3333\rangle}^4$ and $C_{\langle3333\rangle}^3$ of $\langle3333\rangle$ (see \Cref{fig:case3333}). Note now both ends of the gluon exchange are anchored on $\phi\phi$ lines. Comparing the two cases we can conclude that
\begin{subequations}
    \begin{align}
        E^3_{\langle 3333 \rangle}\!+\!E^4_{\langle 3333 \rangle}\!+\!C^3_{\langle 3333 \rangle} &=\lambda\,\Hb_{12,34}\,d_4 \times \frac{t_{12}}{x^2_{12}} \frac{t_{34}}{x^2_{34}}+\lambda\, \Hb_{41,23}\,d_4 \!\times\! \frac{t_{14}}{x^2_{14}} \frac{t_{23}}{x^2_{23}} \, ,\\
        E^3_{\langle 43342 \rangle}\!+\!E^4_{\langle 43342 \rangle}\!+\!C^3_{\langle 43342 \rangle} &=\lambda\,\Hb_{12,34}\,d_5 \!\times\! \frac{t_{12}}{x^2_{12}}\frac{t_{14}}{x^2_{14}}\frac{t_{34}}{x^2_{34}}+\lambda\, \Hb_{41,23}\,d_5 \!\times\! \left(\!\frac{t_{14}}{x^2_{14}}\!\right)^{\!2}\! \frac{t_{23}}{x^2_{23}} \, .
    \end{align}
\end{subequations}
These relations can be viewed as consequences of a third general relation
\begin{align}\label{eq:Hbrule3}
    \parbox{2.3cm}{\tikz[thick]{
        \def\RR{1}
        \coordinate (m0) at (15:\RR);
        \coordinate [label=45:{\scriptsize $i$}] (m1) at (45:\RR);
        \coordinate (m2) at (75:\RR);
        \coordinate (m3) at (105:\RR);
        \coordinate [label=135:{\scriptsize $j$}] (m4) at (135:\RR);
        \coordinate (m5) at (165:\RR);
        \coordinate (m6) at (195:\RR);
        \coordinate [label=225:{\scriptsize $k$}] (m7) at (225:\RR);
        \coordinate (m8) at (255:\RR);
        \coordinate (m9) at (285:\RR);
        \coordinate [label=315:{\scriptsize $l$}] (m10) at (315:\RR);
        \coordinate (m11) at (345:\RR);
        \coordinate (t) at ($(m1)!.5!(m4)+(0,-0.2)$);
        \coordinate (b) at ($(m7)!.5!(m10)+(0,0.2)$);
        \draw (m0) -- (m1) -- (m2) (m3) -- (m4) -- (m5) (m6) -- (m7) -- (m8) (m9) -- (m10) -- (m11);
        \draw [dotted] (m2) -- (m3) (m5) -- (m6) (m8) -- (m9) (m11) -- (m0);
        \midarrow{m11}{.7}{m10}\midarrow{m10}{.8}{m9}\midarrow{m8}{.7}{m7}\midarrow{m7}{.8}{m6}\midarrow{m5}{.7}{m4}\midarrow{m4}{.8}{m3}\midarrow{m2}{.7}{m1}\midarrow{m1}{.8}{m0}
        \draw [dashed] (m1) .. controls +(-150:.5) and +(-30:.5) .. (m4) (m10) .. controls +(150:.5) and +(30:.5) .. (m7);
        \draw [decorate,decoration={snake}] (t) -- (b);
        \foreach \i in {1,4,7,10} \draw [fill=white] (m\i) circle [radius=2pt];
        \foreach \i in {t,b} \fill (\i) circle [radius=2pt];
    }}+
    \parbox{2.3cm}{\tikz[thick]{
        \def\RR{1}
        \coordinate (m0) at (15:\RR);
        \coordinate [label=45:{\scriptsize $i$}] (m1) at (45:\RR);
        \coordinate (m2) at (75:\RR);
        \coordinate (m3) at (105:\RR);
        \coordinate [label=135:{\scriptsize $j$}] (m4) at (135:\RR);
        \coordinate (m5) at (165:\RR);
        \coordinate (m6) at (195:\RR);
        \coordinate [label=225:{\scriptsize $k$}] (m7) at (225:\RR);
        \coordinate (m8) at (255:\RR);
        \coordinate (m9) at (285:\RR);
        \coordinate [label=315:{\scriptsize $l$}] (m10) at (315:\RR);
        \coordinate (m11) at (345:\RR);
        \coordinate (l) at ($(m4)!.5!(m7)+(0.2,0)$);
        \coordinate (r) at ($(m1)!.5!(m10)+(-0.2,0)$);
        \draw (m0) -- (m1) -- (m2) (m3) -- (m4) -- (m5) (m6) -- (m7) -- (m8) (m9) -- (m10) -- (m11);
        \draw [dotted] (m2) -- (m3) (m5) -- (m6) (m8) -- (m9) (m11) -- (m0);
        \midarrow{m11}{.7}{m10}\midarrow{m10}{.8}{m9}\midarrow{m8}{.7}{m7}\midarrow{m7}{.8}{m6}\midarrow{m5}{.7}{m4}\midarrow{m4}{.8}{m3}\midarrow{m2}{.7}{m1}\midarrow{m1}{.8}{m0}
        \draw [dashed] (m4) .. controls +(-60:.5) and +(60:.5) .. (m7) (m1) .. controls +(-120:.5) and +(120:.5) .. (m10);
        \draw [decorate,decoration={snake}] (l) -- (r);
        \foreach \i in {1,4,7,10} \draw [fill=white] (m\i) circle [radius=2pt];
        \foreach \i in {l,r} \fill (\i) circle [radius=2pt];
    }}+
    \parbox{2.3cm}{\tikz[thick]{
        \def\RR{1}
        \coordinate (m0) at (15:\RR);
        \coordinate [label=45:{\scriptsize $i$}] (m1) at (45:\RR);
        \coordinate (m2) at (75:\RR);
        \coordinate (m3) at (105:\RR);
        \coordinate [label=135:{\scriptsize $j$}] (m4) at (135:\RR);
        \coordinate (m5) at (165:\RR);
        \coordinate (m6) at (195:\RR);
        \coordinate [label=225:{\scriptsize $k$}] (m7) at (225:\RR);
        \coordinate (m8) at (255:\RR);
        \coordinate (m9) at (285:\RR);
        \coordinate [label=315:{\scriptsize $l$}] (m10) at (315:\RR);
        \coordinate (m11) at (345:\RR);
        \coordinate (o) (0,0);
        \draw (m0) -- (m1) -- (m2) (m3) -- (m4) -- (m5) (m6) -- (m7) -- (m8) (m9) -- (m10) -- (m11);
        \draw [dotted] (m2) -- (m3) (m5) -- (m6) (m8) -- (m9) (m11) -- (m0);
        \midarrow{m11}{.7}{m10}\midarrow{m10}{.8}{m9}\midarrow{m8}{.7}{m7}\midarrow{m7}{.8}{m6}\midarrow{m5}{.7}{m4}\midarrow{m4}{.8}{m3}\midarrow{m2}{.7}{m1}\midarrow{m1}{.8}{m0}
        \draw [dashed] (m1) -- (m7) (m4) -- (m10);
        \foreach \i in {1,4,7,10} \draw [fill=white] (m\i) circle [radius=2pt];
        \foreach \i in {o} \fill (\i) circle [radius=2pt];
    }}=
    \parbox{2.3cm}{\tikz[thick]{
        \def\RR{1}
        \coordinate (m0) at (15:\RR);
        \coordinate [label=45:{\scriptsize $i$}] (m1) at (45:\RR);
        \coordinate (m2) at (75:\RR);
        \coordinate (m3) at (105:\RR);
        \coordinate [label=135:{\scriptsize $j$}] (m4) at (135:\RR);
        \coordinate (m5) at (165:\RR);
        \coordinate (m6) at (195:\RR);
        \coordinate [label=225:{\scriptsize $k$}] (m7) at (225:\RR);
        \coordinate (m8) at (255:\RR);
        \coordinate (m9) at (285:\RR);
        \coordinate [label=315:{\scriptsize $l$}] (m10) at (315:\RR);
        \coordinate (m11) at (345:\RR);
        \coordinate (t) at ($(m1)!.5!(m4)$);
        \coordinate (b) at ($(m7)!.5!(m10)$);
        \coordinate (o) at ($(t)!.5!(b)$);
        \draw (m0) -- (m1) -- (m2) (m3) -- (m4) -- (m5) (m6) -- (m7) -- (m8) (m9) -- (m10) -- (m11);
        \draw [dotted] (m2) -- (m3) (m5) -- (m6) (m8) -- (m9) (m11) -- (m0);
        \midarrow{m11}{.7}{m10}\midarrow{m10}{.8}{m9}\midarrow{m8}{.7}{m7}\midarrow{m7}{.8}{m6}\midarrow{m5}{.7}{m4}\midarrow{m4}{.8}{m3}\midarrow{m2}{.7}{m1}\midarrow{m1}{.8}{m0}
        \draw [thin,fill=gray!20] (m1) .. controls +(-150:.3) and ($(o)!.6!(t)$) .. (o) .. controls ($(o)!.6!(b)$) and +(150:.3) .. (m10) .. controls +(150:.6) and +(30:.6) .. (m7) .. controls +(30:.3) and ($(o)!.6!(b)$) .. (o) .. controls ($(o)!.6!(t)$) and +(-30:.3) .. (m4) .. controls +(-30:.6) and +(-150:.6) .. cycle;
        \draw [RedOrange,dashed] (m1) .. controls +(175:.5) and +(5:.5) .. (m4) (m10) .. controls +(-175:.5) and +(-5:.5) .. (m7);
        \foreach \i in {1,4,7,10} \draw [fill=white] (m\i) circle [radius=2pt];
    }}+
    \parbox{2.3cm}{\tikz[thick]{
        \def\RR{1}
        \coordinate (m0) at (15:\RR);
        \coordinate [label=45:{\scriptsize $i$}] (m1) at (45:\RR);
        \coordinate (m2) at (75:\RR);
        \coordinate (m3) at (105:\RR);
        \coordinate [label=135:{\scriptsize $j$}] (m4) at (135:\RR);
        \coordinate (m5) at (165:\RR);
        \coordinate (m6) at (195:\RR);
        \coordinate [label=225:{\scriptsize $k$}] (m7) at (225:\RR);
        \coordinate (m8) at (255:\RR);
        \coordinate (m9) at (285:\RR);
        \coordinate [label=315:{\scriptsize $l$}] (m10) at (315:\RR);
        \coordinate (m11) at (345:\RR);
        \coordinate (l) at ($(m4)!.5!(m7)$);
        \coordinate (r) at ($(m1)!.5!(m10)$);
        \coordinate (o) at ($(l)!.5!(r)$);
        \draw (m0) -- (m1) -- (m2) (m3) -- (m4) -- (m5) (m6) -- (m7) -- (m8) (m9) -- (m10) -- (m11);
        \draw [dotted] (m2) -- (m3) (m5) -- (m6) (m8) -- (m9) (m11) -- (m0);
        \midarrow{m11}{.7}{m10}\midarrow{m10}{.8}{m9}\midarrow{m8}{.7}{m7}\midarrow{m7}{.8}{m6}\midarrow{m5}{.7}{m4}\midarrow{m4}{.8}{m3}\midarrow{m2}{.7}{m1}\midarrow{m1}{.8}{m0}
        \draw [thin,fill=gray!20] (m1) .. controls +(-120:.3) and ($(o)!.6!(r)$) .. (o) .. controls ($(o)!.6!(l)$) and +(-60:.3) .. (m4) .. controls +(-60:.6) and +(60:.6) .. (m7) .. controls +(60:.3) and ($(o)!.6!(l)$) .. (o) .. controls ($(o)!.6!(r)$) and +(120:.3) .. (m10) .. controls +(120:.6) and +(-120:.6) .. cycle;
        \draw [RedOrange,dashed] (m1) .. controls +(-85:.5) and +(85:.5) .. (m10) (m4) .. controls +(-95:.5) and +(95:.5) .. (m7);
        \foreach \i in {1,4,7,10} \draw [fill=white] (m\i) circle [radius=2pt];
    }}\;.
\end{align}

Using these rules \cref{eq:Hbrule1}, \cref{eq:Hbrule2} and \cref{eq:Hbrule3}, we observe that the computation of $\langle43342\rangle$ boils down to summing over four effective diagrams involving $\Hb$
\begin{align}\label{eq:full43342}
    \langle43342\rangle&=
    \underbrace{
    \parbox{2.6cm}{\tikz[thick]{
        \def\RR{1}
        \coordinate [label=36:{\small 3}] (m3) at (36:\RR);
        \coordinate [label=108:{\small 4}] (m4) at (108:\RR);
        \coordinate [label=180:{\small 5}] (m5) at (180:\RR);
        \coordinate [label=252:{\small 1}] (m1) at (252:\RR);
        \coordinate [label=324:{\small 2}] (m2) at (324:\RR);
        \coordinate (m14) at ($(m1)!.5!(m4)$);
        \coordinate (m23) at ($(m2)!.5!(m3)$);
        \coordinate (h) at ($(m14)!.4!(m23)$);
        \draw (m1) -- (m2) -- (m3) -- (m4) -- (m5) -- cycle;
        \midarrow{m1}{.7}{m5}\midarrow{m5}{.7}{m4}\midarrow{m4}{.7}{m3}\midarrow{m3}{.7}{m2}\midarrow{m2}{.7}{m1}
        \draw [thin,fill=gray!20] (m1) .. controls +(18+60:.3) and ($(h)!.6!(m14)$) .. (h) .. controls ($(h)!.4!(m23)$) and +(90+40:.2) .. (m2) .. controls +(90+40:.6) and +(-90-40:.6) .. (m3) .. controls +(-90-40:.2) and ($(h)!.4!(m23)$) .. (h) .. controls ($(h)!.6!(m14)$) and +(-18-60:.3) .. (m4) .. controls +(-18-60:.8) and +(18+60:.8) .. (m1);
        \draw [RedOrange,dashed] (m2) .. controls +(90+20:0.6) and +(-90-20:.6) .. (m3) (m1) .. controls +(18+75:.9) and +(-18-75:.9) .. (m4) .. controls +(-18-90:.9) and +(18+90:.9) .. (m1);
        \foreach \i in {1,2,3,4,5} \draw [fill=white] (m\i) circle [radius=2pt];
    }}+
    \parbox{2.6cm}{\tikz[thick]{
        \def\RR{1}
        \coordinate [label=36:{\small 3}] (m3) at (36:\RR);
        \coordinate [label=108:{\small 4}] (m4) at (108:\RR);
        \coordinate [label=180:{\small 5}] (m5) at (180:\RR);
        \coordinate [label=252:{\small 1}] (m1) at (252:\RR);
        \coordinate [label=324:{\small 2}] (m2) at (324:\RR);
        \coordinate (m12) at ($(m1)!.3!(m2)$);
        \coordinate (m34) at ($(m4)!.3!(m3)$);
        \coordinate (h) at ($(m12)!.5!(m34)$);
        \draw (m1) -- (m2) -- (m3) -- (m4) -- (m5) -- cycle;
        \midarrow{m1}{.7}{m5}\midarrow{m5}{.7}{m4}\midarrow{m4}{.7}{m3}\midarrow{m3}{.7}{m2}\midarrow{m2}{.7}{m1}
        \draw [thin,fill=gray!20] (m1) .. controls +(18+45:.4) and ($(h)!.3!(m12)$) .. (h) .. controls ($(h)!.3!(m34)$) and +(-18-45:.4) .. (m4) .. controls +(-18-45:.8) and +(-90-70:.8) .. (m3) .. controls +(-90-70:.8) and ($(h)!.2!(m34)$) .. (h) .. controls ($(h)!.2!(m12)$) and +(90+70:.8) .. (m2) .. controls +(90+70:.8) and +(18+45:.8) .. (m1);
        \draw [RedOrange,dashed] (m3) .. controls +(162+30:.5) and +(-18-30:.5) .. (m4) .. controls +(-126+20:.7) and +(126-20:.7) .. (m1) .. controls +(18+30:.5) and +(-162-30:.5) .. (m2);
        \foreach \i in {1,2,3,4,5} \draw [fill=white] (m\i) circle [radius=2pt];
    }}}_{\text{(a)}}+\underbrace{
    \parbox{2.6cm}{\tikz[thick]{
        \def\RR{1}
        \coordinate [label=36:{\small 3}] (m3) at (36:\RR);
        \coordinate [label=108:{\small 4}] (m4) at (108:\RR);
        \coordinate [label=180:{\small 5}] (m5) at (180:\RR);
        \coordinate [label=252:{\small 1}] (m1) at (252:\RR);
        \coordinate [label=324:{\small 2}] (m2) at (324:\RR);
        \draw (m1) -- (m2) -- (m3) -- (m4) -- (m5) -- cycle;
        \midarrow{m1}{.7}{m5}\midarrow{m5}{.7}{m4}\midarrow{m4}{.7}{m3}\midarrow{m3}{.7}{m2}\midarrow{m2}{.7}{m1}
        \draw [thin,fill=gray!20] (m1) .. controls +(18+60:.3) and ($(h)!.6!(m14)$) .. (h) .. controls ($(h)!.4!(m23)$) and +(90+40:.2) .. (m2) .. controls +(90+40:.6) and +(-90-40:.6) .. (m3) .. controls +(-90-40:.2) and ($(h)!.4!(m23)$) .. (h) .. controls ($(h)!.6!(m14)$) and +(-18-60:.3) .. (m4) .. controls +(-18-60:.8) and +(18+60:.8) .. (m1);
        \draw [RedOrange,dashed] (m3) .. controls +(162+40:.5) and +(-18-40:.5) .. (m4) .. controls +(-126+20:.7) and +(126-20:.7) .. (m1) .. controls +(18+40:.5) and +(-162-40:.5) .. (m2);
        \foreach \i in {1,2,3,4,5} \draw [fill=white] (m\i) circle [radius=2pt];
    }}}_{\text{(b)}}+\underbrace{
    \parbox{2.6cm}{\tikz[thick]{
        \def\RR{1}
        \coordinate [label=36:{\small 3}] (m3) at (36:\RR);
        \coordinate [label=108:{\small 4}] (m4) at (108:\RR);
        \coordinate [label=180:{\small 5}] (m5) at (180:\RR);
        \coordinate [label=252:{\small 1}] (m1) at (252:\RR);
        \coordinate [label=324:{\small 2}] (m2) at (324:\RR);
        \draw (m1) -- (m2) -- (m3) -- (m4) -- (m5) -- cycle;
        \midarrow{m1}{.7}{m5}\midarrow{m5}{.7}{m4}\midarrow{m4}{.7}{m3}\midarrow{m3}{.7}{m2}\midarrow{m2}{.7}{m1}
        \draw [thin,fill=gray!20] (m1) .. controls +(18+45:.4) and ($(h)!.3!(m12)$) .. (h) .. controls ($(h)!.3!(m34)$) and +(-18-45:.4) .. (m4) .. controls +(-18-45:.8) and +(-90-70:.8) .. (m3) .. controls +(-90-70:.8) and ($(h)!.2!(m34)$) .. (h) .. controls ($(h)!.2!(m12)$) and +(90+70:.8) .. (m2) .. controls +(90+70:.8) and +(18+45:.8) .. (m1);
        \draw [RedOrange,dashed] (m2) .. controls +(90+40:0.7) and +(-90-40:.7) .. (m3) (m1) .. controls +(18+75:.9) and +(-18-75:.9) .. (m4) .. controls +(-18-90:.9) and +(18+90:.9) .. (m1);
        \foreach \i in {1,2,3,4,5} \draw [fill=white] (m\i) circle [radius=2pt];
    }}}_{\text{(c)}}\nonumber\\
    &=\lambda\left(\Hb_{12,34}+\Hb_{41,23}\right)d_5\times\left(\left(\frac{t_{14}}{x_{14}^2}\right)^2\frac{t_{23}}{x_{23}^2}+\frac{t_{12}}{x_{12}^2}\frac{t_{14}}{x_{14}^2}\frac{t_{23}}{x_{23}^2}\right).
\end{align}
Just like in four-point correlators, the above effective diagrams show that all $\phi$ fields only contribute as multiplicative factors in the final result given by Wick contractions. Moreover, we explicitly see that each $\phi$ contraction is restricted to be within one of the four white regions  carved out by the shaded region corresponding to $\Hb$. This observation allows us to interpret the above summation of effective diagrams in a different way: We first consider inserting a shaded region for $\Hb$ in all possible ways into the pentagon  defined by the cycle of $\bar{q}q$ contractions. Then in each case we further consider all different $\phi$ contractions such that each contraction sits inside one white region. 

From the point of view of inserting $\Hb$, at five points we can in principle cook up eight more $\Hb$ insertions other than the two in \cref{eq:full43342}, e.g., $\Hb_{12,45}$. However, none of them show up in the case of $\langle 43342\rangle$, since they allow no compatible $\phi$ contractions. For $\Hb_{12,34}$ and $\Hb_{41,23}$, because they are anchored on the same four points, the white regions they carved out are the same, and so they share the same set of compatible $\phi$ contractions. As a result the final expression factorizes as at four points.

\begin{figure}[ht]
    \centering
    \begin{tikzpicture}[thick,decoration={zigzag,segment length=2.4pt,amplitude=.8pt}]
        \begin{scope}
            \def\RR{1.4}
			\coordinate [label=162:{\small $1$}] (m1) at (162:\RR);
			\coordinate [label=234:{\small $2$}] (m2) at (234:\RR);
			\coordinate [label=306:{\small $3$}] (m3) at (306:\RR);
			\coordinate [label=18:{\small $4$}] (m4) at (18:\RR);
			\coordinate [label=90:{\small $5$}] (m5) at (90:\RR);
			\coordinate (m12) at ($(m1)!.5!(m2)$);
			\coordinate (m34) at ($(m3)!.5!(m4)$);
			\coordinate (h) at ($(m12)!.5!(m34)$);
			\draw (m1) -- (m2) -- (m3) -- (m4) -- (m5) -- cycle;
			\midarrow{m1}{.6}{m5}\midarrow{m5}{.6}{m4}\midarrow{m4}{.6}{m3}\midarrow{m3}{.6}{m2}\midarrow{m2}{.6}{m1}
			\draw [thin,fill=gray!20] (m1) .. controls +(-45:.7) and ($(h)!.5!(m12)$) .. (h) .. controls ($(h)!.5!(m34)$) and +(-135:.7) .. (m4) .. controls +(-135:.7) and +(120:.9) .. (m3) .. controls +(120:.6) and ($(h)!.2!(m34)$) .. (h) .. controls ($(h)!.2!(m12)$) and +(60:.6) .. (m2) .. controls +(60:.9) and +(-45:.7) .. (m1);
			\draw [RedOrange,decorate] (m1) .. controls +(36-24:.7) and +(-144+24:.7) .. (m5) .. controls +(-36-24:.7) and +(144+24:.7) .. (m4) .. controls +(-108-20:.7) and +(72+20:.7) .. (m3) .. controls +(180-24:.7) and +(24:.7) .. (m2) .. controls +(108-20:.7) and +(-72+20:.7) .. (m1);
			\draw [NavyBlue,decorate] (m4) .. controls +(-162:1.2) and +(-18:1.2) .. (m1);
			\foreach \i in {1,2,...,5} \draw [fill=white] (m\i) circle [radius=2pt];
			\node [anchor=north] at (-90:\RR+.2) {\small (a)};
		\end{scope}
		\begin{scope}[xshift=5cm]
            \def\RR{1.4}
			\coordinate [label=162:{\small $1$}] (m1) at (162:\RR);
			\coordinate [label=234:{\small $2$}] (m2) at (234:\RR);
			\coordinate [label=306:{\small $3$}] (m3) at (306:\RR);
			\coordinate [label=18:{\small $4$}] (m4) at (18:\RR);
			\coordinate [label=90:{\small $5$}] (m5) at (90:\RR);
			\coordinate (m14) at ($(m1)!.5!(m4)$);
			\coordinate (m23) at ($(m2)!.5!(m3)$);
			\coordinate (h) at ($(m14)!.5!(m23)$);
			\draw (m1) -- (m2) -- (m3) -- (m4) -- (m5) -- cycle;
			\midarrow{m1}{.6}{m5}\midarrow{m5}{.6}{m4}\midarrow{m4}{.6}{m3}\midarrow{m3}{.6}{m2}\midarrow{m2}{.6}{m1}
			\draw [thin,fill=gray!20] (m1) .. controls +(-20:.7) and ($(h)!.4!(m14)$) .. (h) .. controls ($(h)!.3!(m23)$) and +(36:.7) .. (m2) .. controls +(36:.9) and +(144:.9) .. (m3) .. controls +(144:.7) and ($(h)!.3!(m23)$) .. (h) .. controls ($(h)!.4!(m14)$) and +(-160:.7) .. (m4) .. controls +(-160:1) and +(-20:1) .. (m1);
			\draw [RedOrange,decorate] (m1) .. controls +(36-24:.7) and +(-144+24:.7) .. (m5) .. controls +(-36-24:.7) and +(144+24:.7) .. (m4) .. controls +(-108-20:.7) and +(72+20:.7) .. (m3) .. controls +(180-24:.7) and +(24:.7) .. (m2) .. controls +(108-20:.7) and +(-72+20:.7) .. (m1);
			\draw [NavyBlue,decorate] (m4) .. controls +(175:1.2) and +(5:1.2) .. (m1);
			\foreach \i in {1,2,...,5} \draw [fill=white] (m\i) circle [radius=2pt];
			\node [anchor=north] at (-90:\RR+.2) {\small (b)};
		\end{scope}
    \end{tikzpicture}
    \caption{Two inequivalent ways of inserting $\Hb_{ij,kl}$ at the same points: (a) both $(ij)$ and $(kl)$ are adjacent in the planar ordering, (b) only $(ij)$ or $(kl)$ are adjacent. Each zigzag line denotes the summation over $\phi$ contractions between its two end points, which appears in the genrerating function $\mathbb{G}_5$.}
    \label{fig:weak_fig1}
\end{figure}
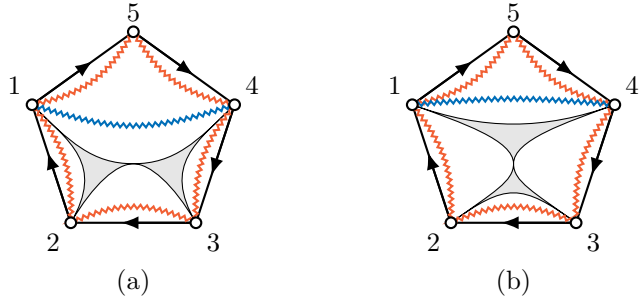

\newpage
\paragraph{\underline{Five-point generating function of master operators}}~\\

The effective rules as summarized above can be applied to five-point correlators of arbitrary meson operators. There are two inequivalent ways of inserting a shaded region $\Hb_{ij,kl}$ up to cyclic permutations, as is illustrated in \Cref{fig:weak_fig1}. The further dressing with $\phi$ contractions in general has multiple options, as long as the contractions do not cross into different white regions. To be concrete, let us denote the number of $\phi$ contractions between the points $i$ and $j$ as $b_{ij}\in\mathbb{N}$. The two diagrams in \Cref{fig:weak_fig1}, with a particular $\phi$ contraction, are then proportional to the factor (assuming the $\phi$ fields are contracted in the same way)
\begin{align}
    \left(\frac{t_{12}}{x_{12}^2}\right)^{b_{12}}\left(\frac{t_{14}}{x_{14}^2}\right)^{b_{14}}\left(\frac{t_{15}}{x_{15}^2}\right)^{b_{15}}\left(\frac{t_{23}}{x_{23}^2}\right)^{b_{23}}\left(\frac{t_{34}}{x_{34}^2}\right)^{b_{34}}\left(\frac{t_{45}}{x_{45}^2}\right)^{b_{45}},
\end{align}
where $b_{ij}$ are constrained by
\begin{align}\label{piequalnij}
    &p_1=b_{12}+b_{14}+b_{15}+2\;,\quad p_2=b_{12}+b_{23}+2\;,\quad p_3=b_{23}+b_{34}+2\;,\nonumber\\
    &p_4=b_{14}+b_{34}+b_{45}+2\;,\quad p_5=b_{15}+b_{45}+2\;.
\end{align}
For a specific correlator $\langle p_1p_2p_3p_4p_5\rangle$, the different contractions are give by all the solutions of $b_{ij}$ to (\ref{piequalnij}). As in the four-point function case, it is useful to consider a generating function by summing up all the five-point functions
\begin{align}
    \mathbb{G}_5=\sum_{p_1,p_2,p_3,p_4,p_5=2}^\infty\langle p_1p_2p_3p_4p_5\rangle\;,
\end{align}
where the sum over $p_i$ is really with respect to $b_{ij}$. 

To explicitly perform the sum, however, we need to distinguish two different cases. In the first case, the contraction is between two neighboring points $i$ and $i+1$ in the planar ordering. Regardless of how $\bar{\mathcal{H}}$ is inserted, $b_{i,i+1}$ is allowed to take any values starting from 0. Therefore, the sum over $b_{i,i+1}$ contributes a factor
\begin{align}\label{eq:sumn0}
    \sum_{b_{i,i+1}=0}^\infty\left(\frac{t_{i,i+1}}{x_{i,i+1}^2}\right)^{b_{i,i+1}}=\frac{x_{i,i+1}^2}{X_{i,i+1}^2}\;.
\end{align}
These are represented by red zigzag lines in \Cref{fig:weak_fig1}. In the second case, the contraction is between two non-neighboring points $i$ and $j$. An example is the contractions between 1 and 4 in \Cref{fig:weak_fig1} (blue zigzag lines). In this case, we need to be careful about how $\Hb$ is inserted. If the effective exchange has indices $\Hb_{ij,kl}$, such as in \Cref{fig:weak_fig1} (b), then there must be at least one propagator between $i$ and $j$ so that the gluon can be exchanged. Then the sum over $b_{ij}$ starts with 1 and we get  
\begin{align}\label{eq:sumn1}
    \sum_{b_{ij}=1}^\infty\left(\frac{t_{ij}}{x_{ij}^2}\right)^{b_{ij}}=\frac{t_{ij}}{X_{ij}^2}\;.
\end{align}
However, if $i$ and $j$ do not appear in the same pair of indices in $\Hb$, the sum over $b_{ij}$ again starts with 0 and the sum of the geometric series reduces to the neighboring points case \cref{eq:sumn0}. In order to conveniently take care of these two situations, we can define
\begin{equation}
    P_{ij}^{(\kappa)}=\begin{cases}\frac{x_{ij}^2}{X_{ij}^2},&\kappa=0,\\\frac{t_{ij}}{X_{ij}^2},&\kappa=1,\end{cases}
\end{equation}
where $\kappa$ denotes the starting point of the above geometric sums.

Note from the computation of $\langle22222\rangle$ and $\langle43342\rangle$, the five-point correlators always contains a universal factor $d_5$. It is then useful to define the combinations
\begin{equation}\label{eq:defDn}
    D_{ij}\equiv P_{ij}^{(0)} \times \frac{v_{ij}}{x_{ij}^2}=\frac{v_{ij}}{X_{ij}^2} \;,
\end{equation}
and
\begin{equation}\label{eq:defD}
     D_n=D_{12}D_{23}\cdots D_{n-1,n}D_{n1}\;,
\end{equation}
which absorbs all $P_{i,i+1}^{(0)}$'s. Hence the generating function for the five-point correlators reads
\begin{align}\label{eq:5pointmastercorrelator2}
    \mathbb{G}_5=\lambda\,D_5\left(\Hb_{12,34}\,P_{14}^{(0)}+\Hb_{41,23}\,P_{14}^{(1)}\right)+(\text{cyclic})\,.
\end{align}
Expanding $\mathbb{G}_5$ and collecting the relevant structures, it is straightforward to check that the previous examples of $\langle 22222\rangle$ and $\langle 43342\rangle$ are reproduced. The generating function (\ref{eq:5pointmastercorrelator2}) also reveals how the hidden eight-dimensional symmetry is generalized to the one-loop corrections to five-point functions. Here we note that unlike the four-point case, other variables than $X_{ij}^2$ also appear in the generating function. A similar phenomenon was also observed at strong coupling \cite{Huang:2024dxr}.

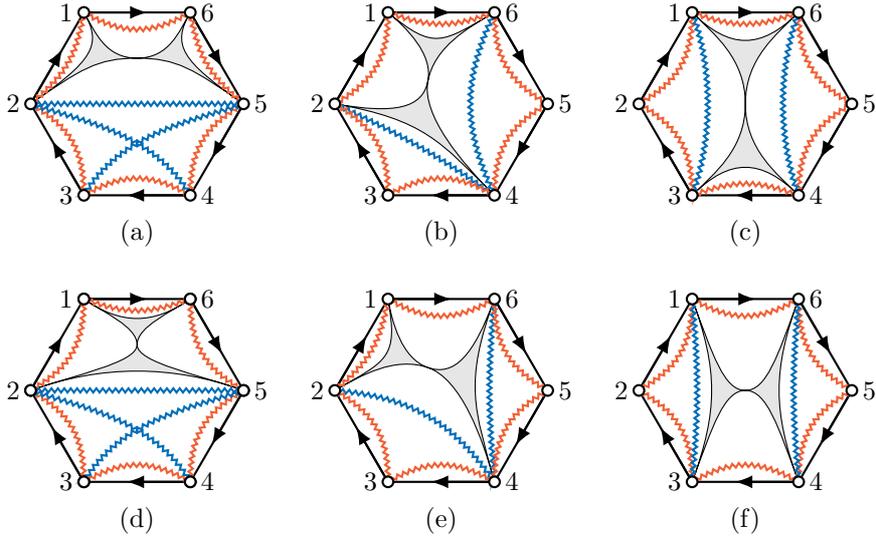
\begin{figure}[ht]
    \centering
    \begin{tikzpicture}[thick,decoration={zigzag,segment length=2.4pt,amplitude=.8pt}]
        \begin{scope}
            \def\RR{1.4}
            \coordinate [label=left:{\small 1}] (m1) at (120:\RR);
            \coordinate [label=left:{\small 2}] (m2) at (180:\RR);
            \coordinate [label=left:{\small 3}] (m3) at (240:\RR);
            \coordinate [label=right:{\small 4}] (m4) at (300:\RR);
            \coordinate [label=right:{\small 5}] (m5) at (0:\RR);
            \coordinate [label=right:{\small 6}] (m6) at (60:\RR);
            \coordinate (m12) at ($(m1)!.5!(m2)$);
            \coordinate (m56) at ($(m5)!.5!(m6)$);
            \coordinate (h) at ($(m12)!.5!(m56)$);
            \draw (m1) -- (m2) -- (m3) -- (m4) -- (m5) -- (m6) -- cycle;
            \midarrow{m1}{.6}{m6}\midarrow{m6}{.6}{m5}\midarrow{m5}{.6}{m4}\midarrow{m4}{.6}{m3}\midarrow{m3}{.6}{m2}\midarrow{m2}{.6}{m1}
            \draw [thin,fill=gray!20] (m1) .. controls +(-60:.3) and ($(h)!.4!(m12)$) .. (h) .. controls ($(h)!.4!(m56)$) and +(-120:.3) .. (m6) .. controls +(-120:.5) and +(145:1) .. (m5) .. controls +(145:.8) and ($(h)!.5!(m56)$) .. (h) .. controls ($(h)!.5!(m12)$) and +(35:.8) .. (m2) .. controls +(35:1) and +(-60:.5) .. cycle;
            \draw [RedOrange,decorate] (m1) .. controls +(-30:.6) and +(-150:.6) .. (m6) .. controls +(-60-20:.6) and +(120+20:.6) .. (m5) .. controls +(-120-30:.6) and +(60+30:.6) .. (m4) .. controls +(180-30:.6) and +(30:.6) .. (m3) .. controls +(120-30:.6) and +(-60+30:.6) .. (m2) .. controls +(60-20:.6) and +(-120+20:.6) .. cycle;
            \draw [NavyBlue,decorate] (m2) -- (m5) .. controls +(-160:.6) and +(60:1) .. (m3) (m2) .. controls +(-20:.6) and +(120:1) .. (m4);
            \foreach \i in {1,2,...,6} \draw [fill=white] (m\i) circle [radius=2pt];
            \node [anchor=north] at (-90:\RR) {\small (a)};
        \end{scope}
        \begin{scope}[xshift=4cm]
            \def\RR{1.4}
            \coordinate [label=left:{\small 1}] (m1) at (120:\RR);
            \coordinate [label=left:{\small 2}] (m2) at (180:\RR);
            \coordinate [label=left:{\small 3}] (m3) at (240:\RR);
            \coordinate [label=right:{\small 4}] (m4) at (300:\RR);
            \coordinate [label=right:{\small 5}] (m5) at (0:\RR);
            \coordinate [label=right:{\small 6}] (m6) at (60:\RR);
            \coordinate (m16) at ($(m1)!.5!(m6)$);
            \coordinate (m24) at ($(m2)!.5!(m4)$);
            \coordinate (h) at ($(m16)!.5!(m24)$);
            \draw (m1) -- (m2) -- (m3) -- (m4) -- (m5) -- (m6) -- cycle;
            \midarrow{m1}{.6}{m6}\midarrow{m6}{.6}{m5}\midarrow{m5}{.6}{m4}\midarrow{m4}{.6}{m3}\midarrow{m3}{.6}{m2}\midarrow{m2}{.6}{m1}
            \draw [thin,fill=gray!20] (m1) .. controls +(-40:.3) and ($(h)!.4!(m16)$) .. (h) .. controls ($(h)!.4!(m24)$) and +(-10:.8) .. (m2) .. controls +(-10:1.2) and +(135:1) .. (m4) .. controls +(135:.6) and ($(h)!.5!(m24)$) .. (h) .. controls ($(h)!.5!(m16)$) and +(-150:.4) .. (m6) .. controls +(-150:.8) and +(-40:.8) .. cycle;
            \draw [NavyBlue,decorate] (m2) .. controls +(-25:1) and +(145:1) .. (m4) .. controls +(115:1) and +(-115:1) .. (m6);
            \draw [RedOrange,decorate] (m1) .. controls +(-20:.6) and +(-160:.6) .. (m6) .. controls +(-60-30:.6) and +(120+30:.6) .. (m5) .. controls +(-120-30:.6) and +(60+30:.6) .. (m4) .. controls +(180-30:.6) and +(30:.6) .. (m3) .. controls +(120-30:.6) and +(-60+30:.6) .. (m2) .. controls +(60-30:.6) and +(-120+30:.6) .. cycle;
            \foreach \i in {1,2,...,6} \draw [fill=white] (m\i) circle [radius=2pt];
            \node [anchor=north] at (-90:\RR) {\small (b)};
        \end{scope}
        \begin{scope}[xshift=8cm]
            \def\RR{1.4}
            \coordinate [label=left:{\small 1}] (m1) at (120:\RR);
            \coordinate [label=left:{\small 2}] (m2) at (180:\RR);
            \coordinate [label=left:{\small 3}] (m3) at (240:\RR);
            \coordinate [label=right:{\small 4}] (m4) at (300:\RR);
            \coordinate [label=right:{\small 5}] (m5) at (0:\RR);
            \coordinate [label=right:{\small 6}] (m6) at (60:\RR);
            \coordinate (m16) at ($(m1)!.5!(m6)$);
            \coordinate (m34) at ($(m3)!.5!(m4)$);
            \coordinate (h) at ($(m16)!.5!(m34)$);
            \draw (m1) -- (m2) -- (m3) -- (m4) -- (m5) -- (m6) -- cycle;
            \midarrow{m1}{.6}{m6}\midarrow{m6}{.6}{m5}\midarrow{m5}{.6}{m4}\midarrow{m4}{.6}{m3}\midarrow{m3}{.6}{m2}\midarrow{m2}{.6}{m1}
            \draw [thin,fill=gray!20] (m1) .. controls +(-45:.4) and ($(h)!.5!(m16)$) .. (h) .. controls ($(h)!.5!(m34)$) and +(45:.4) .. (m3) .. controls +(45:.7) and +(135:.7) .. (m4) .. controls +(135:.4) and ($(h)!.5!(m34)$) .. (h) .. controls ($(h)!.5!(m16)$) and +(-135:.4) .. (m6) .. controls +(-135:.7) and +(-45:.7) .. cycle;
            \draw [RedOrange,decorate] (m1) .. controls +(-20:.6) and +(-160:.6) .. (m6) .. controls +(-60-30:.6) and +(120+30:.6) .. (m5) .. controls +(-120-30:.6) and +(60+30:.6) .. (m4) .. controls +(180-20:.6) and +(20:.6) .. (m3) .. controls +(120-30:.6) and +(-60+30:.6) .. (m2) .. controls +(60-30:.6) and +(-120+30:.6) .. cycle;
            \draw [NavyBlue,decorate] (m1) .. controls +(-70:.8) and +(70:.8) .. (m3) (m6) .. controls +(-110:.8) and +(110:.8) .. (m4);
            \foreach \i in {1,2,...,6} \draw [fill=white] (m\i) circle [radius=2pt];
            \node [anchor=north] at (-90:\RR) {\small (c)};
        \end{scope}
        \begin{scope}[yshift=-3.8cm]
            \def\RR{1.4}
            \coordinate [label=left:{\small 1}] (m1) at (120:\RR);
            \coordinate [label=left:{\small 2}] (m2) at (180:\RR);
            \coordinate [label=left:{\small 3}] (m3) at (240:\RR);
            \coordinate [label=right:{\small 4}] (m4) at (300:\RR);
            \coordinate [label=right:{\small 5}] (m5) at (0:\RR);
            \coordinate [label=right:{\small 6}] (m6) at (60:\RR);
            \coordinate (m16) at ($(m1)!.5!(m6)$);
            \coordinate (m25) at ($(m2)!.5!(m5)$);
            \coordinate (h) at ($(m16)!.5!(m25)$);
            \draw (m1) -- (m2) -- (m3) -- (m4) -- (m5) -- (m6) -- cycle;
            \midarrow{m1}{.6}{m6}\midarrow{m6}{.6}{m5}\midarrow{m5}{.6}{m4}\midarrow{m4}{.6}{m3}\midarrow{m3}{.6}{m2}\midarrow{m2}{.6}{m1}
            \draw [thin,fill=gray!20] (m1) .. controls +(-35:.3) and ($(h)!.4!(m16)$) .. (h) .. controls ($(h)!.4!(m25)$) and +(20:.6) .. (m2) .. controls +(20:1) and +(160:1) .. (m5) .. controls +(160:.6) and ($(h)!.4!(m25)$) .. (h) .. controls ($(h)!.4!(m16)$) and +(-145:.3) .. (m6) .. controls  +(-145:.6) and +(-35:.6) .. cycle;
            \draw [RedOrange,decorate] (m1) .. controls +(-20:.6) and +(-160:.6) .. (m6) .. controls +(-60-30:.6) and +(120+30:.6) .. (m5) .. controls +(-120-30:.6) and +(60+30:.6) .. (m4) .. controls +(180-30:.6) and +(30:.6) .. (m3) .. controls +(120-30:.6) and +(-60+30:.6) .. (m2) .. controls +(60-30:.6) and +(-120+30:.6) .. cycle;
            \draw [NavyBlue,decorate] (m2) -- (m5) .. controls +(-160:.6) and +(60:1) .. (m3) (m2) .. controls +(-20:.6) and +(120:1) .. (m4);
            \foreach \i in {1,2,...,6} \draw [fill=white] (m\i) circle [radius=2pt];
            \foreach \i in {1,2,...,6} \draw [fill=white] (m\i) circle [radius=2pt];
            \node [anchor=north] at (-90:\RR) {\small (d)};
        \end{scope}
        \begin{scope}[xshift=4cm,yshift=-3.8cm]
            \def\RR{1.4}
            \coordinate [label=left:{\small 1}] (m1) at (120:\RR);
            \coordinate [label=left:{\small 2}] (m2) at (180:\RR);
            \coordinate [label=left:{\small 3}] (m3) at (240:\RR);
            \coordinate [label=right:{\small 4}] (m4) at (300:\RR);
            \coordinate [label=right:{\small 5}] (m5) at (0:\RR);
            \coordinate [label=right:{\small 6}] (m6) at (60:\RR);
            \coordinate (m12) at ($(m1)!.5!(m2)$);
            \coordinate (m46) at ($(m4)!.5!(m6)$);
            \coordinate (h) at ($(m12)!.5!(m46)$);
            \draw (m1) -- (m2) -- (m3) -- (m4) -- (m5) -- (m6) -- cycle;
            \midarrow{m1}{.6}{m6}\midarrow{m6}{.6}{m5}\midarrow{m5}{.6}{m4}\midarrow{m4}{.6}{m3}\midarrow{m3}{.6}{m2}\midarrow{m2}{.6}{m1}
            \draw [thin,fill=gray!20] (m1) .. controls +(-80:.3) and ($(h)!.4!(m12)$) .. (h) .. controls ($(h)!.4!(m46)$) and +(-110:.8) .. (m6) .. controls +(-110:1.2) and +(105:1) .. (m4) .. controls +(105:.6) and ($(h)!.5!(m46)$) .. (h) .. controls ($(h)!.5!(m12)$) and +(30:.4) .. (m2) .. controls +(30:.8) and +(-80:.8) .. cycle;
            \draw [NavyBlue,decorate] (m2) .. controls +(-5:1) and +(125:1) .. (m4) .. controls +(95:1) and +(-95:1) .. (m6);
            \draw [RedOrange,decorate] (m1) .. controls +(-30:.6) and +(-150:.6) .. (m6) .. controls +(-60-30:.6) and +(120+30:.6) .. (m5) .. controls +(-120-30:.6) and +(60+30:.6) .. (m4) .. controls +(180-30:.6) and +(30:.6) .. (m3) .. controls +(120-30:.6) and +(-60+30:.6) .. (m2) .. controls +(60-20:.6) and +(-120+20:.6) .. cycle;
            \foreach \i in {1,2,...,6} \draw [fill=white] (m\i) circle [radius=2pt];
            \node [anchor=north] at (-90:\RR) {\small (e)};
        \end{scope}
        \begin{scope}[xshift=8cm,yshift=-3.8cm]
            \def\RR{1.4}
            \coordinate [label=left:{\small 1}] (m1) at (120:\RR);
            \coordinate [label=left:{\small 2}] (m2) at (180:\RR);
            \coordinate [label=left:{\small 3}] (m3) at (240:\RR);
            \coordinate [label=right:{\small 4}] (m4) at (300:\RR);
            \coordinate [label=right:{\small 5}] (m5) at (0:\RR);
            \coordinate [label=right:{\small 6}] (m6) at (60:\RR);
            \coordinate (m13) at ($(m1)!.5!(m3)$);
            \coordinate (m46) at ($(m4)!.5!(m6)$);
            \coordinate (h) at ($(m13)!.5!(m46)$);
            \draw (m1) -- (m2) -- (m3) -- (m4) -- (m5) -- (m6) -- cycle;
            \midarrow{m1}{.6}{m6}\midarrow{m6}{.6}{m5}\midarrow{m5}{.6}{m4}\midarrow{m4}{.6}{m3}\midarrow{m3}{.6}{m2}\midarrow{m2}{.6}{m1}
            \draw [thin,fill=gray!20] (m1) .. controls +(-70:.8) and ($(h)!.4!(m13)$) .. (h) .. controls ($(h)!.4!(m46)$) and +(-110:.8) .. (m6) .. controls +(-110:1) and +(110:1) .. (m4) .. controls +(110:.8) and ($(h)!.4!(m46)$) .. (h) .. controls ($(h)!.4!(m13)$) and +(70:.8) .. (m3) .. controls +(70:1) and +(-70:1) .. cycle;
            \draw [RedOrange,decorate] (m1) .. controls +(-30:.6) and +(-150:.6) .. (m6) .. controls +(-60-30:.6) and +(120+30:.6) .. (m5) .. controls +(-120-30:.6) and +(60+30:.6) .. (m4) .. controls +(180-30:.6) and +(30:.6) .. (m3) .. controls +(120-30:.6) and +(-60+30:.6) .. (m2) .. controls +(60-30:.6) and +(-120+30:.6) .. cycle;
            \draw [NavyBlue,decorate] (m1) .. controls +(-85:1) and +(85:1) .. (m3) (m6) .. controls +(-95:1) and +(95:1) .. (m4);
            \foreach \i in {1,2,...,6} \draw [fill=white] (m\i) circle [radius=2pt];
            \node [anchor=north] at (-90:\RR) {\small (f)};
        \end{scope}
    \end{tikzpicture}
    \caption{Representative diagrams for $\mathbb{G}_6$. The zigzag lines indicate summation over $\phi$ contractions. While the red lines are absorbed into an overall factor $D_6$ defined in \cref{eq:defD}, the blue lines turn into the $W$ functions defined in \cref{eq:defW}. Note that in (a) and (d) the blue lines (24) and (35) cross each other, which means that in the summation $b_{24}$ and $b_{35}$ cannot be simultaneously positive.}
    \label{fig:6pt}
\end{figure}

\subsection{Generating functions at six and higher points}

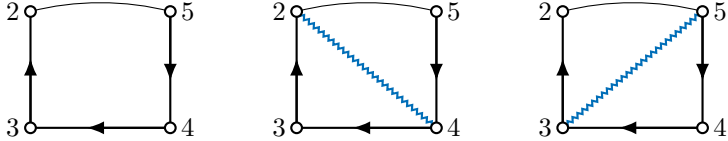
\begin{figure}[ht]
    \centering
    \begin{tikzpicture}[thick,decoration={zigzag,segment length=2.4pt,amplitude=.8pt}]
        \begin{scope}
            \def\RR{1.2}
            \coordinate [label=left:{\small 2}] (m2) at (140:\RR);
            \coordinate [label=left:{\small 3}] (m3) at (-140:\RR);
            \coordinate [label=right:{\small 4}] (m4) at (-40:\RR);
            \coordinate [label=right:{\small 5}] (m5) at (40:\RR);
            \draw (m2) -- (m3) -- (m4) -- (m5);
            \midarrow{m5}{.6}{m4}\midarrow{m4}{.6}{m3}\midarrow{m3}{.6}{m2}
            \draw [thin] (m2) .. controls +(15:.6) and +(165:.6) .. (m5);
            \foreach \i in {2,3,4,5} \draw [fill=white] (m\i) circle [radius=2pt];
        \end{scope}
        \begin{scope}[xshift=3.5cm]
            \def\RR{1.2}
            \coordinate [label=left:{\small 2}] (m2) at (140:\RR);
            \coordinate [label=left:{\small 3}] (m3) at (-140:\RR);
            \coordinate [label=right:{\small 4}] (m4) at (-40:\RR);
            \coordinate [label=right:{\small 5}] (m5) at (40:\RR);
            \draw (m2) -- (m3) -- (m4) -- (m5);
            \midarrow{m5}{.6}{m4}\midarrow{m4}{.6}{m3}\midarrow{m3}{.6}{m2}
            \draw [thin] (m2) .. controls +(15:.6) and +(165:.6) .. (m5);
            \draw [NavyBlue,decorate] (m2) -- (m4);
            \foreach \i in {2,3,4,5} \draw [fill=white] (m\i) circle [radius=2pt];
        \end{scope}
        \begin{scope}[xshift=7cm]
            \def\RR{1.2}
            \coordinate [label=left:{\small 2}] (m2) at (140:\RR);
            \coordinate [label=left:{\small 3}] (m3) at (-140:\RR);
            \coordinate [label=right:{\small 4}] (m4) at (-40:\RR);
            \coordinate [label=right:{\small 5}] (m5) at (40:\RR);
            \draw (m2) -- (m3) -- (m4) -- (m5);
            \midarrow{m5}{.6}{m4}\midarrow{m4}{.6}{m3}\midarrow{m3}{.6}{m2}
            \draw [thin] (m2) .. controls +(15:.6) and +(165:.6) .. (m5);
            \draw [NavyBlue,decorate] (m3) -- (m5);
            \foreach \i in {2,3,4,5} \draw [fill=white] (m\i) circle [radius=2pt];
        \end{scope}
    \end{tikzpicture}
    \caption{Partitions of a 4-gon.}
    \label{fig:partitions}
\end{figure}

Based on our experience at four and five points, we can use a similar strategy to work out the generating function $\mathbb{G}_n$ for $n\geq6$. Starting at six points some new features start to arise regarding the $\phi$ contractions, which needs to be further illustrated. Representative diagrams for $\mathbb{G}_6$ are listed in \Cref{fig:6pt}, up to cyclic permutations.  As the number of external points grows, we see that the number of points included in each white region also increases. In particular, in (a) and (d) we start to encounter a white region with four points. In this region, it is important to note that not all $\phi$ contractions are compatible with each other under planarity: the contractions $(24)$ and $(35)$ cannot coexist. This means that the $b$ summations inside the generating function are not completely independent. Instead they have to obey the condition that $b$ for incompatible contractions should not be simultaneously positive. A convenient way to handle such summations is to treat each white region as a polygon where one segment of its boundary is provided by the shaded region. We then enumerate all possible partitions of the polygon by adding lines between vertices. Each added line represents the summation of $\phi$ contractions $P^{(1)}$. In these partitions, we should include in particular the trivial partition where no lines are added. Take the 4-gon in diagram (a) and (d) as an example, there are three allowed partitions as listed in \Cref{fig:partitions}. The total contribution of these three partitions is
\begin{align}\label{eq:4gonpartition}
    1+P_{24}^{(1)}+P_{35}^{(1)}\,,
\end{align}
where we set the trivial partition to 1. 

On top of this, we also need to be careful with the contractions between neighboring vertices if they are the end points of the segment provided by the shaded region. As we discussed before, if these two vertices belong to different pairs of indices of $\Hb$ (e.g., in (a)), the resumed contractions gives a factor $P^{(0)}\equiv x^2/X^2$ as in \cref{eq:sumn0}. If they belong to the same pair (e.g., in (d)), then there needs to be at least one propagator and the resumed contraction gives $P^{(1)}\equiv t/X^2$ as in \cref{eq:sumn1}. 
Taking these into account, we can write down the following generating function for six-point correlators   
\begin{align}
    \mathbb{G}_6=\lambda\,D_6\,\bigg(&\sum_{\text{cyclic}}\left(\Hb_{12,56}P_{25}^{(0)}+\Hb_{61,25}P_{25}^{(1)}\right)\left(1+P_{24}^{(1)}+P_{35}^{(1)}\right)\nonumber\\
    &+\sum_{\text{cyclic}}\left(\Hb_{61,24}P_{24}^{(1)}P_{46}^{(0)}+\Hb_{12,46}P_{24}^{(0)}P_{46}^{(1)}\right)\nonumber\\
    &+\sum_{\text{cyclic}}\left(\Hb_{61,34}P_{13}^{(0)}P_{46}^{(0)}+\Hb_{13,46}P_{13}^{(1)}P_{46}^{(1)}\right)\bigg)\,.
\end{align}
We also explicitly computed a few low-lying explicit examples using Feynman diagrams. These results are in full agreement with the above generating function.

The generalization to arbitrarily many points now becomes immediate. Let us define the following function which accounts for the total contribution from a white region, with the trivial partition set to 1 
\begin{align}\label{eq:defW}
    W^{(\kappa)}_{i,i+1,\ldots j}=(\text{partition contributions})\times P_{ij}^{(\kappa)}\,.
\end{align}
Here the vertices of the white region are $i,i+1,\ldots,j$ and the segment $(i,j)$ belongs to the shaded region. To make it more explicit, we list here the results for the first few cases 
\begin{subequations}
    \begin{align}
        W^{(\kappa)}_{12}&=1\;,\\
        W^{(\kappa)}_{123}&=P_{13}^{(\kappa)}\;,\\
        W^{(\kappa)}_{1234}&=\left(1+P_{13}^{(1)}+P_{24}^{(1)}\right) P_{14}^{(\kappa)}\;,\\
        W^{(\kappa)}_{12\ldots 5}&=\left(1+P_{13}^{(1)}+P_{14}^{(1)}+P_{24}^{(1)}+P_{25}^{(1)}+P_{35}^{(1)}+P_{13}^{(1)}P_{14}^{(1)}+P_{13}^{(1)}P_{35}^{(1)}\right.\nonumber\\
        &\quad\left.+P_{14}^{(1)}P_{24}^{(1)}+P_{24}^{(1)}P_{25}^{(1)}+P_{25}^{(1)}P_{35}^{(1)}\right)\times P_{15}^{(\kappa)}\;.
    \end{align}
\end{subequations}
For the special case of 2-gons $W_{ij}$, since $(i,j)$ is also adjacent in the planar ordering, its contractions have already been taken into account by $D_n$. The second factor in \cref{eq:defW} is not included to avoid overcounting. In terms of these building blocks, we can write the $n$-point generating function as
\begin{equation}\label{eq:nptfunction}
    \mathbb{G}_n =\lambda\,D_n\sum_{\text{$(ij)(kl)$ pairs} }  {\Hb_{ij,kl}}  \times W^{(1)}_{i,i+1,\ldots j}W^{(0)}_{j,j+1,\ldots k}W^{(1)}_{k,k+1,\ldots l}W^{(0)}_{l,l+1,\ldots i} \, ,
\end{equation}
where the summation is over all different ways of choosing two pairs of points.

Let us make a comment about our result (\ref{eq:nptfunction}) before concluding this section. The chiral algebra condition (\ref{chiralalgebracond}) is not guaranteed by the Feynman diagram computation and therefore is a nontrivial consistency check. Here an important point to note is that for $n>4$ and when all operators are inserted on a 2d plane, some of the building block functions $\Hb$ become linearly dependent. This linear dependence must be taken into account when we check the chiral algebra condition (\ref{chiralalgebracond}). These linear relations can be trivialized by looking at their symbols and this can be conveniently implemented using PolyLogTools \cite{Duhr:2019tlz}. We have explicitly checked the chiral algebra constraint at the level of generating functions up to $n=6$.

\section{The $D_4$ theory}\label{Sec:D4theory}

\begin{table}[t]
\begin{center}
\begin{tabular}{|c|c|c|c|c|c|c|}
\hline
\textrm{Component} & Lorentz & $\SU(2)_a$ &$\SU(2)_{\bar{a}}$ & $\U(1)_R$ & $\USp(2N)$ & $\mathrm{SO}(8)$ \\ \hline
$A^\mu $ & vector & 1 & 1 & $0$ & Adjoint &1 \\
$z^{a\bar{a}}$& scalar& 2 & 2  & $0$ & Antisymmetric & 1 \\ 
$q^{m,a}$& scalar& 2 & 1  & $0$ & Fundamental & 8 \\ \hline
\end{tabular}
\caption{Quantum numbers of bosonic component fields in the $D_4$ theory.
}
\label{fieldrep}
\end{center}
\end{table}

In this section, we consider the other $D_4$ theory described in Section \ref{sec:intro}. In this theory, we have 4d $\mathcal{N}=2$ SYM with gauge group $\USp(2N)$ coupled to matter. The matter content consists of four fundamental hypermultiplets and an antisymmetric hypermultiplet, of which the scalar components we denote as $q$ and $z$ respectively. Since the fundamental representation of $\USp(2N)$ is pseudo-real, the fundamental scalar $q$ and its conjugate $\bar{q}$ can be combined together, enhancing the original $\U(4)$ flavor symmetry to $\SO(8)$. Moreover, there is an $\SU(2)_{\bar{a}}$ flavor symmetry for the antisymmetric fields. The R-symmetry group is $\U(2)_R \sim \SU(2)_a \times \U(1)_R$. Transformation properties of various fields are summarized in Table \ref{fieldrep}. The explicit Lagrangian of the $D_4$ theory can be found in (2.12) of \cite{Bedford:2007qj}.\footnote{We have slightly changed the convention of \cite{Bedford:2007qj} to match the one in our paper. 
We first take $q,z \to q/g_{\rm YM},z/g_{\rm YM}$, then rescale the whole Lagrangian by $2$ and wick rotate to the Euclidean signature. We denote the labels $A,B$ by $\bar{a},\bar{b}$ now, and also $M,N$ by $m,n$. } and we record here the part relevant for computing the one-loop corrections to the meson correlators
\begin{align}
    \mathcal{L} =  &\ \frac{2N}{\lambda} \Big \{ \mathrm{Tr} \left[ \frac{1}{4} F^{\mu\nu} F_{\mu\nu} +\frac{1}{2} D^\mu z_{\bar{a}a}D_\mu z^{a \bar{a}} + \frac{1}{4} [ z^{a \bar{a}},z_{\bar{a} b } ] [z^{b \bar{b}},z_{\bar{b} a }] \right] + \frac{1}{2}  D^\mu q_{m,a}  D_\mu q^{m,a} \nonumber\\
    &+  \frac{1}{4}  q_{m, a} [z^{a\bar{a}},z_{\bar{a} b}] q^{m, b} + \frac{1}{8}\left( (q_{m,a} q^{n,a})(q_{n,b}q^{m,b})  + (q_{m,a} q_{n,b} )(q^{n,a} q^{m,b}) \right) \nonumber\\
    &+(\textrm{\small{other parts}})\Big\} \; .
 \label{D4Lag}
\end{align}
From Table \ref{fieldrep}, it is clear that the $\frac{1}{2}$-BPS meson operators can only be constructed using $q$ and $z$ and they take the following form
\begin{equation}
   \mathcal{M}_p^f(x;v,\bar{v})= \sqrt{\frac{2^{p-2}N}{\lambda^p}} (T^f)_{mn} q_m^{\ (a_1}z^{a_2(\bar{a}_1}\ldots z^{a_{p-1}\bar{a}_{p-2})}q^{a_p)}_n v_{a_1}\ldots v_{a_p}\bar{v}_{\bar{a}_1}\ldots \bar{v}_{\bar{a}_{p-2}}\;,
   \label{normalizeD4}
\end{equation}
which is similar to the operators in \eqref{eq:SUNfmeson}. We should also take $(T^f)^m_{\ \ n}$ to be the generator of the flavor group $\SO(8)$. Similar to (\ref{eq:Nfdecompose}), we  perform a decomposition of the correlator in terms of the traces of $\SO(8)$ generators
\begin{align}
    \langle \mathcal{M}_{p_1}^{f_1}\mathcal{M}_{p_2}^{f_2} \cdots \mathcal{M}_{p_n}^{f_n} \rangle = \sum_{\sigma \in S_n/(Z_n\times Z_2)}\tr(T^{f_{\sigma(1)}}\cdots T^{f_{\sigma(n)}}) G[\sigma] + \cdots\;.
\end{align}
We will refer to $G[\sigma]$ as the single-trace partial correlator and $\cdots$ stands for higher-trace contributions. An important difference compared with the $\SU(N_F)$ case is that the trace of $\SO(2n)$ has an additional reflection symmetry $Z_2$\footnote{Another subtlety is that $\SO(8)$ has a small rank. So the single trace and higher traces of $\SO(8)$ become linear dependent when we consider higher-point correlators. Here we also pretend to work with a general $\SO(2m)$ group and set $m=4$ in the last step. }
\begin{align}
    \tr(T^{f_1}\cdots T^{f_n}) = (-1)^n\tr(T^{f_n}\cdots T^{f_1})\;,
\end{align}
which we need to mod out in the summation. The partial correlator we want to consider is the one with canonical ordering
\begin{align}
    G_n\equiv G[12\cdots n].
\end{align}

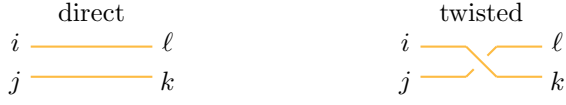
\begin{figure}[t]
    \centering
\begin{tikzpicture}
        \path (-2,-0.6) rectangle +(5,1.2);
        \coordinate  (a) at (-1,0);
        \coordinate  (b) at (1,0);
        \coordinate [label=above:{\small direct}] (c) at (0,0.4);
        \draw [thick,Dandelion] ($(a)+(0.2,0.2)$) -- ($(b)+(-0.2,0.2)$);
        \draw [thick,Dandelion] ($(b)+(-0.2,-0.2)$) -- ($(a)+(0.2,-0.2)$);
        \node [anchor=north] at (a) {\small $j$};
        \node [anchor=north] at (b) {\small $k$};
        \node [anchor=south] at (a) {\small $i$};
        \node [anchor=south] at (b) {\small $\ell$};
    \end{tikzpicture}
   \begin{tikzpicture}
        \path (-2,-0.6) rectangle +(5,1.2);
        \coordinate  (a) at (-1,0);
        \coordinate  (b) at (1,0);
        \coordinate [label=above:{\small twisted}] (c) at (0,0.4);
        \draw [thick,Dandelion] ($(a)+(0.2,0.2)$) -- ($(b)+(-1.2,0.2)$);
         \draw [thick,Dandelion] ($(b)+(-1.2,0.2)$) -- ($(b)+(-0.8,-0.2)$);
          \draw [thick,Dandelion] ($(b)+(-0.8,-0.2)$) -- ($(b)+(-0.2,-0.2)$);
        \draw [thick,Dandelion] ($(b)+(-1.2,-0.2)$) -- ($(a)+(0.2,-0.2)$);
        \draw [thick,Dandelion] ($(b)+(-1.2,-0.2)$) -- ($(b)+(-1.1,-0.1)$);
        \draw [thick,Dandelion] ($(b)+(-0.92,0.08)$) -- ($(b)+(-0.8,0.2)$);
        \draw [thick,Dandelion] ($(b)+(-0.8,0.2)$) -- ($(b)+(-0.2,0.2)$);
        \node [anchor=north] at (a) {\small $j$};
        \node [anchor=north] at (b) {\small $k$};
        \node [anchor=south] at (a) {\small $i$};
        \node [anchor=south] at (b) {\small $\ell$};
    \end{tikzpicture}
    \caption{The color structures in $A^\mu$ and $z$ propagators. }
    \label{fig:twist double line}
\end{figure}

Using the Schouten identity and the cyclicity of the trace, we can rewrite the four-scalar vertices in (\ref{D4Lag}) as 
\begin{align}
    \mathcal{L} =  &\ \frac{2N}{\lambda} \Big \{ \mathrm{Tr} \left[ \frac{1}{4} F^{\mu\nu} F_{\mu\nu} +\frac{1}{2} D^\mu z_{\bar{a}a}D_\mu z^{a \bar{a}} - \frac{1}{4} [ z^{a \bar{a}},z^{b\bar{b}} ] [z_{\bar{a}a},z_{\bar{b} b }] \right] + \frac{1}{2}  D^\mu q_{m,a}  D_\mu q^{m,a} \nonumber\\
    &+  \frac{1}{4}  q_{m, a} [z^{a\bar{a}},z_{\bar{a} b}] q^{m, b} + \frac{1}{4}\left( (q_{m,a} q^{n,a})(q_{n,b}q^{m,b})  -\frac{1}{2} (q_{m,a} q^{n,b} )(q_{n,b} q^{m,a}) \right) \nonumber\\
    &+(\textrm{\small{other parts}})\Big\} \; .
\end{align}
This already brings the Lagrangian to a form much closer to (\ref{previousL}) where the indices are contracted in the same way as in the $\SU(N_f)$ theory except some coefficients are different. To consider the perturbative calculation in more detail, let us look at the color structure of the propagators. In contrast to the case of $\SU(N)$ gauge group, fields in the adjoint and antisymmetric representations of $\USp(2N)$ gauge group have two color structures in their propagators 
\begin{align}
    &\langle (A^\mu)^i_{\ j} (A^\nu)^k_{\ \ell} \rangle \sim \frac{1}{2}\left(\delta^i_{\ \ell} \delta_j^{\ k}-J^{ik}J_{j \ell}\right), \\
    &\langle z^i_{\ j} z^k_{\ \ell} \rangle \sim \frac{1}{2}\left(\delta^{i}_{\  \ell}\delta_{j}^{\  k} + J^{ik} J_{j\ell} -\frac{1}{N} \delta^{i}_{  j} \delta^{k}_{\ \ell}\right)\;.
\end{align}
Here $J$ is the antisymmetric invariant tensor of $\USp(2N)$ which relates the fundamental and anti-fundamental representations. These two color structures are depicted in \Cref{fig:twist double line}. In addition to the direct contraction of indices, we also have a twisted one. Both must be taken into account when we count the number of color loops in correlators. Another subtlety is the aforementioned reflection symmetry of the $\SO(8)$ flavor ordering which also enters the combinatorics. 

Because of these complications, we will not attempt to present in this paper a general proof which maps Feynman diagrams from one theory into the other. Instead, we content ourselves with computing several examples of correlators at small points (i.e., $n=4,5,6$) and finding the results in the $D_4$ theory fully agree with those of the $\SU(N_f)$ theory.\footnote{That the results are identical for $n=4$ is also expected from superconformal symmetry. The similarity between the two Lagrangians means the contributing Feynman diagrams are the same. However, for four-point functions the coefficients of different diagrams can be uniquely fixed by the superconformal Ward identities.}\footnote{There appears to be an additional overall factor $2^{2-\frac{n}{2}}$ which should be attributed to the difference of gauge groups.} These nontrivial cases are sufficient to conclude that the one-loop corrections to the meson correlators are identical and the same eight dimensional hidden structures also exist in the $D_4$ theory.

\section{Discussions}\label{Sec:discussion}
In this paper, we studied meson correlators in $\mathcal{N}=2$ SCFTs in four dimensions at weak coupling. We focused on two theories: $\mathcal{N}=4$ SYM coupled to $\mathcal{N}=2$ hypermultiplets and the $D_4$ theory which is an $\mathcal{N}=2$ $\USp(2N)$ gauge theory with $\SO(8)$ flavor symmetry. Using Feynman diagram techniques, we computed the one-loop corrections to $n$-point correlators. In both theories, we observed that these corrections can be resumed into generating functions manifesting hidden 8d structures. This complements the strong coupling result \cite{Alday:2021odx,Huang:2024dxr} where a similar structure was found, and provides a nice parallel of analogous results in $\mathcal{N}=4$ SYM. There are several interesting directions to explore in the future. 

First, it would be interesting to see how universal the hidden structures are in weakly coupled $\mathcal{N}=2$ SCFTs. In the strong coupling limit where the theories are described as $AdS_5\times S^3$ $\mathcal{N}=1$ SYM by AdS/CFT, it was shown in \cite{Alday:2021odx} (and at loop levels in \cite{Alday:2021ajh,Huang:2023oxf,Huang:2023ppy,Huang:2024rxr}) that the meson correlators are essentially independent of the detailed field theory realization. This comes as a consequence of the fact that the analytic bootstrap method used in these works is agnostic to the flavor group (or color group in AdS).  Different choices of theories only amount to changing the flavor group, of which the dependence is factored out from the dynamic part. Consequently, at strong coupling the emergent 8d structures are a property universal to all models. It is natural to ask if this universality is also shared in the opposite weakly coupled regime for those theories admitting a marginal coupling. In this paper, we investigated two such theories and found that the one-loop corrections to meson correlators are identical. It would be interesting to consider other Lagrangian $\mathcal{N}=2$ SCFTs. 

Second, in this paper we have only computed the corrections to meson correlators at one-loop level. An obvious but important extension would be to study correlations at higher loops and see if the hidden higher dimensional structures persist. In $\mathcal{N}=4$ SYM, there is ample evidence indicating that the hidden structures extend to all loops. As we have seen from the one-loop  calculation, the conspiracies of Feynman diagrams leading to the higher dimensional structures  are very similar in $\mathcal{N}=2$ and $\mathcal{N}=4$. This makes the higher-loop extension of the structure in the $\mathcal{N}=2$ case quite hopeful. However, a direct Feynman diagram analysis would be too complicated. More efficient techniques, such as a generalization of the Lagrangian insertion method in $\mathcal{N}=4$ SYM, will need to be established first in order to make systematic progress at higher loops.   

Finally, via AdS/CFT, strongly coupled correlators are dual to on-shell scattering amplitudes in weakly coupled AdS theories. An AdS double copy relation, highly similar to the flat-space version, has been found in \cite{Zhou:2021gnu} and relates all infinitely many tree-level four-point functions in $AdS_5\times S^5$ IIB supergravity and $AdS_5\times S^3$ SYM where they describe graviton and gluon scattering respectively. This relation establishes a surprising connection between these two seemingly unrelated SCFTs and in particular preserves the hidden higher dimensional structures on both sides. An interesting question is if a ``double copy'' relation, defined in a suitable sense, can also be found at weak coupling which relates loop corrections in $\mathcal{N}=2$ and $\mathcal{N}=4$ correlators. The hidden higher dimensional structures we found in this paper will make it straightforward to generalize the relation to correlators of arbitrary conformal dimensions starting from operators with the lowest dimension 2. Note that the double copy relation at strong coupling is formulated in Mellin space where the scattering amplitude nature is manifested. However, at weak coupling it might be more natural to consider correlators in the original position space.

\acknowledgments
	
The authors would like to thank Yichao Tang for useful discussions. ZH, BW and EYY are supported by National Science Foundation of China under Grant No.~12175197 and Grand No.~12347103. EYY is also supported by National Science Foundation of China under Grant No.~11935013, and by the Fundamental Research Funds for the Chinese Central Universities under Grant No.~226-2022-00216. X.Z. and X.D. are supported by the NSFC Grant No.~12275273, funds from Chinese Academy of Sciences, University of Chinese Academy of Sciences, and the Kavli Institute for Theoretical Sciences. X.Z. is also supported by the Xiaomi Foundation.

\appendix

\section{Notation and conventions}\label{notation}

In this appendix we summarize the notations and conventions used in  Section \ref{Sec:N=4withN=2general} and in Appendix \ref{derive N=2}.

\subsection{Group theory conventions}
 We take the spacetime metric to be 
\begin{align}
    \eta^{\mu\nu} = {\rm diag}(1,-1,-1,-1),
\end{align}
 and use the antisymmetric tensors $\e^{\a\b}$ and  $\e_{\a\b}$
\begin{align}
\epsilon^{\alpha \beta}=\epsilon^{\da \db}=\begin{pmatrix}
0 & 1 \\
-1 & 0
\end{pmatrix},\qquad  \epsilon_{\alpha \beta}=\epsilon_{\dot{\alpha} \dot{\beta}}=\begin{pmatrix}
0 & -1 \\
1 & 0
\end{pmatrix},
\end{align}
to raise and lower the spinor indices
\begin{align}
    \psi^\alpha = \epsilon^{\alpha\beta} \psi_\beta,\qquad\psi_\alpha = \epsilon_{\alpha\beta}\psi^\beta, \qquad \bar{\psi}^{\dot{\alpha}} = \epsilon^{\dot{\alpha}\dot{\beta}}\bar{\psi}_{\dot{\beta}},\qquad \bar{\psi}_{\dot{\alpha}} = \epsilon_{\dot{\alpha}\dot{\beta}}\bar{\psi}^{\dot{\beta}}\,.
\end{align}
The Pauli matrices are defined as 
\begin{align}
    \sigma_1 = \begin{pmatrix} 0&1\\1&0 \end{pmatrix},\quad \sigma_2 = \begin{pmatrix} 0&-i\\i&0 \end{pmatrix},\quad \sigma_3 = \begin{pmatrix} 1&0\\0&-1 \end{pmatrix}.
\end{align}
Accordingly, the 4d Pauli matrices for the Lorentz group $\SO(3,1)$ are defined as
\begin{align}
    (\sigma^\mu)_{\alpha\dot \alpha}\equiv (1,\vec\sigma),\qquad (\bar\sigma^\mu)^{\dot \alpha \alpha }\equiv (1,-\vec\sigma),
\end{align}
and for the group $\SO(4)\sim \SU(2)_R\times \SU(2)_L$ they are defined as 
\begin{align}
    (\sigma^A)^{a\bar{a}}\equiv (i\sigma_3,-1,i\sigma_1,-i\sigma_2),\qquad (\bar\sigma^A)_{\bar{a}a}\equiv (-i\sigma_3,-1,-i\sigma_1,i\sigma_2).
\end{align}
These Pauli matrices satisfy 
\begin{align}
    \sigma^\mu \bar{\sigma}^\nu + \sigma^\nu \bar{\sigma}^\mu = 2\eta^{\mu\nu},&\qquad\qquad (\sigma^\mu)^\dagger = \bar{\sigma}^\mu,\\
    \sigma^A \bar{\sigma}^B + \sigma^B \bar{\sigma}^A = 2\delta^{AB},&\qquad\qquad (\sigma^A)^\dagger = \bar{\sigma}^A.
\end{align}
For the 6d Gamma matrices of $\SO(6)\sim \SU(4)$, we adopt the convention that the contraction with a 6d vector
\begin{align}
    t^I =(\Re z_1, \Im z_1,\Re z_2, \Im z_2,\Re z_3, \Im z_3)\;,
\end{align}
gives
\begin{align}
    t^I (\Gamma^I)^{ij} = i\begin{pmatrix} 0 & z_1 & z_2 & z_3 \\ -z_1 & 0 & z_3^\dagger & -z_2^\dagger \\ -z_2 & -z_3^\dagger & 0 & z_1^\dagger \\ -z_3 & z_2^\dagger & -z_1^\dagger & 0 \end{pmatrix},\quad  t^I (\bar{\Gamma}^I)_{ij} = i\begin{pmatrix} 0 & z_1^\dagger & z_2^\dagger & z_3^\dagger \\ -z_1^\dagger & 0 & z_3 & -z_2 \\ -z_2^\dagger & -z_3 & 0 & z_1 \\ -z_3^\dagger & z_2 & -z_1 & 0 \end{pmatrix}.
    \label{gamma}
\end{align}
These gamma matrices satisfy 
\begin{align}
    \Gamma^I \bar{\Gamma}^J + \Gamma^J\bar{\Gamma}^I = 2\delta^{IJ},\qquad\qquad (\Gamma^I)^\dagger = \bar{\Gamma}^I.
\end{align}
When limited on $I=3,4,5,6$ and $i=1,2,\ j=3,4$, the 6d gamma matrices $(\Gamma^I)^{ij}$ reduce to the 4d Pauli matrices $(\sigma^A)^{a\bar{a}}$. 

The conventions for the generators $T^I$ of the gauge group $\SU(N)$ are
\begin{align}
    [T^I,T^J] = i f^{IJK} T^K,\quad \tr T^I T^J = \delta^{IJ},\quad (T^I)^i_{\ j}(T^I)^k_{\ \ell} = \delta^i_{\ l}\delta^k_{\ j} - \frac{1}{N}\delta^i_{\ j}\delta^k_{\ \ell}\,.
\end{align}
For the generators of $\USp(2N)$, we use the conventions 
\begin{align}
    [T^I,T^J] = i f^{IJK} T^K,\quad \tr T^I T^J = \delta^{IJ},\quad (T^I)^i_{\ j}(T^I)^k_{\ \ell} = \frac{1}{2}\left(\delta^i_{\ \ell} \delta_j^{\ k}-J^{ik}J_{j \ell}\right)\,,
\end{align}
where
\begin{align}
    J_{ij} = J^{ij} = -(J^{-1})^{ij} = i\sigma_2 \otimes 1_{N\times N}\;,
\end{align}
is the invariant tensor of $\USp(2N)$. 

\subsection{Bracket notation}

For the contractions of fermionic spinor variables, we introduce the bracket notation to lighten the expressions
\begin{align}
    \bracketaa{\psi\chi} \equiv \psi^\alpha \chi_\alpha\;,\qquad  \bracketss{\bar{\psi}\bar{\chi}} \equiv \bar{\psi}_{\dot{\alpha}} \bar{\chi}^{\dot{\alpha}}\;.
\end{align}
The conjugations of these spinors are defined as
\begin{align}
    |\psi\rangle^* = |\bar{\psi}]\,,\qquad |\psi\rangle^\dagger = [\bar{\psi}|\,.
\end{align}
We also summarize a few identities which will be useful in the superspace computation 
\begin{align}
    \bracketaa{\psi\chi} =&\ \bracketaa{\chi\psi}\;,\\
    \bracketas{\psi\sigma^\mu\bar{\chi}} =& -\bracketsa{\bar{\chi} \bar{\sigma}^\mu \psi}\;,\\
    \bracketaa{\psi\sigma^\mu\bar{\sigma}^\nu\chi} =&\ \bracketaa{\chi\sigma^\nu\bar{\sigma}^\mu\psi}\;.
\end{align}

\subsection{$\mathcal{N}=1$ superspace and superfields}

Here we introduce our conventions of $\mathcal{N}=1$ superspace and superfields, which will be used in Appendix \ref{derive N=2} to construct the Lagrangian of the $\SU(N_f)$ theory. In $\mathcal{N}=1$ superspace there are two Grassmann coordinates $\t_\a$, $\bar{\t}_\da$ which combine with the usual spacetime coordinate $x^\mu$ to form the superspace coordinates $(x^\mu,\t_\a,\bar{\t}_\da)$. The $\mathcal{N}=1$ superfields are functions on these superspace coordinates and can be used to construct Lagrangians with $\mathcal{N}=1$ supersymmetry in a manifest way.

The Grassmannian coordinates satisfy the relations
\begin{align}
 \theta^2 =&\theta^\alpha\theta_\alpha = -2\theta^1\theta^2 \;,\quad
  \theta^\alpha\theta^\beta =
 -\frac{1}{2}\epsilon^{\alpha\beta} \theta^2\;, \nonumber\\
\bar\theta^2 =& \bar
 \theta_{\dot\alpha} \bar \theta^{\dot\alpha} = 2\bar\theta_{\dot
 1}\bar\theta_{\dot 2} \;, \quad
 \bar\theta_{\dot\alpha}\bar\theta_{\dot\beta} = -\frac{1}{2}
 \epsilon_{\dot\alpha\dot\beta} \bar\theta^2\;, \nonumber\\
  \int\, & d\theta\, \theta= 1 \;,\quad \int\,  d^2\theta\, \theta^\alpha\theta^\beta =  - \frac{1}{2} \epsilon^{\alpha\beta}\,,
\end{align}
and we have the following useful identities
\begin{align}
 \bracketaa{\psi\theta}\bracketaa{\theta\chi} =& -\frac{1}{2}\bracketaa{\psi\chi}\bracketaa{\theta\theta}\;,\\
    \bracketas{\theta\sigma^\mu\bar{\theta}}\bracketas{\theta\sigma^\nu\bar{\theta}} =&\ \frac{1}{2}\bracketaa{\theta\theta}\bracketss{\bar{\theta}\bar{\theta}}\eta^{\mu\nu}\;.
\end{align}

The $\Ncal=1$ chiral superfield consists of a complex scalar $\phi$, a Weyl fermion $\psi_{\alpha}$ and an auxiliary field $F$. In terms of these component fields, the chiral superfield can be written as 
\begin{align}
    \Phi =&\ \phi(y) + \sqrt{2}\, \bracketaa{\theta\, \psi(y)} + \bracketaa{\theta \theta} F(y)\,,
\end{align} 
where 
\begin{align}
    y^\mu =  x^\mu + i\, \bracketas{\theta \sigma^\mu \bar{\theta}}\;,\qquad \bar{y}^\mu = x^\mu - i\, \bracketas{\theta \sigma^\mu \bar{\theta}}\;,
\end{align}
are the chiral coordinates. 

The $\Ncal=1$ vector superfield contains of a Weyl fermion $\lambda_\alpha$ and a vector field $A_\mu$, both in the adjoint representation of the gauge group. Under Wess-Zumino gauge, it contains one additional auxiliary field $D$, and can be written as 
\begin{align}
    V = \bracketas{\theta\sigma^\mu\bar{\theta}}A_\mu(x) + \bracketaa{\theta\theta}\bracketss{\bar{\theta}\bar{\lambda}(x)} + \bracketss{\bar{\theta}\bar{\theta}}\bracketaa{\theta\lambda(x)} + \frac{1}{2} \bracketaa{\theta\theta}\bracketss{\bar{\theta}\bar{\theta}} D(x)\,.
\end{align}
The super field strength can be constructed from $V$ by taking  superderivatives 
\begin{align}
    W_\alpha \equiv& -\frac{1}{8}\bracketss{\bar{D}\bar{D}}e^{-2V}D_\alpha e^{2V} = -\frac{1}{4}\bracketss{\bar{D}\bar{D}}D_\alpha V + \frac{1}{4}\bracketss{\bar{D}\bar{D}}[V,D_\alpha V]\,, \\
   \bar{W}_{\dot{\alpha}} \equiv& -\frac{1}{8} \bracketaa{DD}e^{-2V}\bar{D}_{\dot{\alpha}}e^{2V} = -\frac{1}{4} \bracketaa{DD}\bar{D}_{\dot{\alpha}}V +\frac{1}{4} \bracketaa{DD}  [V,\bar{D}_{\dot{\alpha}}V ]\,,
\end{align}
where the superderivatives are defined as 
\begin{align}
    D_\alpha = \partial_\alpha + i|\sigma^\mu \bar{\theta}]_\alpha \partial_\mu ,\qquad \bar{D}_{\dot{\alpha}} = -\bar{\partial}_{\dot{\alpha}} - i\langle\theta\sigma^\mu|_{\dot{\alpha}} \partial_\mu\,.
\end{align}
The components of $W_\alpha$, in terms of chiral coordinates, are 
\begin{align}
    W_\alpha = |\lambda(y)\rangle_\alpha +|\theta\rangle_\alpha D(y) + |\sigma^{\mu\nu}\theta\rangle_\alpha F_{\mu\nu}(y) -  i\bracketaa{\theta\theta} |\sigma^\mu D_\mu \bar{\lambda}(y) ]_\alpha\,,
\end{align}
and the component of $\bar{W}_{\dot{\alpha}}$ can be obtained by conjugation.

\section{The $\Ncal=2$ Lagrangian of the $\SU(N_f)$ theory}\label{derive N=2}
In this appendix, we give more details of the $\Ncal=2$ Lagrangian of the $\SU(N_f)$ theory studied in this paper. We begin by first writing down the $\mathcal{N}=4$ SYM and the fundamental $\mathcal{N}=2$ hypermultiplets in the $\mathcal{N}=1$ language.

The $\Ncal=2$ vector multiplet contains an $\Ncal=1$ vector multiplet and an $\Ncal=1$ adjoint chiral multiplet, of which the physical degrees of freedom are a 
gauge boson, two Weyl fermions and a complex scalar. The $\Ncal=2$ supersymmetry can be obtained by demanding that the theory is invriant under the $\SU(2)$ which rotates the two Weyl fermions. The Lagrangian of $\Ncal=2$ super-Yang-Mills reads
\begin{align}
    \mathcal{L}^{\mathcal{N}=2}_{\rm YM} =&\ \left(\frac{{\rm Im}\, \tau}{8\pi} \int \dd^2 \theta \tr W^\alpha W_\alpha\ +\  \text{c.c.}\right) +  \frac{2}{g_{\rm YM}^2}\int \dd^2 \theta \dd^2\bar{\theta}\ \tr \Phi^\dagger e^{2[V,\, \cdot\, ]} \Phi\,,
\end{align}
where $ \tau = \frac{\mathcal{\vartheta}}{2\pi}+\frac{4\pi i}{g_{\rm YM}^2}$ is the complexified Yang-Mills coupling. The $\Ncal=2$ hypermultiplet contains an $\Ncal=1$ chiral multiplet $\Phi_n$ and an $\Ncal=1$ antichiral multiplet $\Tilde{\Phi}_n^\dagger$.\footnote{This $n$ labels different hypermultiplets, which we will use later.} Its Lagranian is given by
\begin{align}
    \mathcal{L}^{\mathcal{N}=2}_{{\rm hyper},n} =&\ \int\! \dd^2\theta\dd^2\bar{\theta}\!\left[ \Phi_n^\dagger e^{2V} \Phi_n +  \Tilde{\Phi}_n e^{-2V} \Tilde{\Phi}_n^{\dagger}\right] -i\! \int\! \dd^2\theta \sqrt{2}\Tilde{\Phi}_n \Phi \Phi_n  +i\! \int\! \dd^2\bar{\theta}\sqrt{2}\Phi_n^\dagger \Phi^\dagger \Tilde{\Phi}_n^{\dagger}\, .
\end{align}
The Lagrangian of $\Ncal=4$ SYM can be constructed by coupling the $\Ncal=2$ SYM with an $\Ncal=2$ hypermultiplet in the adjoint representation, or equivalently, coupling $\mathcal{N}=1$ SYM with three adjoint hypermultiplets. The physical fields are six scalar $\phi^I = \sqrt{2}\, (\Re \phi_1, \Im \phi_1,$ $\Re \phi_2, \Im \phi_2,\Re \phi_3, \Im \phi_3)$, four fermions $\lambda^i = (\lambda,\psi_1,\psi_2,\psi_3)$ and one gauge field $A_\mu$. The $\SO(6)$ vector $\phi^I$ can be written in an $\SU(4)$ manner via the 6d gamma matrices \cref{gamma}. In terms of the $\Ncal=1$ superfields, we can write the Lagrangian of $\Ncal=4$ SYM as
\begin{align}
    \mathcal{L}^{\mathcal{N}=4} =&\  \left(\frac{{\rm Im}\, \tau}{8\pi} \int \dd^2 \theta \tr W^\alpha W_\alpha\ +\  \text{c.c.}\right) +  \frac{2}{g_{\rm YM}^2}\int \dd^2 \theta \dd^2\bar{\theta}\ \tr \Phi_{1,2,3}^\dagger e^{2[V,\, \cdot\, ]} \Phi_{1,2,3}\nonumber \\&-i\! \int\! \dd^2\theta \sqrt{2}\tr \Phi_3\, [\Phi_1, \Phi_2]  +i\! \int\! \dd^2\bar{\theta}\sqrt{2}\tr \Phi_2^\dagger\, [\Phi_1^\dagger, \Phi_3^\dagger\,]\,. 
\end{align}

Finally, to get the $\SU(N_f)$ theory we couple $\mathcal{N}=4$ SYM to $N_f$ fundamental $\mathcal{N}=2$ hypermultiplets. Let us denote the $\Ncal=2$ hypermultiplet by
$H^n = (q^{n,a},\psi^n,\bar{\chi}^n)$ where $n=1,2,...,N_f$, $a=1,2$. The hypermultiplet contains the chiral fermion $\psi^n$, the anti-chiral fermion $\bar{\chi}^n$ and the complex scalars $q^{n,a}$.  The Lagrangian of the theory is
\begin{align}
    \mathcal{L} = \mathcal{L}^{\mathcal{N}=4} + \sum_{n=1}^{N_f}\mathcal{L}^{\mathcal{N}=2}_{{\rm hyper},n}\,.
\end{align}
Writing it explicitly in terms of the component fields and integrate out all the auxiliary fields, we get
\begin{align}
    \mathcal{L} =&\ \frac{2N}{\lambda}\left\{ \tr\left[\frac{1}{4}F^{\mu\nu}F_{\mu\nu} + \frac{1}{2}D^\mu \phi^I D_\mu \phi^I  + \bar{\lambda}_i \bar{\sigma}^\mu D_\mu \lambda^i + \sqrt{2} \lambda^i \phi_{ij}  \lambda^j + \sqrt{2} \bar{\lambda}_i \phi^{ij} \bar{\lambda}_j - \frac{1}{4}{\tr}[\phi^I,\phi^J]^2 \right] \right. \nonumber\\[1mm]
    &+ D^{\mu}\bar{q}_{n,a} D_\mu q^{n,a} + \bar{\psi}_n \bar{\sigma}^\mu D_\mu \psi^n + \bar{\chi}^n \bar{\sigma}^\mu D_\mu \chi_n \nonumber\\[2mm]
    &+i\sqrt{2}(-\chi_n \varphi \psi^n +\bar{\psi}_n \varphi^\dagger \bar{\chi}^n - \bar{q}_{n,a} \lambda^a \psi^n +  \bar{\psi}_n \bar{\lambda}_a q^{n,a} + \epsilon^{ab}\bar{q}_{n,a}\bar{\lambda}_b \bar{\chi}^n + \epsilon_{ab} \chi_n\lambda^a q^{n,b} ) \nonumber\\
    &+\bar{q}_{n,a} \phi^{\bar{A}}\phi^{\bar{A}} q^{n,a} + \bar{q}_{n,a} [\phi^{a\bar{a}},\phi_{\bar{a}b}]q^{n,b} + \bar{q}_{m,a} q^{n,a}\ \bar{q}_{n,b} q^{m,b} - \frac{1}{2}\bar{q}_{m,a} q^{n,b}\ \bar{q}_{n,b} q^{m,a} \nonumber\\
    & \left. - \frac{1}{N}\left(\bar{q}_{m,a} q^{m,b}\ \bar{q}_{n,b} q^{n,a} - \frac{1}{2} \bar{q}_{m,a} q^{m,a}\ \bar{q}_{n,b} q^{n,b} \right)\right\}.
\end{align}
with
$ \lambda \equiv g_{\rm YM}^2 N$.

\bibliography{refs}

\providecommand{\href}[2]{#2}\begingroup\raggedright\begin{thebibliography}{10}

\bibitem{Rastelli:2016nze}
L.~Rastelli and X.~Zhou, ``{Mellin amplitudes for $AdS_5\times S^5$},''
  \href{http://dx.doi.org/10.1103/PhysRevLett.118.091602}{{\em Phys. Rev.
  Lett.} {\bfseries 118} no.~9, (2017) 091602},
  \href{http://arxiv.org/abs/1608.06624}{{\ttfamily arXiv:1608.06624
  [hep-th]}}.

\bibitem{Rastelli:2017udc}
L.~Rastelli and X.~Zhou, ``{How to Succeed at Holographic Correlators Without
  Really Trying},'' \href{http://dx.doi.org/10.1007/JHEP04(2018)014}{{\em JHEP}
  {\bfseries 04} (2018) 014}, \href{http://arxiv.org/abs/1710.05923}{{\ttfamily
  arXiv:1710.05923 [hep-th]}}.

\bibitem{Rastelli:2019gtj}
L.~Rastelli, K.~Roumpedakis, and X.~Zhou, ``{$\mathbf{AdS_3\times S^3}$
  Tree-Level Correlators: Hidden Six-Dimensional Conformal Symmetry},''
  \href{http://dx.doi.org/10.1007/JHEP10(2019)140}{{\em JHEP} {\bfseries 10}
  (2019) 140}, \href{http://arxiv.org/abs/1905.11983}{{\ttfamily
  arXiv:1905.11983 [hep-th]}}.

\bibitem{Alday:2020lbp}
L.~F. Alday and X.~Zhou, ``{All Tree-Level Correlators for M-theory on $AdS_7
  \times S^4$},'' \href{http://dx.doi.org/10.1103/PhysRevLett.125.131604}{{\em
  Phys. Rev. Lett.} {\bfseries 125} no.~13, (2020) 131604},
  \href{http://arxiv.org/abs/2006.06653}{{\ttfamily arXiv:2006.06653
  [hep-th]}}.

\bibitem{Alday:2020dtb}
L.~F. Alday and X.~Zhou, ``{All Holographic Four-Point Functions in All
  Maximally Supersymmetric CFTs},''
  \href{http://dx.doi.org/10.1103/PhysRevX.11.011056}{{\em Phys. Rev. X}
  {\bfseries 11} no.~1, (2021) 011056},
  \href{http://arxiv.org/abs/2006.12505}{{\ttfamily arXiv:2006.12505
  [hep-th]}}.

\bibitem{Alday:2021odx}
L.~F. Alday, C.~Behan, P.~Ferrero, and X.~Zhou, ``{Gluon Scattering in AdS from
  CFT},'' \href{http://dx.doi.org/10.1007/JHEP06(2021)020}{{\em JHEP}
  {\bfseries 06} (2021) 020}, \href{http://arxiv.org/abs/2103.15830}{{\ttfamily
  arXiv:2103.15830 [hep-th]}}.

\bibitem{Aprile:2017bgs}
F.~Aprile, J.~M. Drummond, P.~Heslop, and H.~Paul, ``{Quantum Gravity from
  Conformal Field Theory},''
  \href{http://dx.doi.org/10.1007/JHEP01(2018)035}{{\em JHEP} {\bfseries 01}
  (2018) 035}, \href{http://arxiv.org/abs/1706.02822}{{\ttfamily
  arXiv:1706.02822 [hep-th]}}.

\bibitem{Aprile:2017xsp}
F.~Aprile, J.~M. Drummond, P.~Heslop, and H.~Paul, ``{Unmixing Supergravity},''
  \href{http://dx.doi.org/10.1007/JHEP02(2018)133}{{\em JHEP} {\bfseries 02}
  (2018) 133}, \href{http://arxiv.org/abs/1706.08456}{{\ttfamily
  arXiv:1706.08456 [hep-th]}}.

\bibitem{Alday:2018kkw}
L.~F. Alday, ``{On genus-one string amplitudes on $AdS_5 \times S^5$},''
  \href{http://dx.doi.org/10.1007/JHEP04(2021)005}{{\em JHEP} {\bfseries 04}
  (2021) 005}, \href{http://arxiv.org/abs/1812.11783}{{\ttfamily
  arXiv:1812.11783 [hep-th]}}.

\bibitem{Aprile:2019rep}
F.~Aprile, J.~Drummond, P.~Heslop, and H.~Paul, ``{One-loop amplitudes in
  $AdS_5 \times S^5$ supergravity from $ \mathcal{N} $ = 4 SYM at strong
  coupling},'' \href{http://dx.doi.org/10.1007/JHEP03(2020)190}{{\em JHEP}
  {\bfseries 03} (2020) 190}, \href{http://arxiv.org/abs/1912.01047}{{\ttfamily
  arXiv:1912.01047 [hep-th]}}.

\bibitem{Alday:2019nin}
L.~F. Alday and X.~Zhou, ``{Simplicity of AdS Supergravity at One Loop},''
  \href{http://dx.doi.org/10.1007/JHEP09(2020)008}{{\em JHEP} {\bfseries 09}
  (2020) 008}, \href{http://arxiv.org/abs/1912.02663}{{\ttfamily
  arXiv:1912.02663 [hep-th]}}.

\bibitem{Alday:2021ajh}
L.~F. Alday, A.~Bissi, and X.~Zhou, ``{One-loop gluon amplitudes in AdS},''
  \href{http://dx.doi.org/10.1007/JHEP02(2022)105}{{\em JHEP} {\bfseries 02}
  (2022) 105}, \href{http://arxiv.org/abs/2110.09861}{{\ttfamily
  arXiv:2110.09861 [hep-th]}}.

\bibitem{Huang:2021xws}
Z.~Huang and E.~Y. Yuan, ``{Graviton scattering in AdS$_{5}$\texttimes{}
  S$^{5}$ at two loops},''
  \href{http://dx.doi.org/10.1007/JHEP04(2023)064}{{\em JHEP} {\bfseries 04}
  (2023) 064}, \href{http://arxiv.org/abs/2112.15174}{{\ttfamily
  arXiv:2112.15174 [hep-th]}}.

\bibitem{Drummond:2022dxw}
J.~M. Drummond and H.~Paul, ``{Two-loop supergravity on $AdS_{5}\times S^{5}$
  from CFT},'' \href{http://dx.doi.org/10.1007/JHEP08(2022)275}{{\em JHEP}
  {\bfseries 08} (2022) 275}, \href{http://arxiv.org/abs/2204.01829}{{\ttfamily
  arXiv:2204.01829 [hep-th]}}.

\bibitem{Goncalves:2019znr}
V.~Gon{\c c}alves, R.~Pereira, and X.~Zhou, ``{$20'$ Five-Point Function from
  $AdS_5\times S^5$ Supergravity},''
  \href{http://dx.doi.org/10.1007/JHEP10(2019)247}{{\em JHEP} {\bfseries 10}
  (2019) 247}, \href{http://arxiv.org/abs/1906.05305}{{\ttfamily
  arXiv:1906.05305 [hep-th]}}.

\bibitem{Alday:2022lkk}
L.~F. Alday, V.~Gon\c{c}alves, and X.~Zhou, ``{Supersymmetric Five-Point Gluon
  Amplitudes in AdS Space},''
  \href{http://dx.doi.org/10.1103/PhysRevLett.128.161601}{{\em Phys. Rev.
  Lett.} {\bfseries 128} no.~16, (2022) 161601},
  \href{http://arxiv.org/abs/2201.04422}{{\ttfamily arXiv:2201.04422
  [hep-th]}}.

\bibitem{Goncalves:2023oyx}
V.~Gon\c{c}alves, C.~Meneghelli, R.~Pereira, J.~Vilas~Boas, and X.~Zhou,
  ``{Kaluza-Klein five-point functions from AdS$_{5}$\texttimes{}S$^{5}$
  supergravity},'' \href{http://dx.doi.org/10.1007/JHEP08(2023)067}{{\em JHEP}
  {\bfseries 08} (2023) 067}, \href{http://arxiv.org/abs/2302.01896}{{\ttfamily
  arXiv:2302.01896 [hep-th]}}.

\bibitem{Alday:2023kfm}
L.~F. Alday, V.~Gon\c{c}alves, M.~Nocchi, and X.~Zhou, ``{Six-point AdS gluon
  amplitudes from flat space and factorization},''
  \href{http://dx.doi.org/10.1103/PhysRevResearch.6.L012041}{{\em Phys. Rev.
  Res.} {\bfseries 6} no.~1, (2024) L012041},
  \href{http://arxiv.org/abs/2307.06884}{{\ttfamily arXiv:2307.06884
  [hep-th]}}.

\bibitem{Cao:2023cwa}
Q.~Cao, S.~He, and Y.~Tang, ``{Constructibility of AdS Supergluon
  Amplitudes},'' \href{http://dx.doi.org/10.1103/PhysRevLett.133.021605}{{\em
  Phys. Rev. Lett.} {\bfseries 133} no.~2, (2024) 021605},
  \href{http://arxiv.org/abs/2312.15484}{{\ttfamily arXiv:2312.15484
  [hep-th]}}.

\bibitem{Cao:2024bky}
Q.~Cao, S.~He, X.~Li, and Y.~Tang, ``{Supergluon scattering in AdS:
  constructibility, spinning amplitudes, and new structures},''
  \href{http://arxiv.org/abs/2406.08538}{{\ttfamily arXiv:2406.08538
  [hep-th]}}.

\bibitem{Huang:2024dxr}
Z.~Huang, B.~Wang, E.~Y. Yuan, and J.~Zhang, ``{All Five-point Kaluza-Klein
  Correlators and Hidden 8d Symmetry in $\rm AdS_5\times S^3$},''
  \href{http://arxiv.org/abs/2408.12260}{{\ttfamily arXiv:2408.12260
  [hep-th]}}.

\bibitem{Bissi:2022mrs}
A.~Bissi, A.~Sinha, and X.~Zhou, ``{Selected topics in analytic conformal
  bootstrap: A guided journey},''
  \href{http://dx.doi.org/10.1016/j.physrep.2022.09.004}{{\em Phys. Rept.}
  {\bfseries 991} (2022) 1--89},
  \href{http://arxiv.org/abs/2202.08475}{{\ttfamily arXiv:2202.08475
  [hep-th]}}.

\bibitem{Caron-Huot:2018kta}
S.~Caron-Huot and A.-K. Trinh, ``{All Tree-Level Correlators in
  AdS${}_5\times$S${}_5$ Supergravity: Hidden Ten-Dimensional Conformal
  Symmetry},'' \href{http://dx.doi.org/10.1007/JHEP01(2019)196}{{\em JHEP}
  {\bfseries 01} (2019) 196}, \href{http://arxiv.org/abs/1809.09173}{{\ttfamily
  arXiv:1809.09173 [hep-th]}}.

\bibitem{Zhou:2021gnu}
X.~Zhou, ``{Double Copy Relation in AdS Space},''
  \href{http://dx.doi.org/10.1103/PhysRevLett.127.141601}{{\em Phys. Rev.
  Lett.} {\bfseries 127} no.~14, (2021) 141601},
  \href{http://arxiv.org/abs/2106.07651}{{\ttfamily arXiv:2106.07651
  [hep-th]}}.

\bibitem{Zhou:2020ptb}
X.~Zhou, ``{How to Succeed at Witten Diagram Recursions without Really
  Trying},'' \href{http://dx.doi.org/10.1007/JHEP08(2020)077}{{\em JHEP}
  {\bfseries 08} (2020) 077}, \href{http://arxiv.org/abs/2005.03031}{{\ttfamily
  arXiv:2005.03031 [hep-th]}}.

\bibitem{Behan:2021pzk}
C.~Behan, P.~Ferrero, and X.~Zhou, ``{More on holographic correlators: Twisted
  and dimensionally reduced structures},''
  \href{http://dx.doi.org/10.1007/JHEP04(2021)008}{{\em JHEP} {\bfseries 04}
  (2021) 008}, \href{http://arxiv.org/abs/2101.04114}{{\ttfamily
  arXiv:2101.04114 [hep-th]}}.

\bibitem{Caron-Huot:2021usw}
S.~Caron-Huot and F.~Coronado, ``{Ten dimensional symmetry of $ \mathcal{N} $ =
  4 SYM correlators},'' \href{http://dx.doi.org/10.1007/JHEP03(2022)151}{{\em
  JHEP} {\bfseries 03} (2022) 151},
  \href{http://arxiv.org/abs/2106.03892}{{\ttfamily arXiv:2106.03892
  [hep-th]}}.

\bibitem{Intriligator:1998ig}
K.~A. Intriligator, ``{Bonus symmetries of N=4 superYang-Mills correlation
  functions via AdS duality},''
  \href{http://dx.doi.org/10.1016/S0550-3213(99)00242-4}{{\em Nucl. Phys. B}
  {\bfseries 551} (1999) 575--600},
  \href{http://arxiv.org/abs/hep-th/9811047}{{\ttfamily arXiv:hep-th/9811047}}.

\bibitem{Eden:2011we}
B.~Eden, P.~Heslop, G.~P. Korchemsky, and E.~Sokatchev, ``{Hidden symmetry of
  four-point correlation functions and amplitudes in N=4 SYM},''
  \href{http://dx.doi.org/10.1016/j.nuclphysb.2012.04.007}{{\em Nucl. Phys. B}
  {\bfseries 862} (2012) 193--231},
  \href{http://arxiv.org/abs/1108.3557}{{\ttfamily arXiv:1108.3557 [hep-th]}}.

\bibitem{Karch:2002sh}
A.~Karch and E.~Katz, ``{Adding flavor to AdS / CFT},''
  \href{http://dx.doi.org/10.1088/1126-6708/2002/06/043}{{\em JHEP} {\bfseries
  06} (2002) 043}, \href{http://arxiv.org/abs/hep-th/0205236}{{\ttfamily
  arXiv:hep-th/0205236}}.

\bibitem{Fayyazuddin:1998fb}
A.~Fayyazuddin and M.~Spalinski, ``{Large N superconformal gauge theories and
  supergravity orientifolds},''
  \href{http://dx.doi.org/10.1016/S0550-3213(98)00545-8}{{\em Nucl. Phys. B}
  {\bfseries 535} (1998) 219--232},
  \href{http://arxiv.org/abs/hep-th/9805096}{{\ttfamily arXiv:hep-th/9805096}}.

\bibitem{Aharony:1998xz}
O.~Aharony, A.~Fayyazuddin, and J.~M. Maldacena, ``{The Large N limit of N=2,
  N=1 field theories from three-branes in F theory},''
  \href{http://dx.doi.org/10.1088/1126-6708/1998/07/013}{{\em JHEP} {\bfseries
  07} (1998) 013}, \href{http://arxiv.org/abs/hep-th/9806159}{{\ttfamily
  arXiv:hep-th/9806159}}.

\bibitem{Behan:2024vwg}
C.~Behan, S.~M. Chester, and P.~Ferrero, ``{Towards Bootstrapping F-theory},''
  \href{http://arxiv.org/abs/2403.17049}{{\ttfamily arXiv:2403.17049
  [hep-th]}}.

\bibitem{Behan:2023fqq}
C.~Behan, S.~M. Chester, and P.~Ferrero, ``{Gluon scattering in AdS at finite
  string coupling from localization},''
  \href{http://dx.doi.org/10.1007/JHEP02(2024)042}{{\em JHEP} {\bfseries 02}
  (2024) 042}, \href{http://arxiv.org/abs/2305.01016}{{\ttfamily
  arXiv:2305.01016 [hep-th]}}.

\bibitem{Alday:2024yax}
L.~F. Alday, S.~M. Chester, T.~Hansen, and D.-l. Zhong, ``{The AdS Veneziano
  amplitude at small curvature},''
  \href{http://dx.doi.org/10.1007/JHEP05(2024)322}{{\em JHEP} {\bfseries 05}
  (2024) 322}, \href{http://arxiv.org/abs/2403.13877}{{\ttfamily
  arXiv:2403.13877 [hep-th]}}.

\bibitem{Drukker:2008pi}
N.~Drukker and J.~Plefka, ``{The Structure of n-point functions of chiral
  primary operators in N=4 super Yang-Mills at one-loop},''
  \href{http://dx.doi.org/10.1088/1126-6708/2009/04/001}{{\em JHEP} {\bfseries
  04} (2009) 001}, \href{http://arxiv.org/abs/0812.3341}{{\ttfamily
  arXiv:0812.3341 [hep-th]}}.

\bibitem{Beem:2013sza}
C.~Beem, M.~Lemos, P.~Liendo, W.~Peelaers, L.~Rastelli, and B.~C. van Rees,
  ``{Infinite Chiral Symmetry in Four Dimensions},''
  \href{http://dx.doi.org/10.1007/s00220-014-2272-x}{{\em Commun. Math. Phys.}
  {\bfseries 336} no.~3, (2015) 1359--1433},
  \href{http://arxiv.org/abs/1312.5344}{{\ttfamily arXiv:1312.5344 [hep-th]}}.

\bibitem{Nirschl:2004pa}
M.~Nirschl and H.~Osborn, ``{Superconformal Ward identities and their
  solution},'' \href{http://dx.doi.org/10.1016/j.nuclphysb.2005.01.013}{{\em
  Nucl. Phys. B} {\bfseries 711} (2005) 409--479},
  \href{http://arxiv.org/abs/hep-th/0407060}{{\ttfamily arXiv:hep-th/0407060}}.

\bibitem{Beisert:2002bb}
N.~Beisert, C.~Kristjansen, J.~Plefka, G.~W. Semenoff, and M.~Staudacher,
  ``{BMN correlators and operator mixing in N=4 superYang-Mills theory},''
  \href{http://dx.doi.org/10.1016/S0550-3213(02)01025-8}{{\em Nucl. Phys. B}
  {\bfseries 650} (2003) 125--161},
  \href{http://arxiv.org/abs/hep-th/0208178}{{\ttfamily arXiv:hep-th/0208178}}.

\bibitem{Usyukina:1992jd}
N.~I. Usyukina and A.~I. Davydychev, ``{An Approach to the evaluation of three
  and four point ladder diagrams},''
  \href{http://dx.doi.org/10.1016/0370-2693(93)91834-A}{{\em Phys. Lett. B}
  {\bfseries 298} (1993) 363--370}.

\bibitem{Duhr:2019tlz}
C.~Duhr and F.~Dulat, ``{PolyLogTools \textemdash{} polylogs for the masses},''
  \href{http://dx.doi.org/10.1007/JHEP08(2019)135}{{\em JHEP} {\bfseries 08}
  (2019) 135}, \href{http://arxiv.org/abs/1904.07279}{{\ttfamily
  arXiv:1904.07279 [hep-th]}}.

\bibitem{Bedford:2007qj}
J.~Bedford, C.~Papageorgakis, and K.~Zoubos, ``{Twistor Strings with
  Flavour},'' \href{http://dx.doi.org/10.1088/1126-6708/2007/11/088}{{\em JHEP}
  {\bfseries 11} (2007) 088}, \href{http://arxiv.org/abs/0708.1248}{{\ttfamily
  arXiv:0708.1248 [hep-th]}}.

\bibitem{Huang:2023oxf}
Z.~Huang, B.~Wang, E.~Y. Yuan, and X.~Zhou, ``{AdS super gluon scattering up to
  two loops: a position space approach},''
  \href{http://dx.doi.org/10.1007/JHEP07(2023)053}{{\em JHEP} {\bfseries 07}
  (2023) 053}, \href{http://arxiv.org/abs/2301.13240}{{\ttfamily
  arXiv:2301.13240 [hep-th]}}.

\bibitem{Huang:2023ppy}
Z.~Huang, B.~Wang, E.~Y. Yuan, and X.~Zhou, ``{Simplicity of AdS super
  Yang-Mills at one loop},''
  \href{http://dx.doi.org/10.1007/JHEP01(2024)190}{{\em JHEP} {\bfseries 01}
  (2024) 190}, \href{http://arxiv.org/abs/2309.14413}{{\ttfamily
  arXiv:2309.14413 [hep-th]}}.

\bibitem{Huang:2024rxr}
Z.~Huang, B.~Wang, and E.~Y. Yuan, ``{All Next-Next-to-Extremal One-Loop
  Correlators of AdS Supergluons and Supergravitons},''
  \href{http://arxiv.org/abs/2407.03408}{{\ttfamily arXiv:2407.03408
  [hep-th]}}.

\end{thebibliography}\endgroup
\bibliographystyle{utphys}

\end{document}